\documentclass[11pt,a4paper]{article} 
\pdfoutput=1
\usepackage{jheppub}
\usepackage{enumerate}
\usepackage{epsfig} 
\usepackage{float} 
\usepackage[caption = false]{subfig}
\usepackage{wasysym} 
\usepackage{mathrsfs} 
\usepackage{amsfonts} 
\usepackage{amsbsy} 
\usepackage{amscd} 
\DeclareUnicodeCharacter{200B}{{\hskip 0pt}}
\DeclareUnicodeCharacter{2212}{{\hskip 0pt}}
\usepackage{multirow} 
\usepackage{tikz}
\usepackage{color}
\usepackage{slashed}
\usepackage{array}
\usepackage{tablefootnote}
\usetikzlibrary{arrows,positioning,shapes.geometric} 
\usepackage[compat=1.1.0]{tikz-feynman}          
\usepackage[font=small,labelfont=bf]{caption}
\usepackage{slashed}
\usepackage{multirow}
\usepackage{tabularx}
\usepackage{soul}  
\usepackage{listings} 
\usepackage{cancel}
\usepackage{orcidlink}           
\usepackage{color}

\usepackage{amsthm}

\title{Collider fingerprints of freeze-in dark matter produced during the fast expansion phase of Universe}

\author[a,b]{Anupam Ghosh\,\orcidlink{0000-0003-4163-4491},}
\author[a]{Partha Konar\,\orcidlink{0000-0001-8796-1688},}
\author[c]{Sudipta Show\,\orcidlink{0000-0003-0436-6483} }

\affiliation[a]{Theoretical Physics Division, Physical Research Laboratory, Ahmedabad, 380009, Gujarat, India}
\affiliation[b]{Department of Physics, Indian Institute of Technology Guwahati, North Guwahati, 781039, Assam, India}
\affiliation[c]{Department of Physics, Indian Institute of Technology Kanpur, Kanpur, 208016, UP, India}

\emailAdd{anupamg@rnd.iitg.ac.in}
\emailAdd{konar@prl.res.in}
\emailAdd{sudiptas@iitk.ac.in}

\abstract{
We examine a simple dark sector extension where the observed dark matter (DM) abundance arises from a freeze-in process through the decay of heavy vector-like quarks into a scalar dark matter candidate. The detection prospects of such DM are challenging due to the feeble nature of the interactions, but these vector-like quarks can be produced copiously at the LHC, where they decay to Standard Model quarks along with DM. Depending on the decay rate, this scenario is typically probed through long-lived particle or displaced vertex signatures, assuming a radiation-dominated cosmological background. An alternative hypothesis suggests that the Universe may have experienced a rapid expansion phase instead of the standard radiation-dominated one during freeze-in. This would significantly alter the dark matter phenomenology, requiring a substantial increase in the interaction rate to match the observed relic density, resulting in the rapid decay of the parent particle. As a result, much of the parameter space for this scenario is beyond the reach of traditional long-lived particle and displaced vertex searches. Due to this non-standard cosmic evolution, existing constraints do not cover the expanded dark matter parameter space. We propose a complementary search strategy to explore this scenario, offering additional limits alongside searches for long-lived particles and displaced vertices. In our search, we investigate the feebly interacting massive particle (FIMP) dark matter model at the LHC using boosted fatjets and significant missing transverse momentum. To improve precision, we include one-loop QCD corrections for LHC production processes and employ a boosted decision tree multivariate analysis, leveraging jet substructure variables to explore a vast parameter space for this minimally extended FIMP dark matter model at the 14 TeV LHC.
}
\preprint{\today}

\keywords{Dark Matter, Non Standard Cosmology,  LHC, Vector-like quark}
\begin{document}
\maketitle
\flushbottom

\section{Introduction}   \label{Intro}

Extensive studies with astrophysical and cosmological observations \cite{Sofue:2000jx, Clowe:2006eq, Hinshaw_2013, Planck:2018vyg} have jointly established the existence of a yet unidentified non-luminous form of matter (aka dark matter) as a dominant matter component of the Universe that exerts its presence over a vast range of scales, from galaxies to the cosmological length. In particular, the observation of cosmic microwave background (CMB) has provided a crucial input for studying the properties and evolution of dark matter by quantifying its contribution around 27$\%$ of the total energy budget of the Universe, featuring the relic density, $\Omega h^2=0.12$ \cite{Planck:2018vyg, Drees:2018hzm}.  
Among numerous proposals on particle candidates of dark matter, weakly interacting dark matter particle (WIMP) \cite{Kolb:1990vq,Arcadi:2017kky,Ghosh:2021noq,Ghosh:2024boo,Ghosh:2022rta, Ghosh:2023xhs,Green:2002ht,Bae:2014rfa,Chang:2017gla,Ghosh:2025agw,Srivastava:2025oer,Visinelli:2017qga,Arcadi:2017wqi,Reinert:2017aga, Evans:2017kti,Garny:2018icg,Blanco:2019hah,Bhardwaj:2018lma,Bhardwaj:2019mts,Konar:2020wvl,Konar:2020vuu, Heurtier:2019beu,Habermehl:2020njb,Xing:2021pkb,Borah:2022byb,Belanger:2022qxt,Bernal:2022wck,Kundu:2021cmo,Medina:2021ram,Tallman:2022nts,Kang:2022zqv,Dutta:2022wdi, Bernal:2023ura, Bernal:2024yhu,Silva-Malpartida:2024emu} snatches most of the attention due to its rich phenomenology and search prospects at various frontiers owing to its simple production mechanism and also miraculous weak scale interactions with the standard model (SM) particles. Unfortunately, after decades of rigorous searches using direct \cite{Akerib:2016vxi, Cui:2017nnn, Zhang:2018xdp, Aprile:2018dbl} and indirect \cite{MAGIC:2016xys} detection experiments, as well as collider experiments \cite{Chatrchyan:2012xdj, Aad:2012tfa, ATLAS:2020kdi}, the elusive nature of dark matter persists. Despite their apparent ubiquity, the fundamental properties and evolution mechanism of dark matter particles still remain obscure. However, such null results significantly constrained different DM models, including a favourable range of electro-weak mass scale and the coupling of the dark matter.
That resulted in vigorous scientific efforts to look for possible dark matter signatures outside the electro-weak mass scale and investigate different unconventional dark matter scenarios where the production mechanism supports a minuscule SM interaction, unlike the WIMP paradigm.

One such exciting alternative dark matter candidate is a feebly interacting massive particle (FIMP) \cite{Hall:2009bx, Co:2015pka, Hessler:2016kwm, Ghosh:2017vhe, No:2019gvl, Konar:2021oye, Ghosh:2021wrk, DeRomeri:2020wng, Kim:2017mtc, Kim:2018xsp, Im:2019iwd, Hambye:2018, Banerjee_2019, Dutra:2018gmv, Chu:2011be, Shakya:2015xnx, Coy:2022unt, Chakrabarty:2022bcn, Cheung:2011mg, Medina:2014bga, Bernal:2017kxu, Chowdhury:2023jft, Banerjee:2024caa, Barman:2020plp, Barman:2020plp, Barman:2022njh, Barman:2024lxy, Freese:2024ogj, Sakurai:2024apm, Barman:2024nhr, Barman:2024tjt}, which naturally answers the non-observation of any signature at experimental grounds owing to its tiny interaction with the visible particles. Due to its feeble interaction, FIMP has never been in thermal contact with the SM particles, and the scattering and decay of some particles at the thermal bath populates the FIMP gradually via the so-called freeze-in mechanism. Early cosmological history is crucial since such dark matter production occurs in the early Universe. Typically, studies of such dark matter computation assume a radiation-dominated background, although there is no direct evidence of the energy budget prior to the Big Bang nucleosynthesis (BBN). It is worth investigating the scenario without such particular prejudice that radiation dominated the early Universe. It is even more critical since such an assumption not only changes the evolution of dark matter at early times but also modifies its imprint on our present experiments and how one would scrutinize and constrain such DM models.

Recently, studies of dark matter phenomenology in the presence of non-trivial evolution of cosmological background have become the domain of growing interest \cite{Heurtier:2019beu, Konar:2020vuu, Bernal:2022wck, Bernal:2023ura, Bernal:2024yhu, Silva-Malpartida:2024emu, Ghosh:2021wrk, Banerjee_2019, Das:2023owa}.
One such exciting non-standard cosmological picture is a fast expansion \cite{DEramo:2017gpl, DEramo:2017ecx} where the energy budget is dominated via a species $\phi$ with $\rho_\phi\propto a^{-(4+n)}$ provided $n>0$ and $a(t)$ being the scale factor, redshifting faster than the radiation. Although $\phi$ drives the early Universe's expansion, the subsequent radiation-dominated background can be realized naturally since the energy density of $\phi$ dilutes faster than radiation. Several works have investigated the effect of fast expansion on freeze-in production of dark matter; see, for example, Refs. \cite{DEramo:2017ecx, Ghosh:2021wrk, Das:2023owa}.

One of the main difficulties of freeze-in dark matter is its detection prospects due to the feeble nature of the interaction with SM. However, it is possible to probe freeze-in dark matter at a collider if the DM is directly produced from the decay of a heavy dark sector mediator that can substantially interact with SM. This Yukawa-type interaction involves a heavy particle beyond the standard model (BSM), an SM particle, and a DM. Such dark matter particles can be as light as keV, but one must remember that it can not be lower than 12 keV, and this strong lower bound comes from the Lyman-$\alpha$ constraint \cite{Viel:2013fqw, Yeche:2017upn, Irsic:2017ixq}. Here, the heavy mediator can be produced abundantly at the collider and then decays further to SM particles and DM through a feeble interaction.

Interestingly, the same decay process also enters the freeze-in dark matter production mechanism and helps to realize the desired relic density of DM. Hence, this picture has an exciting natural connection between the collider signature and the cosmological background. For a heavy mediator mass of around a few hundred GeV, signatures with long-lived particle (LLP) searches or displaced vertex (DV) provide us with an excellent avenue to probe this picture. At least, that was the setup in almost all of the previous works while considering DM evolution in a radiation-dominated background. As mentioned earlier, this picture is bound to change if the freeze-in occurs in a fast-expanding Universe. In that case, such modified cosmology demands a larger interaction strength than a conventional radiation-dominated case, resulting in the prompt decay of heavy mediators and poor performance by LLP or DV searches at the LHC.

In this work, we study the impact of non-standard cosmology on dark matter production and demonstrate how to probe such DM at colliders in the context of the simplified hadro-philic model. Here, dark matter generation occurs via the vector-like heavy quark where alternative cosmology mandates a considerable interaction strength than usual to satisfy the relic density constraints. Naturally, the probing methodology at the collider also changes since prompt decay occurs instead of offering a DV signature. 
The hadro-philic model is the minimal extension of the SM and contains a single vector-like quark (VLQ) that mediates the interaction between the FIMP dark matter and the SM quarks. VLQ is an isospin singlet, but depending on its hypercharge, it can have either $2/3e$ or $-1/3e$ electromagnetic charge. The present study offered a collider analysis of this model for $2/3e$ charge VLQ, a pair of which can be copiously produced at the LHC. Interestingly, this study probes the model entirely independent of the BSM couplings \footnote{Model independence is ensured since the decay of the VLQ is prompt owing to the impact of non-standard cosmology, as considered in this work. Hence, the projected constraint from this analysis is expected to be insensitive to the BSM coupling, as can be followed in the final result. This is contrary to the complementary searches through LLP or DV, which are most effective when the BSM coupling between the VLQ with the SM quark and dark matter is minuscule, producing displaced decay VLQs at a specific range of detectors. \label{footnote_1}} and solely depends on the QCD coupling constant ($\alpha_S$) and the mass of VLQ. After pair production, each VLQ promptly decays into a top quark along with the dark matter. The resulting products from top quarks in the final state can form top-like boosted fatjets because of the significant mass difference between VLQ and DM. We utilize jet substructure variables to exploit the internal structure within the fatjets that help to segregate three-pronged top fatjets from the vast SM backgrounds, generating two boosted fatjets with a significant missing transverse momentum (MET). Furthermore, our investigations incorporate one-loop QCD corrections of the VLQ pair production processes for an accurate event rate and differential distributions. We leverage an advanced multivariate analysis, employing a boosted decision tree to optimize our event selection, further enhancing signal and background discrimination.

This article is designed as follows. In Section \ref{sec2}, we discuss the simplified dark matter model. Section \ref{sec3} describes a brief sketch of the non-standard cosmological picture. Dark matter phenomenology and the impact of modified cosmology are further demonstrated in Section \ref{sec4}. Sections \ref{sec5} and \ref{sec6} are devoted to the proposed collider analysis and the current and expected reach of the LHC based on our analysis. Finally, we conclude in Section \ref{sec7}.

\section{The Model}\label{sec2}
\begin{table}[tb!]
	\begin{center}
		\begin{tabular}{|c|c|c|c|c|}
			\hline
			& $H$ & $S$ & $\Psi$   \\
			\hline\hline
			$SU(3)_C$ & 1 & 1& 3  \\  
			\hline
			$SU(2)_L$ & 2 & 1& 1 \\  
			\hline
			$U(1)_Y$ & 1/2 & 0& 2/3 \\    
			\hline
			$\mathcal{Z}_2$  & Even & Odd & Odd \\    
			\hline
		\end{tabular} 
		\caption{Charge assignments for some relevant fields under different symmetry groups.}
		\label{tab_particles}
	\end{center}
\end{table}

We extend the SM by a real singlet scalar dark matter candidate $S$ of mass $m_S$, neutral under SM symmetry group $SU(3)_c\times SU(2)_L\times U(1)_Y$, along with a vector-like up-type quark $\Psi$ with mass $m_{\Psi}$, an $SU(2)$ singlet. The dark sector particles transform odd under the discrete $\mathcal{Z}_2$ symmetry, while the SM particles transform trivially under it. The unbroken $\mathcal{Z}_2$ symmetry, not only forbids the mixing between heavy VLQ with SM quarks or $S$ with SM Higgs doublet ($H$), but also it ensures the stability of DM provided $m_S<m_{\Psi}$. Charge assignments for the fields relevant to our study under different symmetry groups are displayed in Table \ref{tab_particles}. In this minimal freeze-in dark matter scenario, DM interacts with the SM particles through the Yukawa-type interaction involving a left-handed component of the vector-like quark and a right-handed up-type quark. The Lagrangian of the scalar can be written as
\begin{align}
\mathcal{L}=|D^\mu H|^2+\frac{1}{2}(\partial^\mu S)^2-V(H, S)
\label{model_eq1}
\end{align}
with,
\begin{align}
D^\mu=\partial^\mu-ig\frac{\sigma^a}{2}W^{a \mu}-ig^\prime Y B^\mu
\label{model_eq2}
\end{align}
where, $g$ and $g^\prime$ refer to the $SU(2)_L$ and $U(1)_Y$ gauge couplings, respectively. General form of the scalar potential can be given as,
\begin{align}
V(H, S)=-\mu_H^2(H^\dagger H)+\lambda_H(H^\dagger H)^2+\frac{\mu_S^2}{2}S^2+\frac{\lambda_S}{4}S^4+\lambda_{SH}S^2(H^\dagger H),
\label{model_eq3}
\end{align}
In the limit, $\mu_H^2, \mu_S^2 >0$, the vacuum expectation value (vev) of both the scalars can be obtained by minimizing the scalar potential as,
\begin{align}
\langle H\rangle=v,~~~~\langle S\rangle=0,
\label{model_eq4}
\end{align}
Note that, the parameter $\mu_S$ is related to the DM mass as followed from Eq.~(\ref{model_eq3}),
\begin{align}
m_S^2=\mu_S^2+\lambda_{SH} v^2.
\label{model_eq5}
\end{align}
The Lagrangian for the BSM fermion is given by
\begin{align}
\mathcal{L}_f=i\bar\Psi\gamma_\mu D^\mu\Psi-m_\Psi\bar\Psi\Psi-\sum_f\tilde y_f (S\bar\Psi P_R f+ \text{h.c.}),
\label{model_eq6}
\end{align}
where, $f=\{u,c,t\}$, $P_R(=\frac{1+\gamma_5}{2})$ is the right handed projection operator and, $\tilde y_f$ represents the Yukawa coupling. Similar simplified models were considered in the Refs. \cite{Belanger:2018sti, Garny:2018icg, Das:2023owa}

In this work, we are interested in dark matter production through freeze-in processes via the BSM Yukawa interaction. To simplify our demonstration, we set the Higgs portal interaction strength $\lambda_{SH}=0$ that prohibits the additional channels. The self-coupling, $\lambda_S$, is also irrelevant to our study. We only consider the top quark interaction in the rest of our analysis\footnote{Such a setup naturally leads to minimal flavour violation, and these effects are subdued as a result of the smallness of the freeze-in coupling. \label{footnote_2}}. In this case, the model is termed as top-philic model. So, the relevant parameters for the dark matter phenomenology are, 
$\{m_S, m_\Psi, \tilde y_t\}.$

\section{Fast Expanding Universe}    \label{sec3}
The most favoured concept of the origin and evolution of our Universe is encapsulated in the form of the standard model of cosmology that assumes the radiation component dominated the energy budget of the early Universe. However, cosmological observations have yet to rule out the possibility that some other species have dominated the energy density of the early Universe instead of radiation  before the onset of Big Bang Nucleosynthesis (BBN). The new energy component can redshift faster (or slower) than the radiation energy density $\rho_R$, which evolves with  $\rho_R\propto a(t)^{-4}$, $a(t)$ being the scale factor. Several prior works have demonstrated the impact of such alternative cosmology on DM phenomenology \cite{DEramo:2017ecx, Ghosh:2021wrk, Das:2023owa}.

Here, we consider a scenario where a new species, $\phi$, dominantly contributes to the total energy density along with the radiation component of the Universe. It is important to mention that we assume that $\phi$ starts dominating the energy budgets of the early Universe once the inflationary reheating gets over\footnote{We have assumed that the reheating temperature is much larger than all the mass scales involved in our theory. It is essential to mention that in the case of early matter domination, the matter field has to decay to start radiation domination. However, in the case of fast expansion, radiation domination comes naturally. As a result, entropy injection occurs in an early matter domination scenario, which is absent in the expanding universe case. \label{footnote_3}}. The energy density of the field $\phi$ redshifts faster than radiation as $\rho_\phi\propto a(t)^{-(4+n)}$ with the equation of state $\omega=\frac{1}{3}(n+1)$ provided $n>0$. Note that, $\omega=\frac{1}{3}$ for $n=0$, recreates the behaviour like a radiation. In the presence of $\phi$, the expansion in the early Universe is driven jointly by both the energy components, so the modified Hubble rate can be expressed by using the Friedmann equation as
\begin{equation}
H^2=\frac{\rho_R+\rho_\phi}{3 M_P^2},
\label{FE_1}
\end{equation}
where, $M_P(=2.4\times 10^{18}~\text{GeV})$ refers to the reduced Planck mass. Now, the total energy density of the Universe as a function of temperature, can be described as \cite{DEramo:2017gpl}
\begin{align}
\rho(T)=\rho_R(T)+\rho_\phi(T)
=\rho_R(T)\bigg[1+\frac{g_*(T_r)}{g_*(T)}\bigg(\frac{g_{*s}(T)}{g_{*s}(T_r)}\bigg)^{(4+n)/3}\bigg(\frac{T}{T_r}\bigg)^n\bigg],
\label{FE_2}
\end{align}
where, $\rho_R(T)=\frac{\pi^2}{30}g_*(T)T^4$, $T_r$ represents the temperature at which the fast expansion phase ends and, after which the Universe enters to the radiation dominated era. $g_*(T)$ and, $g_{*s}(T)$ refer to the relativistic degrees of freedom of energy and, entropy respectively. In this case, the modified Hubble rate can be approximated as \cite{DEramo:2017gpl}
\begin{equation}
H(T) \simeq \frac{\pi \sqrt{g_*}}{3\sqrt{10}}\frac{T^2}{M_P}\bigg(\frac{T}{T_r}\bigg)^{n/2},~~~~~~~~(T\gg T_r).
\label{FE_3}
\end{equation}

It is clear from the expression that the Hubble rate at any point of the fast expansion stage is always larger than the same for the radiation-dominated background. Interestingly, the impact of non-standard cosmology can be understood by the simple parametrization of two parameters $n$ and $T_r$. Increasing $n$ (or decreasing $T_r$) points towards an even faster expansion of the Universe. Parametrizing the effect of the extra species $\phi$ by the effective number of relativistic degrees of freedom and then considering the observational bound on the same, one can obtain a bound on the reference temperature as $T_r\ge(15.4)^{1/n}$ MeV. In addition, the temperature smaller than BBN temperature ($T_{\text{BBN}}\simeq~4$ MeV) is in tension with the cosmological observation, and it also puts a stringent lower bound on $T_r$. 

The scenario with $0<n\le2$ can be reached by introducing a positive scalar potential, whereas a negative potential is associated with $n>2$. The $n=2$ case is particularly interesting, referring to the well-motivated quintessence fluid \cite{DEramo:2017gpl}. This regime is also known as the kination era as kinetic energy dominates over potential energy where the energy density falls as $\rho_\phi\propto a(t)^{-6}$. Throughout our present analysis, we primarily focus on the dark matter production during the kination phase (i.e., $\omega=1$) to showcase the DM evolution and collider signature exploring the dark sector.

\section{Dark Matter Phenomenology}\label{sec4}

In our present setup, the vector-like quark ($\Psi$) remains part of the primordial thermal bath in addition to the SM particles owing to its gauge interaction with them. However, the dark matter particle $S$ always remains out of thermal equilibrium due to the tiny Yukawa interaction ($\tilde y_t\ll 1$), which connects it with the SM quarks via the vector-like fermion $\Psi$. Hence, starting from zero or negligibly small initial number density of DM in the early Universe, it gradually populates from the decay and scattering of the bath particles. The freeze-in process attains the final DM number density. However, that essentially depends on the background of cosmology in the early Universe. In particular, it depends on the parameters $n$ and $T_r$ in the case of an expanding early Universe.

\begin{figure}[tb!]
	\centering
	\includegraphics[height=6.3cm,width=11cm]{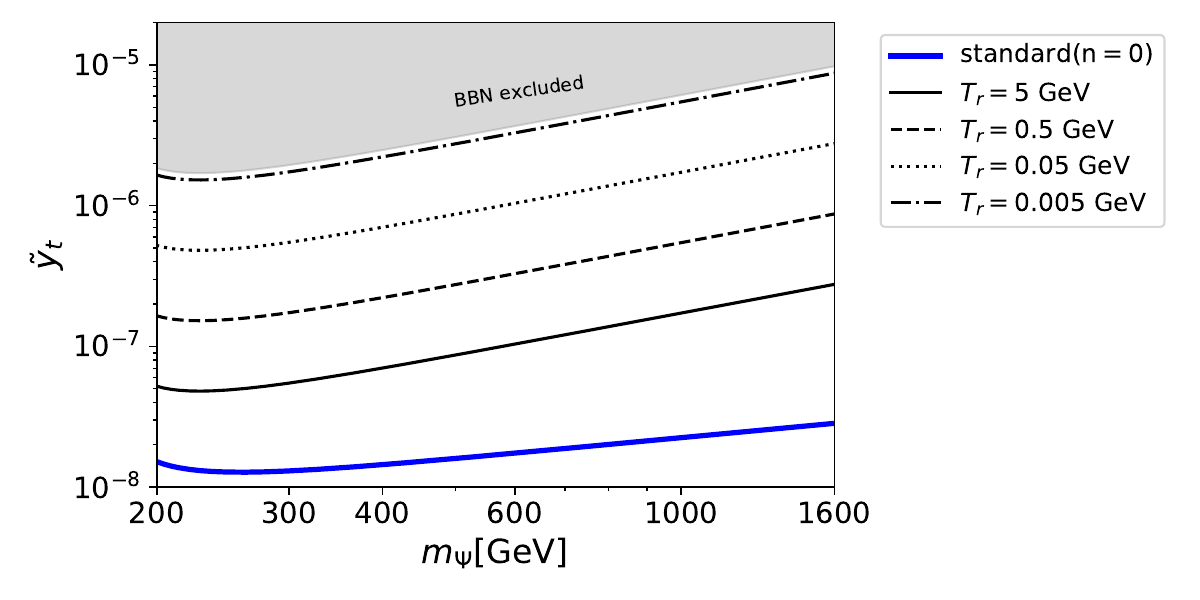}
	\caption{
	In different early Universe cosmological evolution scenarios, contours that satisfy the dark matter relic density constraint $\Omega h^2= 0.12$ are shown in the Yukawa coupling $(\tilde{y}_t)$ and vector-like quark mass $m_\Psi$ plane for the freeze-in dark matter model. The thick blue solid line refers to the standard radiation-dominated (or n=0) background picture. In contrast, the thin (solid, dashed, dotted, and dash-dotted) lines correspond to different values of the non-standard cosmological parameter $T_r\,(=5,0.5,0.05,0.005$ GeV respectively) for a fixed $n\,(=2)$. The dark matter mass is set at $m_S = 12$ keV.
	}
	\label{coupling_mass}
\end{figure}

DM production can occur via the scatterings ($t\bar t\rightarrow SS,~ \Psi \Psi\rightarrow SS$) and\footnote{The scattering via Higgs mediation can also contribute, but for simplicity we do not consider this contribution by setting the Higgs portal coupling ($\lambda_{SH}$) to zero. \label{footnote_4}} the decay $\Psi \rightarrow tS$. Note that the scattering cross-sections and decay rate are proportional to $\tilde y_t^4$ and $\tilde y_t^2$, respectively. For $\tilde y_t\ll 1$, the scattering remains subdominant compared to the decay, so the DM production mainly proceeds via the decay of $\Psi$. The following Boltzmann equation governs the freeze-in production of DM through the decay process,
\begin{equation}
\frac{dn}{dt}+3Hn=\frac{g_\Psi^{} m_\Psi^2 \Gamma_\Psi}{2\pi^2}TK_1\bigg(\frac{m_\Psi}{T}\bigg).
\label{DM_phen1}
\end{equation}
Decay width of vector-like fermion, 
\begin{equation}
\Gamma_\Psi=\frac{\tilde y_t^2}{32\pi}\frac{(m_\Psi^2+m_t^2)(m_\Psi^2-m_t^2)}{m_\Psi^3},
\label{DM_phen2}
\end{equation}
and, $n(t)$ represents the number density of DM, $g_\Psi^{}$ is the internal degrees of freedom of $\Psi$ and $K_1(..)$ refers to the modified Bessel function of first kind. Rewriting the DM abundance in terms of $Y=n/s$, $s$ being the entropy density, and the dimensionless variable $x=m/T$, one can recast the Equation~(\ref{DM_phen1}) in the following form
\begin{equation}
\frac{dY}{dx}=\frac{\langle\Gamma_\Psi\rangle}{Hx}Y_\Psi^{\text{eq}}(x),
\label{DM_phen3}
\end{equation}
where, $Y_\Psi^{\text{eq}}(x)$ denotes equilibrium abundance of the vector-like fermion. $\langle\Gamma_\Psi\rangle$  is the thermally averaged decay rate, expressed as $\Gamma_\Psi\frac{K_1(x)}{K_2(x)}$;  $K_2(x)$ being the modified Bessel function of second kind. Now, one can obtain the relic density after evaluating the final freeze-in DM abundance by solving Eq.~(\ref{DM_phen3}) numerically and plugging it in the following equation
\begin{equation}
\Omega h^2=2.755\times10^8\times m_S\times Y(x\rightarrow\infty),
\label{DM_phen4}
\end{equation} 
In addition, one can also calculate the relic density of DM by using the following expression, deduced by solving Eq.~(\ref{DM_phen3}) analytically,
\begin{equation}
\Omega h^2\simeq \frac{2.755*10^8*45}{4\pi^4*1.66} \times \frac{M_P g_\Psi^{} \Gamma_\Psi T_r^{\frac{n}{2}}}{\sqrt{g_*} g_{*s}} \times \frac{m_S*2^{2+\frac{n}{2}}}{m_\Psi^{2+\frac{n}{2}}} ~ \Gamma\bigg(\frac{6+n}{4}\bigg) ~ \Gamma\bigg(\frac{10+n}{4}\bigg).
\label{DM_phen5}
\end{equation}
$g_*$ and $g_{*s}$ are the effective number of relativistic degrees of freedom contributed to the energy and entropy density, respectively. Here, $\Gamma(..)$ denotes the standard notation for the gamma function. In Figure \ref{coupling_mass}, we show the relic density satisfied contours for the radiation-dominated (standard, $n=0$) and modified cosmology with different values of $T_r$ for a fixed $n=2$. Here, the thick blue solid line represents a radiation-dominated scenario and $T_r=5, 0.5,0.05,0.005$ GeV contours are denoted by the thin solid, dashed, dotted, and dot-dashed lines, respectively. The grey-shaded region is ruled out since the temperature $T_r$, if lower than the BBN temperature ($\simeq$ 4 MeV) conflicts with the cosmological observation. These contours clearly demonstrate the requirement for a larger coupling for a fixed mass as one decreases the value of $T_r$. Lowering $T_r$ means that expansion will be faster, necessitating larger interaction, i.e., larger coupling to compensate for this effect and satisfy the relic density. It is imperative to know whether the DM is still out of equilibrium for such a large coupling. One can check that the coupling required to reach thermal equilibrium with the thermal bath is almost two orders of magnitude larger than the same to satisfy correct order relic density, and it is further discussed in detail in Appendix~(\ref{upper_bound}).

Before discussing the specifics of the collider searches, we elucidate a few representative benchmark points, as outlined in Table \ref{tab:Bemchmark}. These benchmarks satisfy all relevant constraints, including the correct relic density for the permitted temperatures ($T_r$). As already discussed, lower $T_r$ implies a faster Universe expansion rate, as per Eq.~(\ref{FE_3}). Corresponding decay lengths of the VLQ are provided in the final column of Table \ref{tab:Bemchmark}. For a clear comparison, one reference benchmark point (BP0) is also added considering the standard radiation-dominated (or n=0) background, which clearly demonstrates the usefulness of a probe with a long-live particle or displaced vertex signature. However, such a recipe is ineffective for the remaining BPs with alternative cosmology ($n > 0$), which is our primary motivation for finding alternative signatures in this study.

\begin{table}[tb!]
\begin{center}
 \scriptsize
 \begin{tabular}{|c|c|c|c|c|c|c|c|c|}
\hline
Benchmark & \multirow{2}{*}{$\tilde{y}_t$} & $m_\Psi$ & $m_S$ & \multirow{2}{*}{n} & $T_r$ & \multirow{2}{*}{$\Omega_S h^2$} & $\Gamma_\Psi$ & $c\tau_\Psi$  \\ 
points & & [GeV] & [KeV] &  & [GeV] &  & [GeV]  & [mm] \\
\hline\hline
\multirow{2}{*}{BP0} & \multirow{2}{*}{$2.25 \times 10^{-8}$} & \multirow{2}{*}{1000} & \multirow{2}{*}{12} & \multirow{2}{*}{0} & \multirow{2}{*}{--} & \multirow{2}{*}{0.12} & \multirow{2}{*}{$5.03\times 10^{-15}$} & \multirow{2}{*}{39.35}  \\
 &  &  &  &  & &  &  &     \\
\hline\hline
\multirow{2}{*}{BP1} & \multirow{2}{*}{$1.0 \times 10^{-6}$} & \multirow{2}{*}{800} & \multirow{2}{*}{12} & \multirow{2}{*}{2} & \multirow{2}{*}{0.095} & \multirow{2}{*}{0.12} & \multirow{2}{*}{$7.94\times 10^{-12}$} & \multirow{2}{*}{0.025}  \\
 &  &  &  &  & &  &  &     \\
\hline
\multirow{2}{*}{BP2} & \multirow{2}{*}{$3.0 \times 10^{-7}$} & \multirow{2}{*}{1080} & \multirow{2}{*}{12} & \multirow{2}{*}{2} & \multirow{2}{*}{1.929} & \multirow{2}{*}{0.12} & \multirow{2}{*}{$9.66\times 10^{-13}$} & \multirow{2}{*}{0.205}  \\
 &  &  &  &  & &  &  &     \\
\hline
\multirow{2}{*}{BP3} & \multirow{2}{*}{$5.0 \times 10^{-7}$} & \multirow{2}{*}{1500} & \multirow{2}{*}{12} & \multirow{2}{*}{2} & \multirow{2}{*}{1.339} & \multirow{2}{*}{0.12} & \multirow{2}{*}{$3.73\times 10^{-12}$} & \multirow{2}{*}{0.053}  \\
 &  &  &  &  & &  &  &     \\
\hline
 \end{tabular} 
\caption{
A list of benchmark points (BP1 - BP3) is presented with the corresponding parameters chosen for our study on the freeze-in dark matter model under the influence of non-standard cosmology. For an explicit comparison, one additional reference benchmark point (BP0) is added for the standard radiation-dominated (or $n=0$) background. Note the remarkably tiny coupling needed in BP0 to satisfy the correct relic density and demonstrate the usefulness of a probe with a long-live particle or displaced vertex signature in this case. But that is ineffective for the rest of the BPs with alternative cosmology ($n > 0$), and we propose an alternative search strategy in the present work.
}
\label{tab:Bemchmark}
\end{center}
\end{table}

\section{Probing Dark Sector at the LHC}  \label{sec5}

\begin{figure}[tb!]
\centering
\includegraphics[scale=0.32]{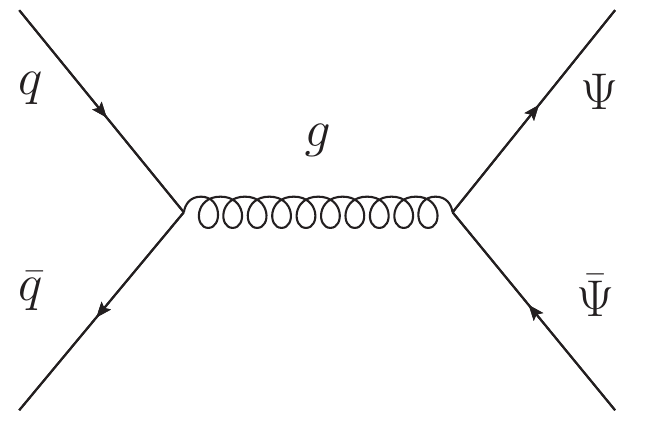}
\includegraphics[scale=0.32]{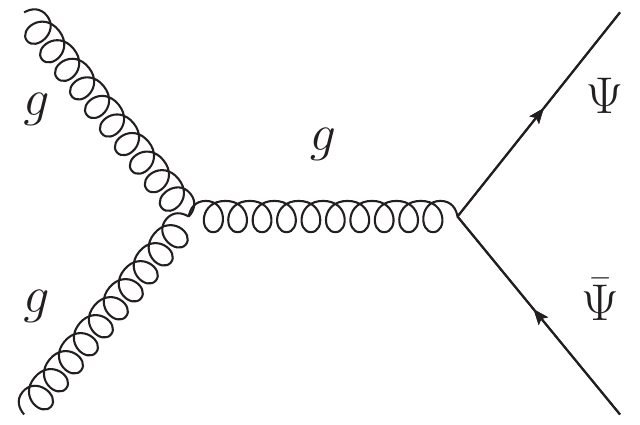}
\includegraphics[scale=0.32]{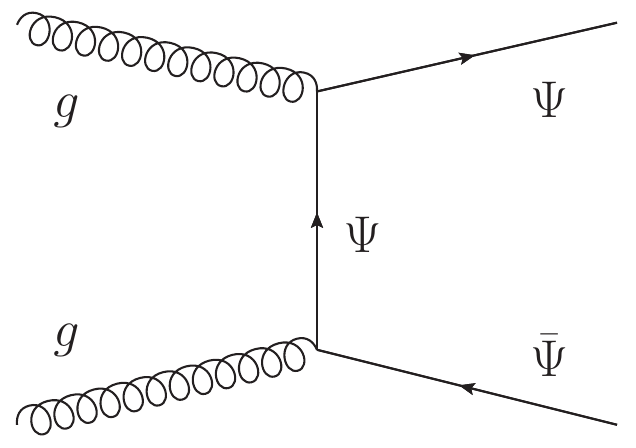}\\
\vspace{0.4cm}
\includegraphics[scale=0.32]{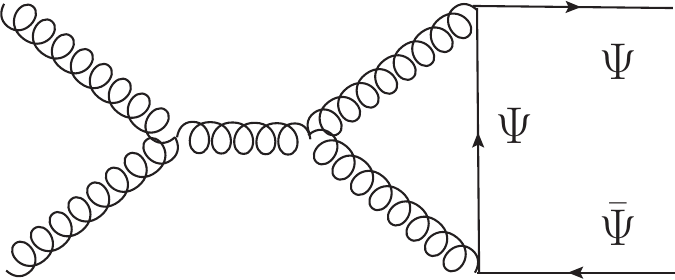}
\includegraphics[scale=0.32]{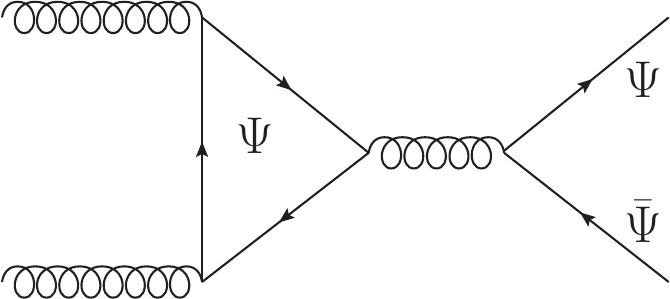}
\includegraphics[scale=0.32]{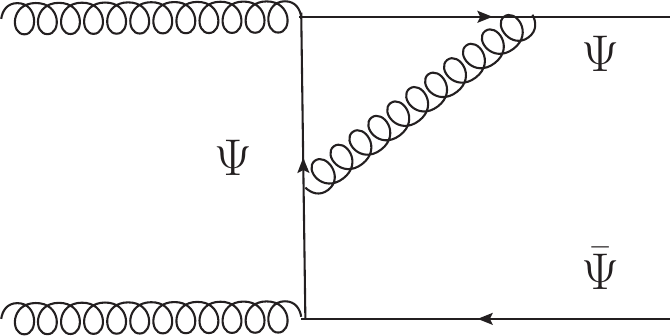}
\includegraphics[scale=0.32]{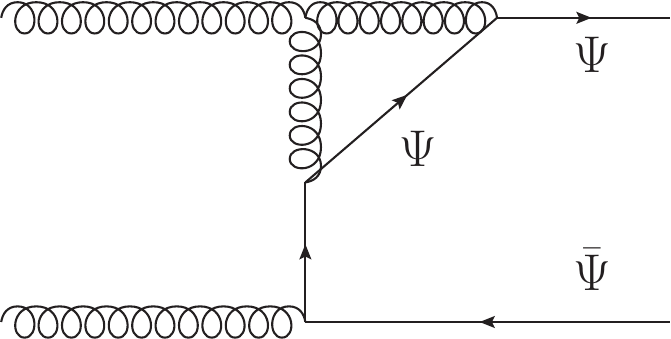}\\
\vspace{0.2cm}
\includegraphics[scale=0.32]{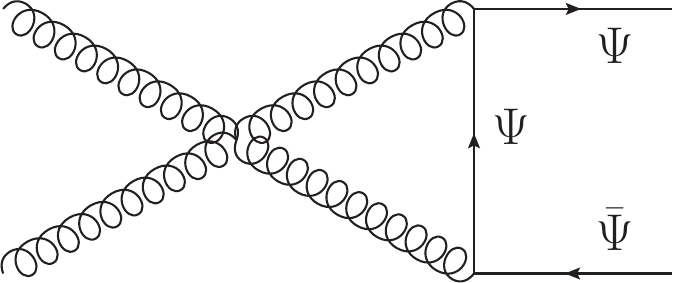}
\includegraphics[scale=0.32]{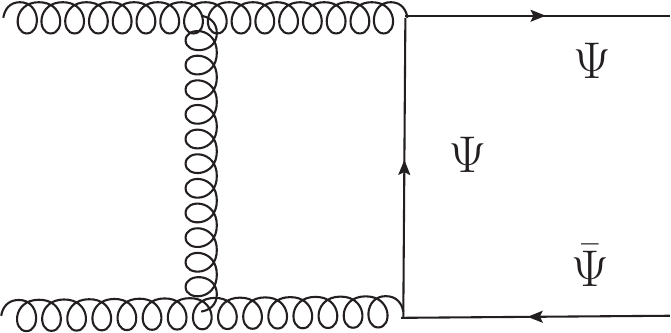}
\includegraphics[scale=0.32]{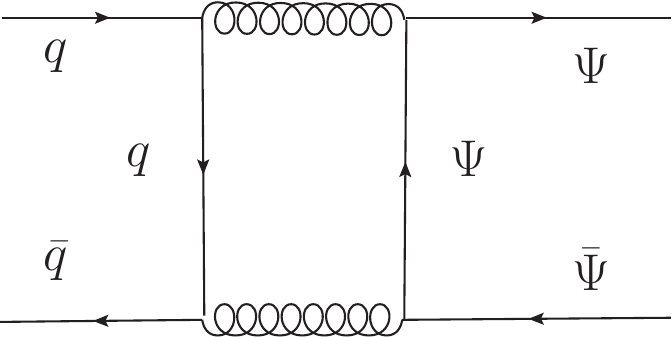}
\includegraphics[scale=0.32]{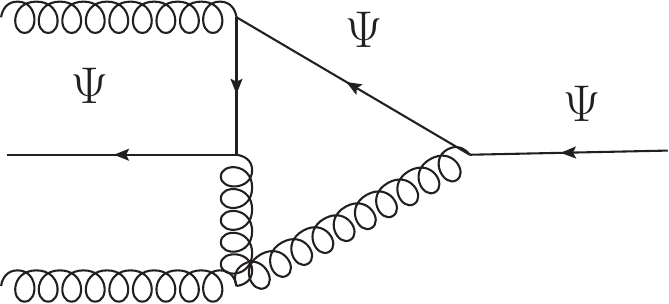}
\caption{
The upper row displays the parton level leading-order Feynman diagrams for VLQ pair production. In our study, we consider one loop correction in $\alpha_S$ of all these diagrams, and some of the representative virtual diagrams are shown in the lower two rows.
}
\label{Feynman_dia}
\end{figure}

The presence of a Vector-Like Quark (VLQ) and its stronger Yukawa interaction with the FIMP candidate opens up intriguing opportunities for exploring the dark sector through an alternative approach at the collider. The coloured $2/3e$ electromagnetic charged VLQ, in principle, can interact with all up-type standard model quarks and dark matter via Yukawa interactions. As a $\mathcal{Z}_2$-odd particle, the VLQ does not mix with SM quarks. We focus on a scenario where the VLQ couples predominantly to the top quark, while its interactions with lighter up- and charm quarks are either absent or negligibly small. Due to this top-philic nature, VLQ production at the LHC occurs primarily through pair production, making their searches at the LHC largely model-independent\footnote{Please note in case of the VLQ couples to the first two generation SM quarks, a DM-mediated t-channel process via quark-antiquark annihilation can also contribute to VLQ pair production, consequently imparting the BSM model dependence in the production cross-section. \label{footnote_5}}. The production of VLQ with dark matter at the LHC is not feasible in this top-philic scenario\footnote{If the VLQ has non-zero interaction with light SM quarks, then associated production of VLQ with the DM is possible. The production cross-section is $\sigma(p p \rightarrow \overset{ \scalebox{0.4}{$(\mkern-3mu-\mkern-3mu)$}}{\Psi} S)_{\text{LO}}\sim \mathcal{O}(\alpha_S~ \tilde y_t^2)$. Given that $\tilde y_t<<1$, the contribution of associated production remains negligible compared to the dominant pair-production cross-section of the VLQ. \label{footnote_6}}. The leading-order (LO) production channels, illustrated in the top row of Figure \ref{Feynman_dia}, depend solely on the strong coupling constant ($\alpha_S$) and the mass of the VLQ. 

Hence, displaced vertex searches at the LHC have been a primary approach for probing FIMP dark matter models. However, in our alternative cosmological scenario, the requirement for a comparatively larger Yukawa coupling reduces the decay length of the VLQ. The decay length in all sample benchmark points is less than a centimetre, indicating prompt decay. This suggests that, for a significant mass gap between the VLQ and dark matter, prompt searches for FIMP dark matter models at the LHC are feasible under this modified cosmology. Such searches could offer better sensitivity compared to displaced vertex searches. To further improve the reach of our analysis, we will employ jet substructure techniques and advanced multivariate analysis methods to enhance the signal.

\begin{figure}[tb!]
\centering
\includegraphics[scale=0.62]{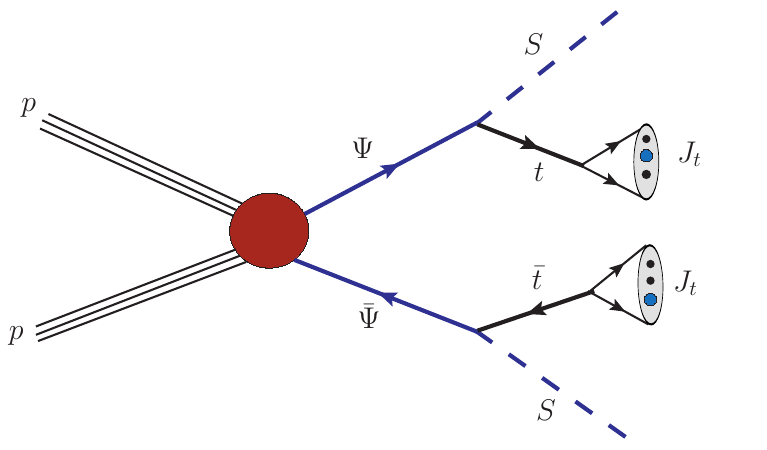}
\caption{
A diagrammatic representation of two top-like fatjets and MET signature through pair production of vector-like quarks at the LHC. Many a times one or both of these fatjets can be mimicked by the QCD radiation, especially for the background processes.
}
\label{FJ_diag}
\end{figure}

To investigate this model at the LHC, we focus on producing the VLQ pair. After production, each VLQ decays into a top quark and a dark matter particle. Consequently, the final state consists of two top quarks and a pair of dark matter candidates. The signal topology is as follows.
\begin{equation}
p p \rightarrow \Psi \bar{\Psi}\rightarrow (t,S), (\bar{t},S) \equiv 2J_t +\text{MET}
\end{equation}
The non-interacting DM particles do not leave any detectable traces in the detector, contributing to some measurable signature termed missing transverse momentum (also customarily labelled as MET) from the momentum imbalance of all the visible products at the detector. If the mass gap between the VLQ and the DM is substantial, produced top quarks from the decay of heavy VLQs are highly boosted, causing their hadronic outcomes to form large-radius, three-prong fatjets. Thus, the final signature includes two such top-like fatjets and significant MET \footnote{Interested readers can find that the same final state can be realized in other BSM scenarios, such as searches for heavy scalar leptoquark \cite{Ghosh:2023ocz,Ghosh:2025gue}. A comprehensive collider analysis, combined with polarization variables, is utilized to exclude some model parameters and distinguish between different scalar leptoquark models. \label{footnote_7}} (see, Figure \ref{FJ_diag} for a diagrammatic representation). The leading-order (LO) cross-section for VLQ pair production is presented in Table \ref{tab:crosssection}. To improve the precision of our estimate, we also account for one-loop QCD corrections to the VLQ pair production processes, providing more accurate cross-sections and differential distributions of various kinematic observables. Some representative virtual one-loop QCD-corrected Feynman diagrams are shown in the last two rows of Figure \ref{Feynman_dia}.

The total next-to-leading-order (NLO) QCD cross-section is also provided in Table \ref{tab:crosssection} for the central choices of renormalization ($\mu_R$) and factorization ($\mu_F$) scales. $\mu_R$ and $\mu_F$ are fictitious scales that appear through higher-order calculations and parton distribution functions. The central choice is set at the dynamical partonic centre-of-mass energy scale of the process, {\it, i.e.} $\mu_R = \mu_F = \sqrt{\hat{s}}$. The scales are further varied as $\mu_R= \zeta_1 \sqrt{\hat{s}}$ and $\mu_F= \zeta_2 \sqrt{\hat{s}}$, with $\zeta_1$ and $\zeta_2$ taking values from the set $\{1/2, 1, 2\}$, generating a total of nine data sets around the central scale. The superscripts and subscripts in the cross-sections represent the envelope of these nine data sets, denoting the theoretical scale uncertainty associated with each cross-section. As shown in Table \ref{tab:crosssection}, the NLO cross-section is substantially higher than the LO prediction, and the scale uncertainties are significantly reduced by implementing the NLO calculation. The K-factor, defined as the NLO to LO cross-section ratio, is also provided in Table \ref{tab:crosssection}.

\subsection{Simulation Details and Backgrounds}
\label{setup}
We implement the extended standard model, as represented in different terms of the Lagrangian in Eqs.~\ref{model_eq1} and \ref{model_eq6} in {\sc FeynRules} \cite{Alloul:2013bka} and use the {\sc NLOCT} \cite{Degrande:2014vpa} package to generate the {\sc UV} and {\sc $R_2$} counter-terms and generate the NLO UFO model file. The signal and background events are generated utilizing the {\sc MadGraph5\_aMC@NLO} \cite{Alwall:2014hca} environment. Our analysis includes all background processes that can mimic the signal. Background events are generated with two to four additional QCD jets using the {\sc MLM} matching \cite{Mangano:2006rw, Hoeche:2005vzu} scheme. All standard model backgrounds are further scaled using available higher-order QCD cross-sections. The generated parton level events are further passed through {\sc Pythia8} \cite{Sjostrand:2001yu, Sjostrand:2014zea} for parton showering, fragmentation, and hadronization. The \emph{NNPDF3.0} \cite{NNPDF:2014otw} LO and NLO PDF sets are utilized to generate leading-order and next-to-leading-order events, respectively. These showered events are also processed through {\sc Delphes3} \cite{deFavereau:2013fsa} to simulate detector effects. Jets with a radius parameter 0.4 are constructed using the $\mbox{anti-k}_T$ \cite{Cacciari:2008gp} clustering algorithm. Fatjets are formed using Delphes towers with a radius of 1.5, clustered with the $\mbox{anti-k}_T$ algorithm. The radius parameter, R, of a fatjet scales approximately as $R=\dfrac{2M_J}{p_T(J)}$, where $M_J$ is the fatjet mass (top quark mass in our case), and $p_T(J)$ is its transverse momentum. {\sc Fastjet 3.2.2} \cite{Cacciari:2011ma} is used for clustering, requiring each fatjet to have a minimum transverse momentum of $p_T \geq 200$ GeV. Subsequently, we further recluster the constituents of each fatjet using the Cambridge-Aachen (C/A) algorithm \cite{Dokshitzer:1997in} to perform pairwise clustering for the soft drop procedure \cite{Larkoski:2014wba}. Soft drop grooming eliminates soft and wide-angle unassociated constituents of the fatjets. Finally, we utilize the adaptive Boosted Decision Tree (BDT) algorithm \cite{Roe:2004na, FREUND1995256, Freund:1997xna} for multivariate analysis (MVA) within the {\sc TMVA} framework \cite{Hocker:2007ht}.

\begin{table}[tb!]
\begin{center}
\scriptsize
 \begin{tabular}{|c|c|p{0.2\textwidth}|p{0.2\textwidth}|c|}
\hline
\multirow{2}{1em}{} &  $m_\Psi$  & \multicolumn{3}{c|}{$\sigma(p p \rightarrow \Psi \bar{\Psi})$ (fb)} \\
\cline{3-5}
&  [GeV] & LO, $~~\mathcal{O}(\alpha_S^2)$   & NLO, $~~\mathcal{O}(\alpha_S^3)$   &K-fac\\
\hline\hline
\multirow{2}{*}{BP1}  & \multirow{2}{*}{800} & \multirow{2}{*}{$129.7^{+28.6\%}_{-20.8\%}$} & \multirow{2}{*}{$187.9^{+8.8\%}_{-10.6\%}$} & \multirow{2}{*}{1.45} \\
  &  &  &  & \\
\hline
\multirow{2}{*}{BP2}  & \multirow{2}{*}{1080} & \multirow{2}{*}{$18.95^{+28.7\%}_{-20.9\%}$} & \multirow{2}{*}{$26.62^{+10.0\%}_{-11.3\%}$} & \multirow{2}{*}{1.40} \\
  &  &  &  & \\
\hline
\multirow{2}{*}{BP3}  & \multirow{2}{*}{1500} & \multirow{2}{*}{$1.787^{+29.2\%}_{-21.3\%}$} & \multirow{2}{*}{$2.327^{+10.9\%}_{-11.9\%}$} & \multirow{2}{*}{1.30} \\
  &  &  &  & \\
  \hline
 \end{tabular} 
\caption{
The cross-section for the VLQ pair production at the 14 TeV LHC at the leading order (LO) and next-to-leading order (NLO) with integrated K-factor are shown for 14 TeV LHC. The superscripts and subscripts indicate the theoretical scale uncertainties (in percentages) of the total cross-section considering an envelope of renormalization ($\mu_R$) and factorization ($\mu_F$) scale variation around the central scale as discussed in the text.
}
\label{tab:crosssection}
\end{center}
\end{table}

The primary standard model background for the signal arises from semi-leptonic top quark pair production $t \bar{t}$+jets. In semi-leptonic decay, one top quark decays hadronically, resulting in a fatjet, while the other decays leptonically. The misidentified charged lepton and the neutrino from the top decay contribute to the MET. Conversely, when both top quarks decay hadronically, producing two fatjets, the requirement for high missing transverse momentum reduces this background significantly. This reduction occurs because the MET, in this case, primarily arises from jet mismeasurements, and imposing a cut of $\text{MET} \geq 150~ \text{GeV}$ further diminishes the background. The semi-leptonic $t \bar{t}$+jets background is generated with two additional QCD jets using an MLM matching scheme, with one of these QCD jets mimicking another fatjet.

Other significant backgrounds include $V$+jets, where $V$ represents either the $Z$ or $W^\pm$ boson. In the case of $Z$+jets, two neutrinos from the invisible decay of the $Z$ boson contribute to a large MET. In contrast, for $W$+jets, the MET arises from the neutrino and the misidentified charged lepton produced in the leptonic decay of the $W$ boson. A generation-level cut on $\text{MET} > 80~ \text{GeV}$ is applied to both backgrounds. Additionally, each background is generated with four extra QCD jets using an MLM matching scheme, and these QCD jets imitate fatjets in both cases.

$tW$+jets provides another significant SM background that mimics the signal. In this process, either the top quark or the $W$ boson decays leptonically, resulting in a large MET, while the other decays hadronically and is reconstructed as a fatjet. This background is generated with up to two additional jets using MLM matching, and the second fatjet arises from these QCD jets.

We also consider the diboson+jets background, which includes $p p\to W^\pm Z, W^+W^-, Z Z$ productions. In these processes, one of the weak bosons decays either invisibly or leptonically, leading to a significant MET, while the other boson decays hadronically, forming a fatjet. All diboson backgrounds are generated with two additional QCD jets using an MLM matching scheme. The second fatjet arises from these QCD jets.

All the backgrounds mentioned above are normalized using the available higher-order QCD-corrected cross-sections \cite{Muselli:2015kba, Kidonakis:2015nna, Campbell:2011bn, Catani:2009sm, Balossini:2009sa}.

\subsection{High-level variables}
\label{variables}
We now define some of the reconstructed variables that probe the internal structure within the fatjets and thus have a close connection with the original coloured particle from which these final state hadrons originated. They are immensely powerful in reducing the unassociated soft and wide-angle QCD contributions, leading to improved prediction of fatjet properties and masses.

\noindent
\underline{$N$-subjettiness ratio:}\hspace*{3mm} In the previous section, we noted that one or more QCD radiation mimics fatjets in the background. At the same time, for the signal, both fatjets are the result of the hadronic decay of boosted top quarks. Therefore, analyzing how the constituents of a fatjet are distributed within both the leading and subleading fatjets can help distinguish the signal from the background. $N$-subjettiness is one such powerful variable employed for this purpose; it seeks to identify $N$ lobes within a boosted fatjet that corresponds to $N$-subjets. For example, let $J$ be a boosted fatjet and $a_1, a_2, \cdots a_N$ are sets of $N$-axes within the fatjet. The unnormalized $N$-subjettiness \cite{Thaler:2010tr} is then defined as follows.
\begin{equation}
\tau_n = \, \sum_i \,  p_{T_{i}} \, \text{min} \{ \Delta R_{i,1}^\beta,\Delta R_{i,2}^\beta,...,\Delta R_{i,n}^\beta\}.
\label{EQ:tau_n}
\end{equation}
The summation includes all constituent particles within the fatjet, with $p_{T_{i}}$ representing the transverse momentum of the $i$-th constituent. $\Delta R_{i,j}$ denotes the separation of the $i$-th constituent from the $j$-th axis in the rapidity-azimuth plane. The minimization in Eq.~(\ref{EQ:tau_n}) is performed by varying the axis directions. The choice of axes is not unique; one minimization procedure that uses a variant of the k-means clustering algorithm is described in Ref. \cite{Lloyd:1982zni}. For our analysis, we consider winner-takes-all-$k_T$-axes \cite{Larkoski:2014uqa, Bertolini:2013iqa, Stewart:2015waa} and choose $\beta=1.0$ in Eq.~(\ref{EQ:tau_n}). We define two variables for each fatjet: $\tau_{32}$ and $\tau_{21}$. $\tau_{32}$ is the ratio of $\tau_{3}$ to $\tau_{2}$ and is particularly effective in distinguishing three-prong top-like fatjet from those with one or two prongs fatjet \cite{Thaler:2011gf}.\\

\noindent
\underline{Soft Drop:}\hspace*{3mm} Jet mass is another good variable for discrimination between the top-like fatjets from the QCD jets. The large radius fatjets are pronged to capture soft and wide-angle constituents that are not part of the original jet. Soft drop grooming/tagging \cite{Larkoski:2014wba} is a procedure to groom away those unassociated constituents and correctly predict the fatjet mass. We use a soft drop grooming method to remove those contaminants. We construct fatjets of radius $R_0 = 1.5$ using the  $\mbox{anti-k}_T$ algorithm. After that, we recluster the constituents of each fatjet with the Cambridge-Aachen (C/A) algorithm to create a pairwise angular-ordered clustering tree. The subsequent steps involve applying the Soft Drop algorithm as follows.
\begin{enumerate}
\item After clustering the jet, $j$, using the CA algorithm, decluster its last recombination stage and label them as two subjets, $j_1$ and $j_2$.
\item Now check the following condition: 
\begin{equation}
 \dfrac{\text{min}\{p_{T_{j_1}}, p_{T_{j_2}}\}}{p_{T_{j_1}}+p_{T_{j_2}}} > z_{\text{cut}}\Bigl( \dfrac{\Delta R(j_1,j_2)}{R_0} \Bigr)^\beta
\end{equation} 
$p_{T_{j_1}}$ and $p_{T_{j_2}}$ are the subjets' transverse momentum, and their distance in the rapidity-azimuth plane is $\Delta R(j_1,j_2)=\sqrt{(y_1-y_2)^2+(\phi_1-\phi_2)^2}$.
\item If the condition is satisfied, set the jet $j$ as a final soft-drop jet. If not, discard the softer subjet and prompt the harder subjet as $j$. Iterate the procedure from step 1.
\end{enumerate}

The soft drop procedure relies on two parameters: a soft threshold, $z_{\text{cut}}$, and an angular exponent, $\beta$. The procedure returns the original ungroomed jet when $z_{\text{cut}}\to 0$ or $\beta\to \infty $. Soft drop grooming mode is infrared and collinear (IRC) safe for $\beta > 0$. In our analysis, we set $\beta = 1.0$ and $z_{\text{cut}} = 0.1$ \cite{Larkoski:2014wba}. For $\beta = 0$, the method reduces to the modified mass drop tagger (mMDT). In this scenario, grooming depends exclusively on the momentum fraction $z_{\text{cut}}$ and is independent of the angular separation. However, as $\beta$ increases, greater emphasis is placed on wide-angle radiations, whereas smaller values of $\beta$ suppress their influence.

\subsection{Event selection}
\label{selection}
The following baseline selection criteria are used to select the events for our analysis.
\begin{enumerate}
\item Each event must contain at least two fatjets of radius 1.5, having substantial transverse momentum $p_T(J_0),~p_T(J_1)>200$ GeV within a rapidity range of $|\eta|<2.5$. $J_0$ and $J_1$ are leading and subleading fatjets, respectively. 
\item Events contain considerable missing transverse momentum,  $\text{MET} > 80$ GeV.
\item To reduce the impact of jet mismeasurements on the missing transverse momentum, an event selection criteria is applied on the azimuthal separation between any fatjet and the MET, $\Delta \phi(J_i,\text{MET})>0.2~(i=0,1)$.
\end{enumerate}

In addition to these baseline selection criteria, we applied the additional event selection criteria to reduce overwhelming backgrounds before the multivariate analysis (MVA). 
\begin{itemize}
\item A lepton veto is applied since no isolated lepton is expected from the signal. We discard any event that contains a lepton ($l=e^\pm, \mu^\pm$) with $p_{T}\geq 10$ GeV and within $|\eta_e|<2.1$, and $|\eta_\mu|<2.4$.
\item The signal consists of two top quarks in the final state, each decaying into a $b$ quark and hadronic products from $W$ boson. Tagging a $b$-jet within either fatjet enhances the signal-to-background ratio. We require a $b$-tagged subjet within the leading or subleading fatjet ($J_0$ or $J_1$), utilizing the default b-tagging efficiency provided by the Delphes CMS card, as described in Ref. \cite{CMS:2012feb}.   
\item MET has further increased from 80 GeV to 150 GeV. 
\item Since both the fatjets in the signal originate from the top quarks, we demand a larger soft-drop jet (defined in Subsection \ref{variables}) mass for both fatjets, $M_{J_0},~M_{J_1}>100$ GeV. 
\end{itemize}

Table \ref{tab:cut_flow} presents the cut flow for the signal (BP1 and BP2) and all SM background processes, including the cut efficiency and the number of events at an integrated luminosity of 300 $\text{fb}^{-1}$ at the 14 TeV LHC. Notably, the cut efficiencies for BP3 and BP2 are nearly identical. From the table, the following observations can be made:
	\begin{itemize}
		\item The lepton veto cut significantly reduces the $t\bar{t}$+jets, $W$+jets, and $WW$+jets backgrounds while preserving nearly $100\%$ of the signal.
		\item A stringent MET cut further suppresses the background effectively.
		\item B-tagging within a fatjet substantially reduces the $V$+jets and $VV$+jets backgrounds ($V=Z,~W$).
		\item Cuts on the soft-drop mass of the two leading fatjets further suppress backgrounds that do not originate from top-quark decays.
	\end{itemize}

\begin{figure}[htbp!]
\centering
\subfloat[] {\label{fig:met} \includegraphics[width=0.30\textwidth]{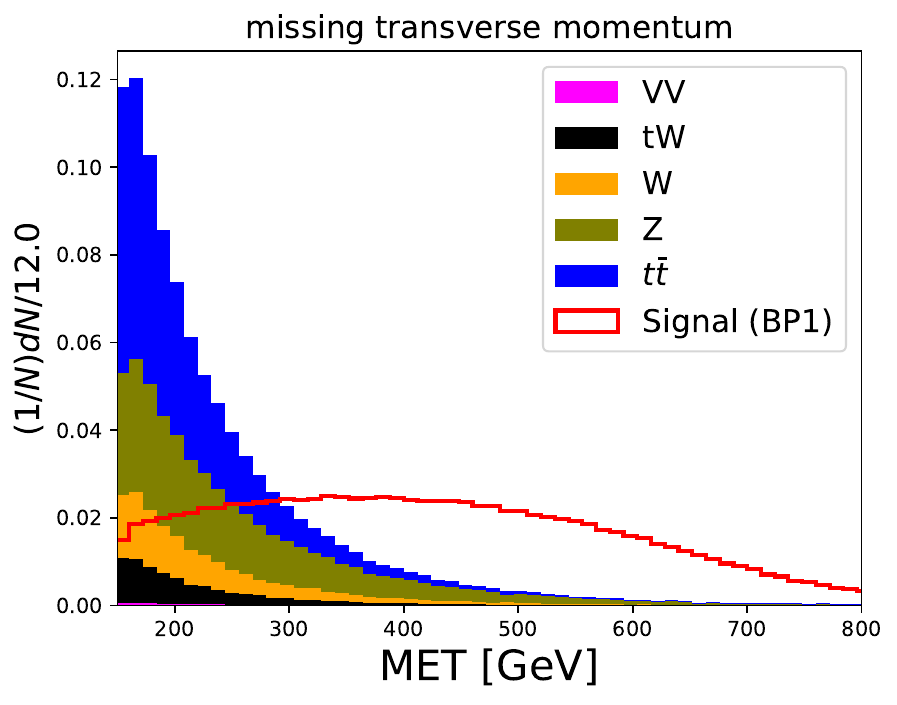}}
\subfloat[] {\label{fig:MJ0} \includegraphics[width=0.30\textwidth]{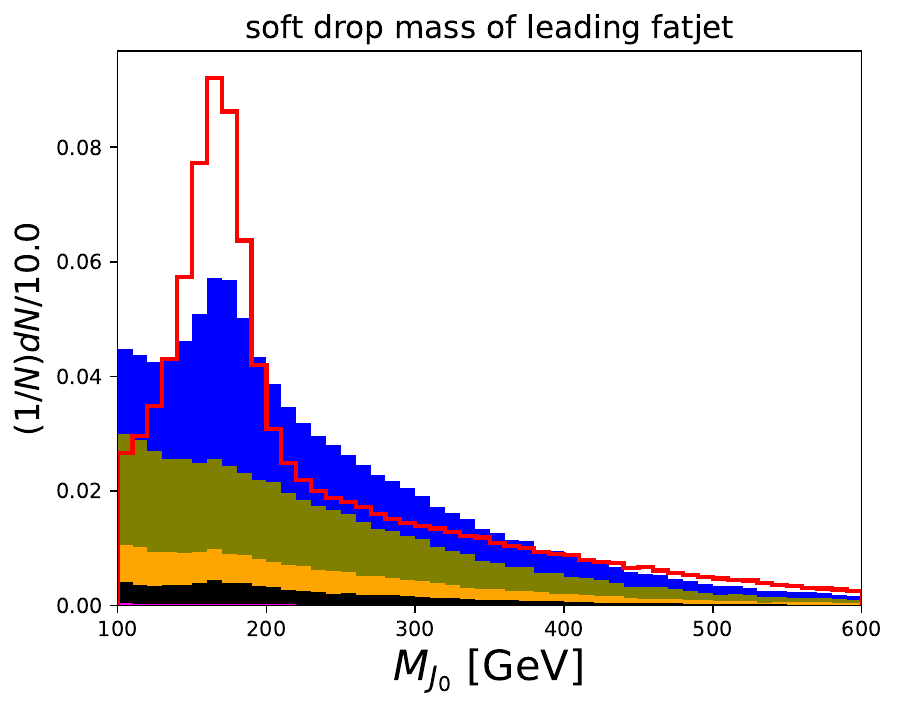}} 
\subfloat[] {\label{fig:MJ1} \includegraphics[width=0.30\textwidth]{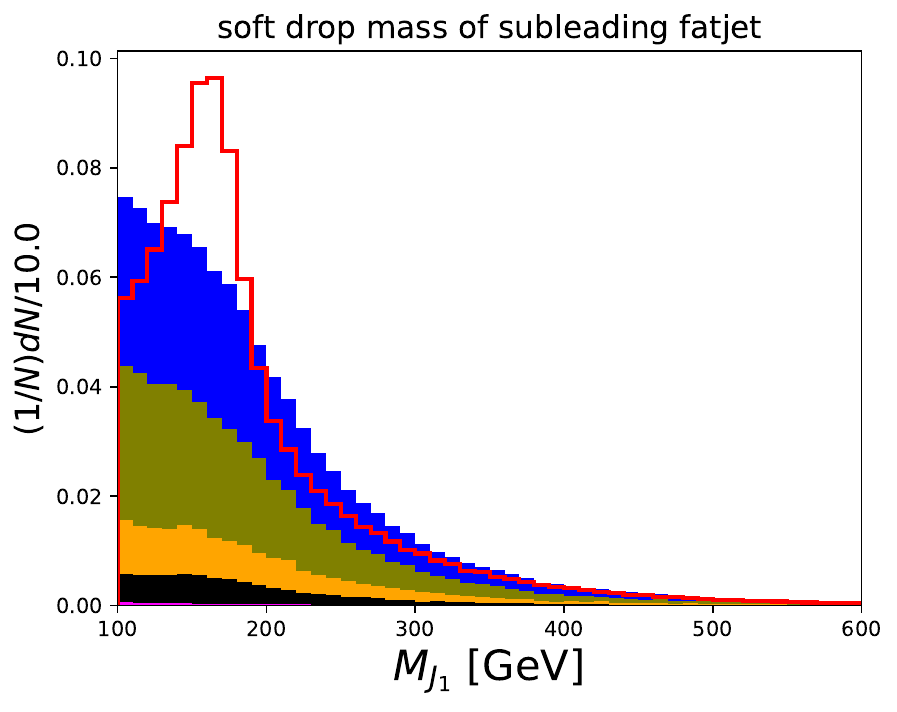}}\\ 
\subfloat[] {\label{fig:DPhi_J0met} \includegraphics[width=0.30\textwidth]{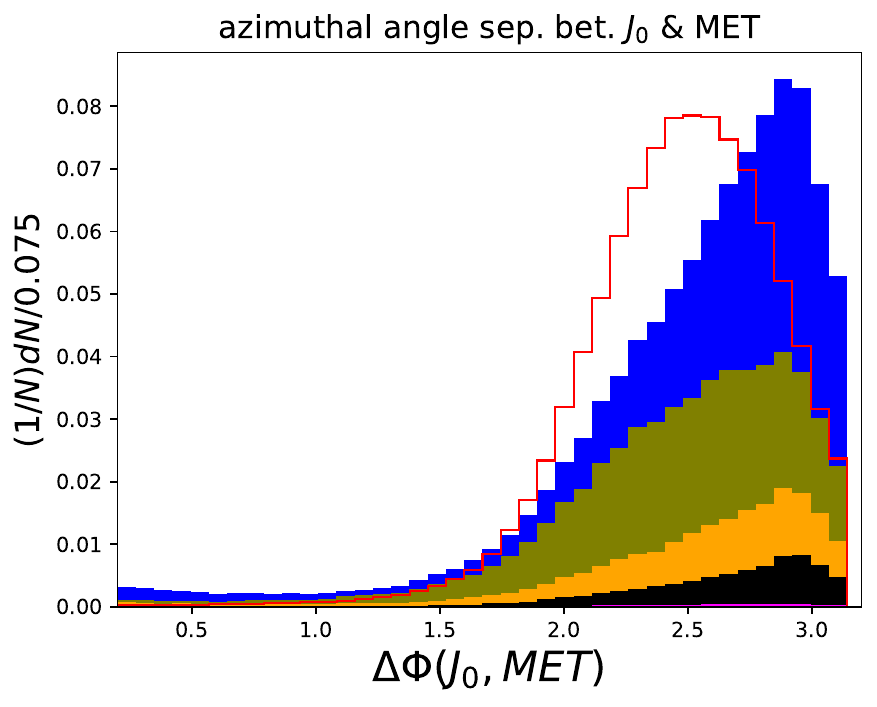}} 
\subfloat[] {\label{fig:DPhi_J1met} \includegraphics[width=0.30\textwidth]{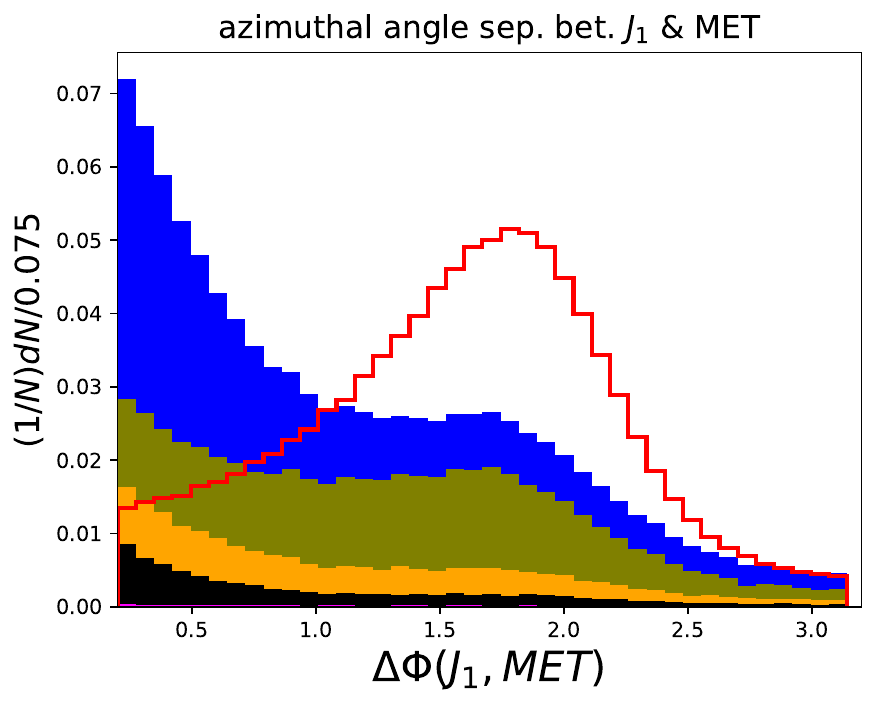}}
\subfloat[] {\label{fig:DRJ0J1} \includegraphics[width=0.30\textwidth]{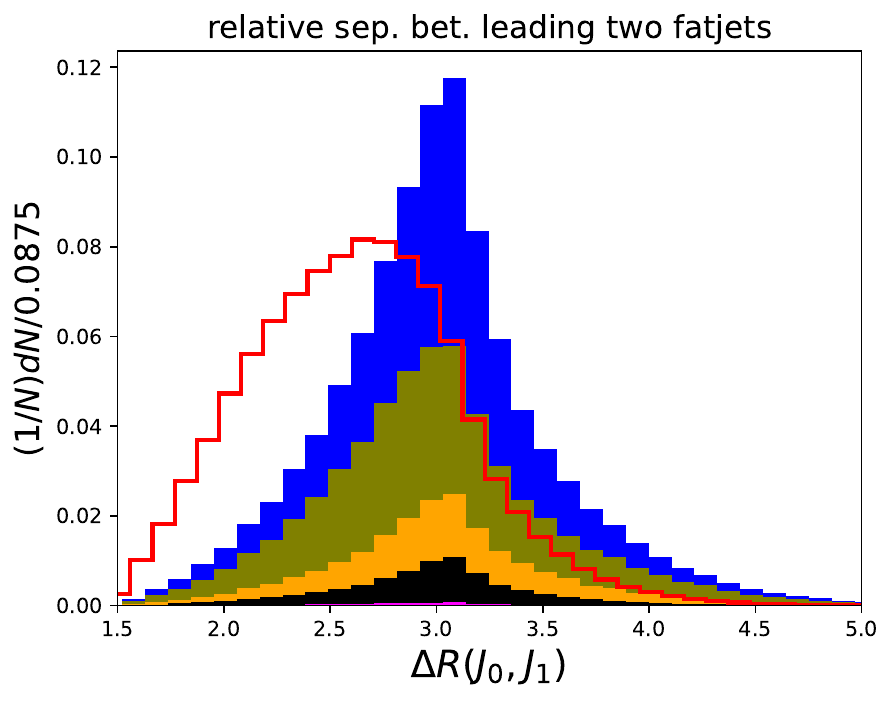}}\\
\subfloat[] {\label{fig:PT_J0} \includegraphics[width=0.30\textwidth]{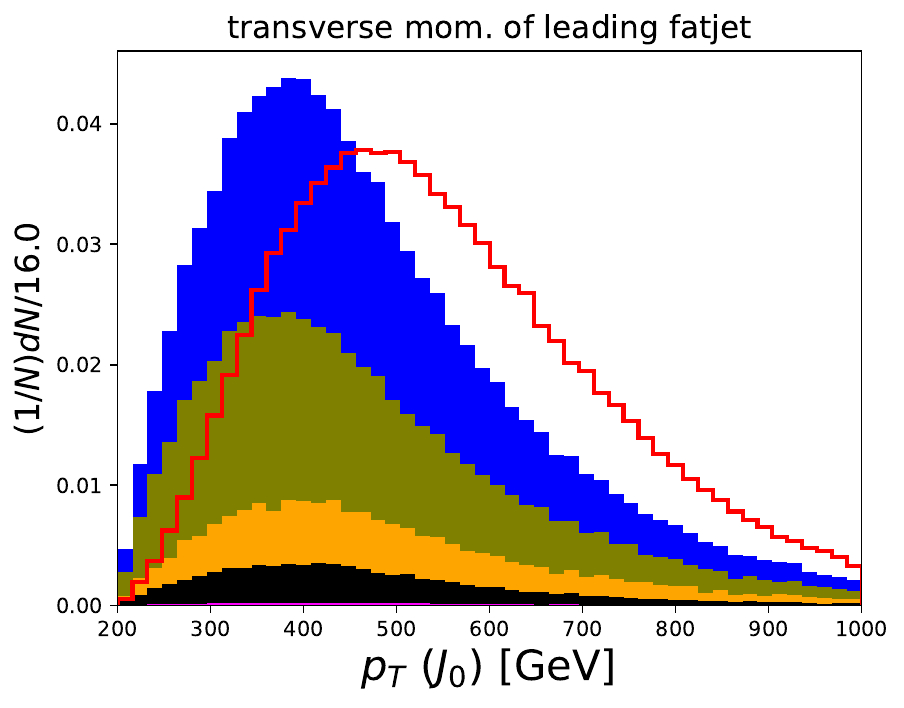}} 
\subfloat[] {\label{fig:PT_J1} \includegraphics[width=0.30\textwidth]{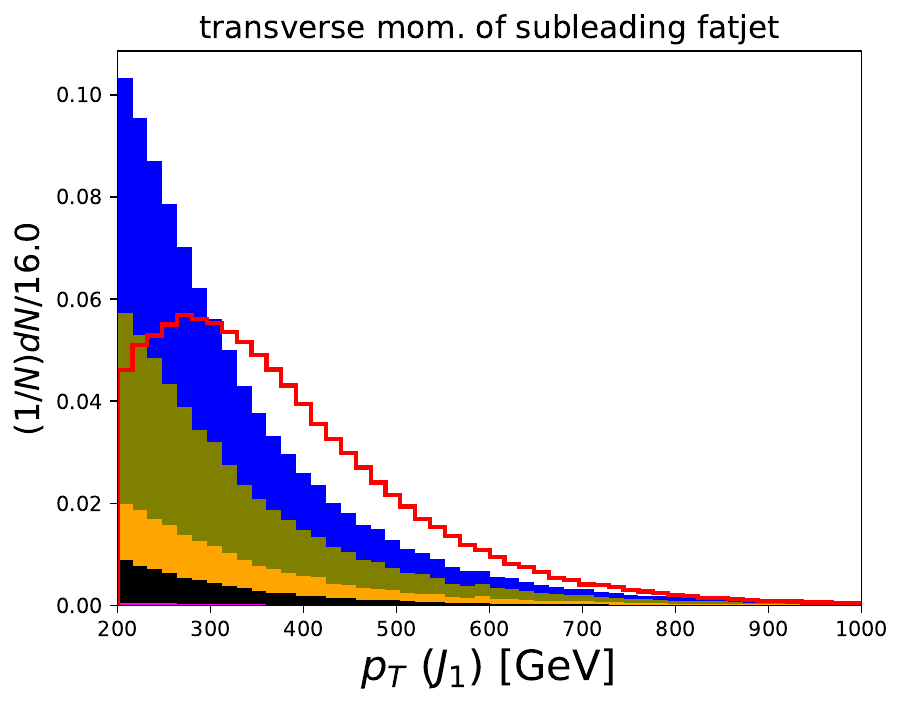}}
\subfloat[] {\label{fig:HT} \includegraphics[width=0.30\textwidth]{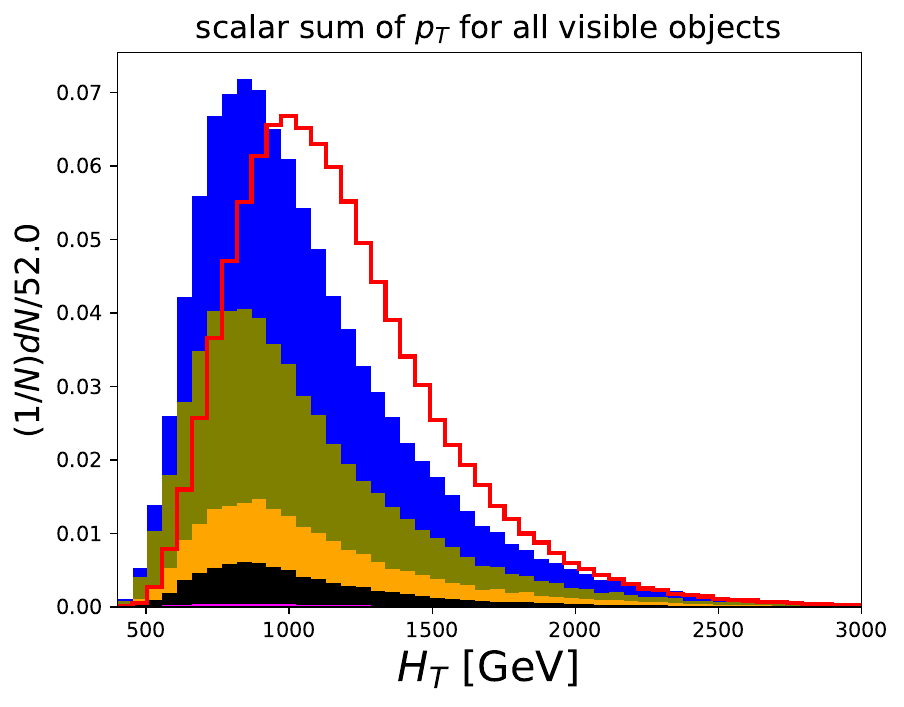}}\\
\subfloat[] {\label{fig:Meff} \includegraphics[width=0.30\textwidth]{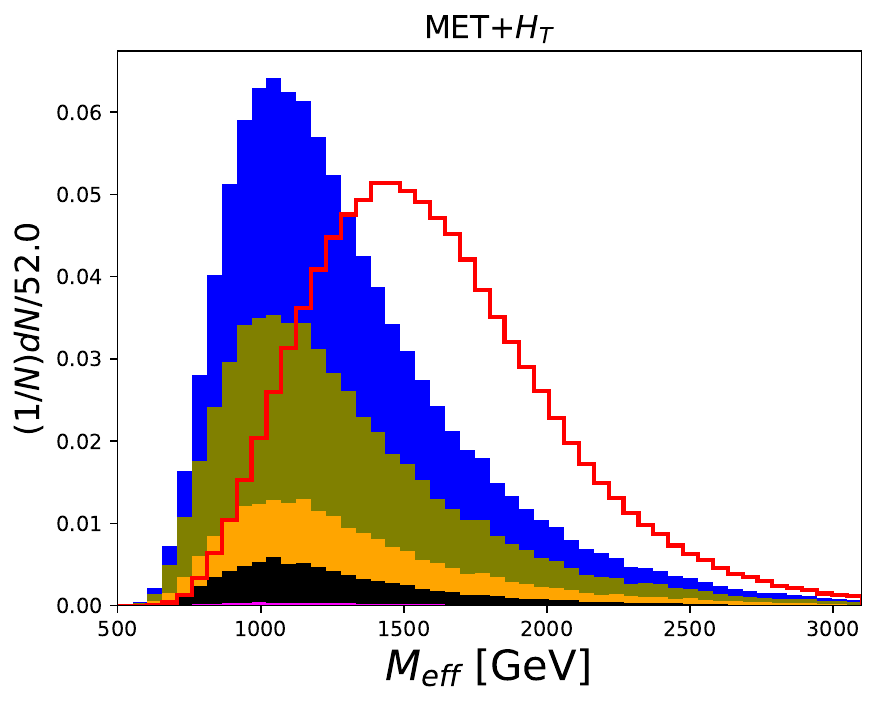}}
\subfloat[] {\label{fig:shat} \includegraphics[width=0.30\textwidth]{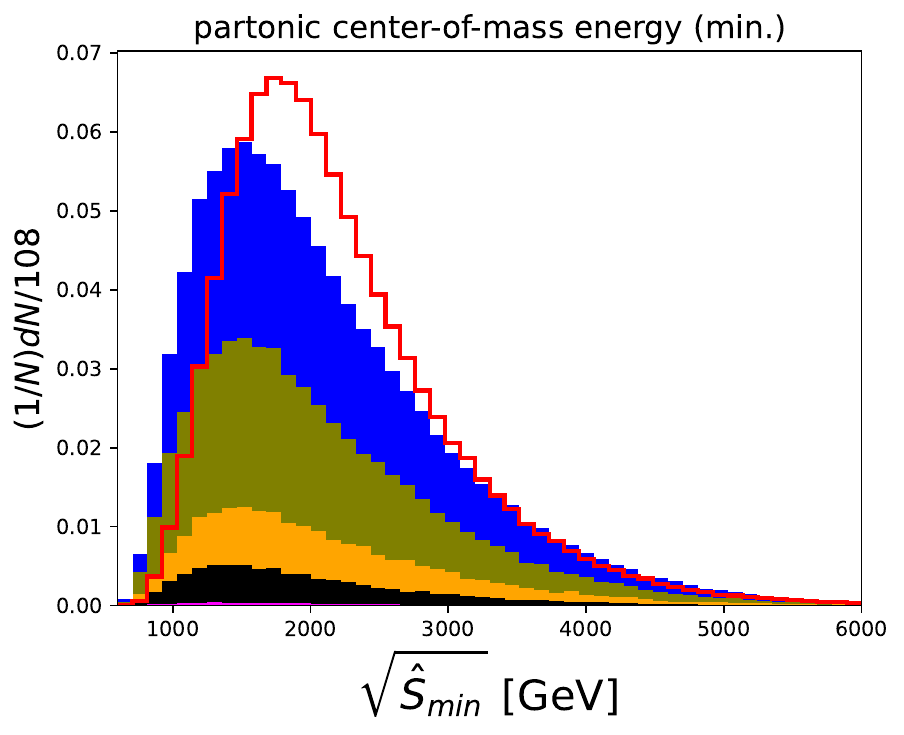}}
\subfloat[] {\label{fig:tau32_J0} \includegraphics[width=0.30\textwidth]{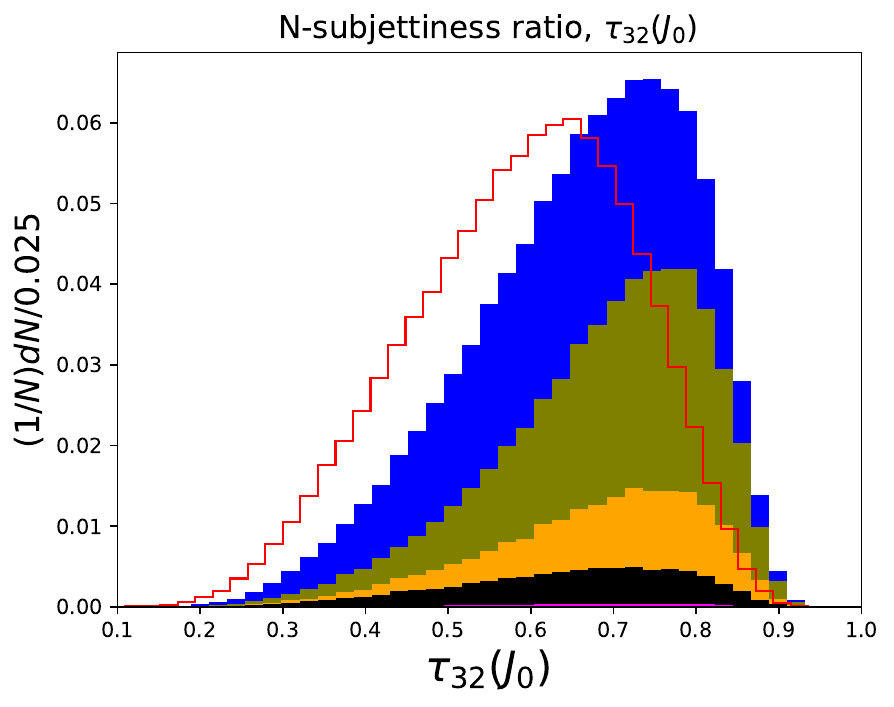}}\\
\subfloat[] {\label{fig:tau32_J1} \includegraphics[width=0.30\textwidth]{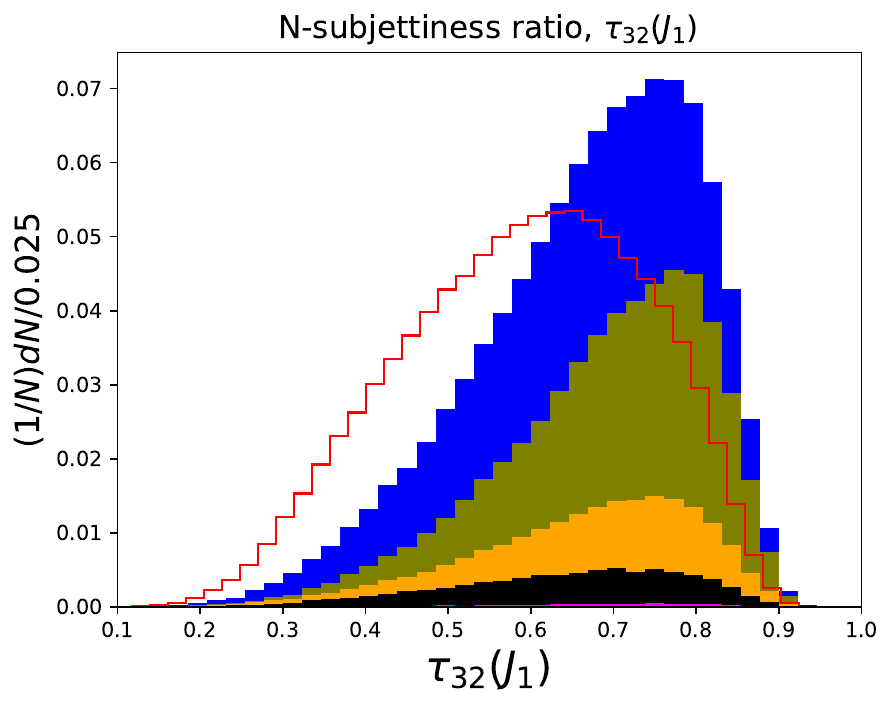}}
\subfloat[] {\label{fig:tau21_J0} \includegraphics[width=0.30\textwidth]{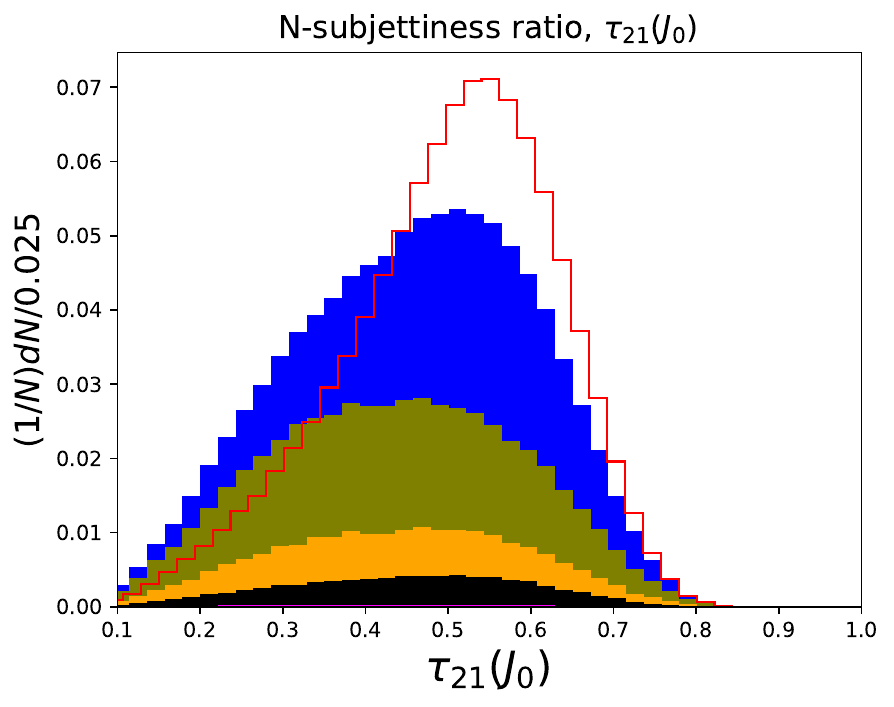}}
\subfloat[] {\label{fig:tau21_J1} \includegraphics[width=0.30\textwidth]{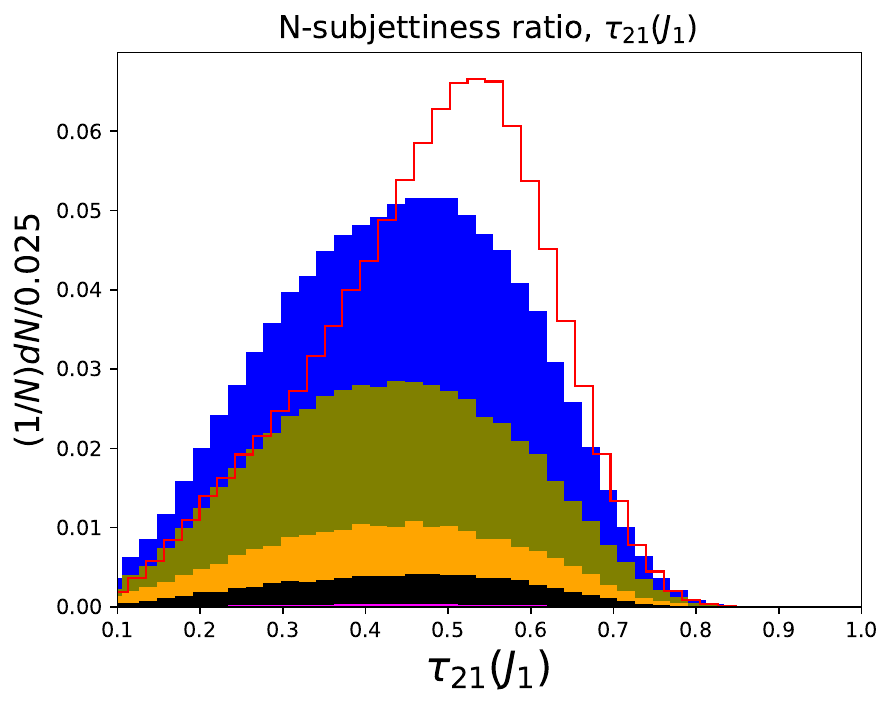}} 
\caption{Stacked histograms of various kinematic observables for SM backgrounds and the signal (BP1) in the fatjets plus MET final state are shown after applying all the event selection criteria as discussed in the text.}
\label{fig:sig_bg_1}
\end{figure}

Figure \ref{fig:sig_bg_1} presents the normalized distributions of certain kinematic variables for the signal (BP1, solid red) and background after applying all the event selection criteria as discussed before. The filled blue, olive, orange, black, and magenta areas represent contributions from different background processes: $t \bar{t}$+jets, $Z$+jets, $W$+jets, $t W$+jets, and diboson+jets, respectively. After applying the baseline selection cuts (outlined at the beginning of subsection \ref{selection}), some of these kinematic variables are shown in Appendix \ref{Appen:baseline} in Figure \ref{fig:sig_bg_pre}.

\begin{table}[tb!]
	\begin{center}
		\scriptsize
		\setlength\tabcolsep{2.7pt} 
				\begin{tabular}{|c|c|c||c|c|c|c|c|c|c|c|}
					\hline				
					\multirow{2}{*}{Cuts} & \multirow{2}{*}{BP1} & \multirow{2}{*}{BP2} &  $t \bar{t}+$ & $Z+$& $W+$ & $tW+$ & $WZ+$ & $WW+$ & $ZZ+$ & total \\
					&  &  & jets & jets & jets &jets   & jets & jets & jets & BG  \\
					\hline \hline				
					2 fatjets, $|\eta_i|<2.5$  & 15411.3  & 2546 &  $1.2\times 10^7$ &  $6.94\times 10^6$  & $9.0\times 10^6$ & $3.3\times 10^6$ & 105196 & 491801 & 30503  & $3.16\times 10^7$ \\
					$p_T(J_i)\geq 200$ GeV	& [$100\%$]  & [$100\%$] & [$100\%$] & [$100\%$] &  [$100\%$] & [$100\%$] & [$100\%$] & [$100\%$]  &[$100\%$] &[$100\%$] \\
					$(i=0,1)$ & & & & & & & & & & \\
					\hline
					\multirow{2}{*}{lepton veto}  & 15394.3  & 2543.5 &  $4.27\times 10^6$ &  $6.94\times 10^6$  & $3.14\times 10^6$ & $2.67\times 10^6$ & 105151 & 154454 & 30493  & $1.73\times 10^7$ \\
					& [$99.9\%$]  & [$99.9\%$] &  [$36.2\%$] & [$99.9\%$]& [$34.9\%$] & [$81.7\%$] & [$99.9\%$] & [$31.4\%$]  &[$99.9\%$] &[$54.7\%$] \\
					\hline
					\multirow{2}{*}{$\rm MET > 80$ GeV}  & 14853  & 2492.4 &  $1.91\times 10^6$ &  $6.3\times 10^6$  & $2.6\times 10^6$ & 528670 & 74475 & 58003 & 20401  & $1.16\times 10^7$ \\
					& [$96.4\%$]  & [$97.9\%$] &  [$16.2\%$] & [$90.7\%$]& [$29.9\%$] & [$16.2\%$] & [$70.8\%$] & [$11.8\%$]  &[$66.9\%$] &[$36.6\%$] \\
					\hline
					$\Delta \phi(J_i,\text{MET})>0.2$  & 14255.5  & 2395.5 &  $1.59\times 10^6$ &  $5.8\times 10^6$  & $2.35\times 10^6$ & 385210 & 66241 & 46757 & 18331  & $1.03\times 10^7$ \\
					& [$92.5\%$]  & [$94.1\%$] &  [$13.5\%$] & [$83.5\%$]& [$26.1\%$] & [$11.8\%$] & [$62.9\%$] & [$9.5\%$]  &[$60.1\%$] &[$32.5\%$] \\
					\hline
					1 b-tagging   & 12029 & 2006.3 &  $1.28\times 10^6$ & $1.04\times 10^6$  & 419492 & 290469 & 13799 & 9643 & 3471.5  & $3.06\times 10^6$ \\
					(within $J_0$ or $J_1$) & [$78.1\%$]  & [$78.8\%$]  & [$10.9\%$] & [$15.0\%$]& [$4.66\%$] & [$8.91\%$] & [$13.1\%$] & [$1.96\%$]  &[$11.4\%$] &[$9.67\%$] \\
					\hline
					\multirow{2}{*}{$\rm MET > 150$ GeV}  & 11130  & 1918 &  410084 &  633069  & 194518 & 88681 & 8072 & 3298.5 & 1935  & $1.34\times 10^6$ \\
					& [$72.2\%$]  & [$75.3\%$] &  [$3.48\%$] & [$9.11\%$]& [$2.16\%$] & [$2.72\%$] & [$7.67\%$] & [$0.67\%$]  &[$6.34\%$] &[$4.24\%$] \\
					\hline
					SD mass  & 6852& 1233 &  175971 & 142991  & 52094.3 & 29635.4 & 1382 & 765.5 & 382  & 403221 \\
					$M_{J_i}\geq100$ GeV & [$44.5\%$]   & [$48.4\%$] & [$1.5\%$] & [$2.0\%$]& [$0.58\%$] & [$0.9\%$] & [$1.3\%$] & [$0.16\%$]  &[$1.3\%$] &[$1.27\%$] \\
					\hline 
				\end{tabular}     
				\caption{After applying the corresponding cuts, the expected number of events, the cut flow for both signal (BP1 and BP2) and background processes for an integrated luminosity of 300 $\text{fb}^{-1}$ are presented. The total background is given in the last column. }
				\label{tab:cut_flow}
		\end{center}
	\end{table}

The missing transverse momentum distribution is shown in Figure \ref{fig:met}. The background decreases sharply at high MET values while the signal distribution remains relatively flat. The source of MET in the signal arises from two dark matter particles produced in the cascade decay of a heavy VLQ pair. This allows the dark matter particles to avail nearly the entire phase space, resulting in a nearly uniform MET distribution up to a certain range. The soft drop mass distributions for the leading and subleading fatjets are presented in Figures \ref{fig:MJ0} and \ref{fig:MJ1}, respectively. In both distributions, the signal exhibits a peak near the top-quark mass. In contrast, the background shows a peak near the top mass for the leading fatjet, while the subleading fatjet does not exhibit a similar peak. This absence indicates that the subleading fatjet primarily originates from QCD radiation in most events. The azimuthal separations between each fatjet and MET are shown in Figures \ref{fig:DPhi_J0met} and \ref{fig:DPhi_J1met}, while Figure \ref{fig:DRJ0J1} illustrates the distance between the two fatjets in the $\eta-\phi$ plane, $\Delta R (J_0,J_1)=\sqrt{(\Delta \eta)^2+(\Delta \phi)^2}$. From these distributions, we observe that the two fatjets are nearly back-to-back in most background events, which is not the case for the signal.

The transverse momentum distributions of the fatjets are illustrated in Figures \ref{fig:PT_J0} and \ref{fig:PT_J1}. The distributions of other inclusive variables- $H_{T}$, $M_{\text{eff}}$, and $\sqrt{\hat{s}_{\text{min}}}$ are presented in Figures \ref{fig:HT}, \ref{fig:Meff}, and \ref{fig:shat}, respectively, with their definitions provided below.
\begin{equation}
H_{T}=\sum_{i\in \text{visible}} |\overrightarrow{p_{T_{i}}}| \,,  
\label{EQ-HT}
\end{equation}
\begin{equation}
M_{\text{eff}}=\text{MET}+\sum_{i\in \text{visible}} |\overrightarrow{p_{T_{i}}}| \,,  
\label{EQ-Meff}
\end{equation}
\begin{equation}
\sqrt{\hat{s}_{\text{min}}} = \sqrt{E^2-P_Z^2}+ \text{MET} \, ,
\label{EQ-Shat}
\end{equation}
Here, $p_{T_{i}}$ is the transverse momentum of the $i$-th AK4 jet, and the summation runs over all visible jets in the event. $E$ and $P_Z$ represent the total energy and longitudinal momentum component of all visible jets in the event. The distributions of the N-subjettiness ratios $\tau_{32}=\tau_{3}/\tau_{2}$ and $\tau_{21}=\tau_{2}/\tau_{1}$ for both leading and subleading fatjets are shown in Figures \ref{fig:tau32_J0}, \ref{fig:tau32_J1}, \ref{fig:tau21_J0}, and \ref{fig:tau21_J1}. As expected, the signal yields a smaller $\tau_{32}$ value than the background. N-subjettiness, $\tau_{n}$, defined in Subsection \ref{variables}, quantifies the number of subjets within a fatjet. Signal fatjets are predominantly three-pronged, yielding a low $\tau_{3}$ and high $\tau_{2}$ values, thus a smaller $\tau_{32}$ ratio. In contrast, background fatjets often appear one- or two-pronged, leading to lower $\tau_{3}$ and $\tau_{2}$ values and, consequently, a higher $\tau_{32}$ ratio than the signal. As a result, $\tau_{32}$ can be a good variable for signal-background discrimination. In contrast, $\tau_{21}$ is less effective for this purpose, as signal fatjets are predominantly three-pronged.

The expected number of signals and total background events that survive after applying all the cuts discussed in Subsection \ref{selection} are presented in the second column of Table \ref{tab:BDT} for the 14 TeV LHC with an integrated luminosity of 300 $\text{fb}^{-1}$. The event yields for individual background processes can be found in the last row of Table \ref{tab:cut_flow}. The last column of Table \ref{tab:BDT} shows the statistical significance achieved through this traditional cut-and-count search method. Although refining kinematic cuts could enhance statistical significance, we prioritize Multivariate Analysis in the next section, where more sophisticated techniques will be employed for better signal-background discrimination.

\subsection{Analysis based on the Multivariate Gradient Boosting Technique}
\label{MVA}

We perform a multivariate analysis utilizing a gradient-boosted decision tree (BDT) method to enhance our collider search. Several characteristics of the reconstructed fatjets and the missing transverse momentum are employed as inputs for the BDT. Those variables include the transverse momentum of the fatjets after soft drop, $p_T(J_i)~(i=0,1)$ and their corresponding masses, $M_{J_i}$, missing transverse momentum (MET), the azimuthal separation between each fatjet and the MET, $\Delta \phi(J_i,\text{MET})$, the separation between fatjets in the rapidity-azimuth plane, $\Delta R (J_0,J_1)$, transverse momentum of the b-jet ($p_T(b)$) that is tagged inside $J_0$ or $J_1$, N-subjettiness ratio $\tau_{32}(J_i)$, $\tau_{21}(J_i)$  and other inclusive variables $H_{T}$, $\text{M}_{eff}$, $\sqrt{\hat{s}_{\text{min}}}$. 

Not all of the variables discussed above are included in the multivariate analysis, as some exhibit significant (anti)correlations with one another. It is important to note that using variables with high (anti)correlation does not enhance efficiency. Therefore, we carefully select only those with low to moderate (anti)correlation. For example, $\text{M}_{eff}$ and $\sqrt{\hat{s}_{\text{min}}}$, defined in Eqs.~(\ref{EQ-Meff}) and (\ref{EQ-Shat}), are highly correlated in both signal and background classes. Similarly, $\text{M}_{eff}$ and $p_T(J_1)$ also show a strong correlation in both classes. We retain $\text{M}_{eff}$ while excluding $p_T(J_1)$ and $\sqrt{\hat{s}_{\text{min}}}$, as $\text{M}_{eff}$ demonstrates greater relative separation power between the signal and background classes. This relative separation power indicates that the variable is more effective in discriminating between the two classes than the others. The background represents the weighted sum of the individual background processes discussed in Subsection \ref{setup}. The final set of variables utilized in our MVA analysis is presented in Table \ref{relative_imp}, along with their relative separation power in distinguishing between signal and background.

\begin{table}[tb!]
\centering
\scriptsize
\setlength\tabcolsep{2.5pt} 
 \begin{tabular}[b]{|c|c|c|c|c|c|c|c|c|c|c|c|}
\hline
 &  \multirow{2}{*}{ MET} &  \multirow{2}{*}{$\text{M}_{eff}$} &  \multirow{2}{*}{$\Delta R (J_0,J_1)$} &  \multirow{2}{*}{$\Delta \Phi(J_1,MET)$} &  \multirow{2}{*}{$\tau_{32}(J_0)$}  &  \multirow{2}{*}{$\tau_{32}(J_1)$} &  \multirow{2}{*}{$\Delta \Phi(J_0,MET)$} &  \multirow{2}{*}{$\tau_{21}(J_0)$} &  \multirow{2}{*}{$M_{J_0}$} &  \multirow{2}{*}{$\tau_{21}(J_1)$} &  \multirow{2}{*}{$M_{J_1}$} \\ %
 &  &  &  &  & &  &  &  &   & & \\
\hline\hline
\multirow{2}{*}{ BP1} & \hspace{0.08cm}  \multirow{2}{*}{34.62}  & \hspace{0.08cm} \multirow{2}{*}{14.28}  & \multirow{2}{*}{13.33} &  \multirow{2}{*}{12.92} & \multirow{2}{*}{7.39} & \multirow{2}{*}{7.10} & \multirow{2}{*}{5.26}  & \multirow{2}{*}{3.16}  & \multirow{2}{*}{2.34} & \multirow{2}{*}{2.21}  & \multirow{2}{*}{1.55}  \\
 &  &  &  &  & &  &  &  &   & & \\

\hline
 \end{tabular} 
\caption{Method unspecific relative ranking of input variables based on their separation power in discriminating the signal (BP1) class from the background before MVA. The same variables are used in all other benchmark scenarios.}\label{relative_imp}
\end{table}

\begin{figure}[tb!]
\centering
  \subfloat {\label{MVA:correlationS}\includegraphics[width=0.495\textwidth]{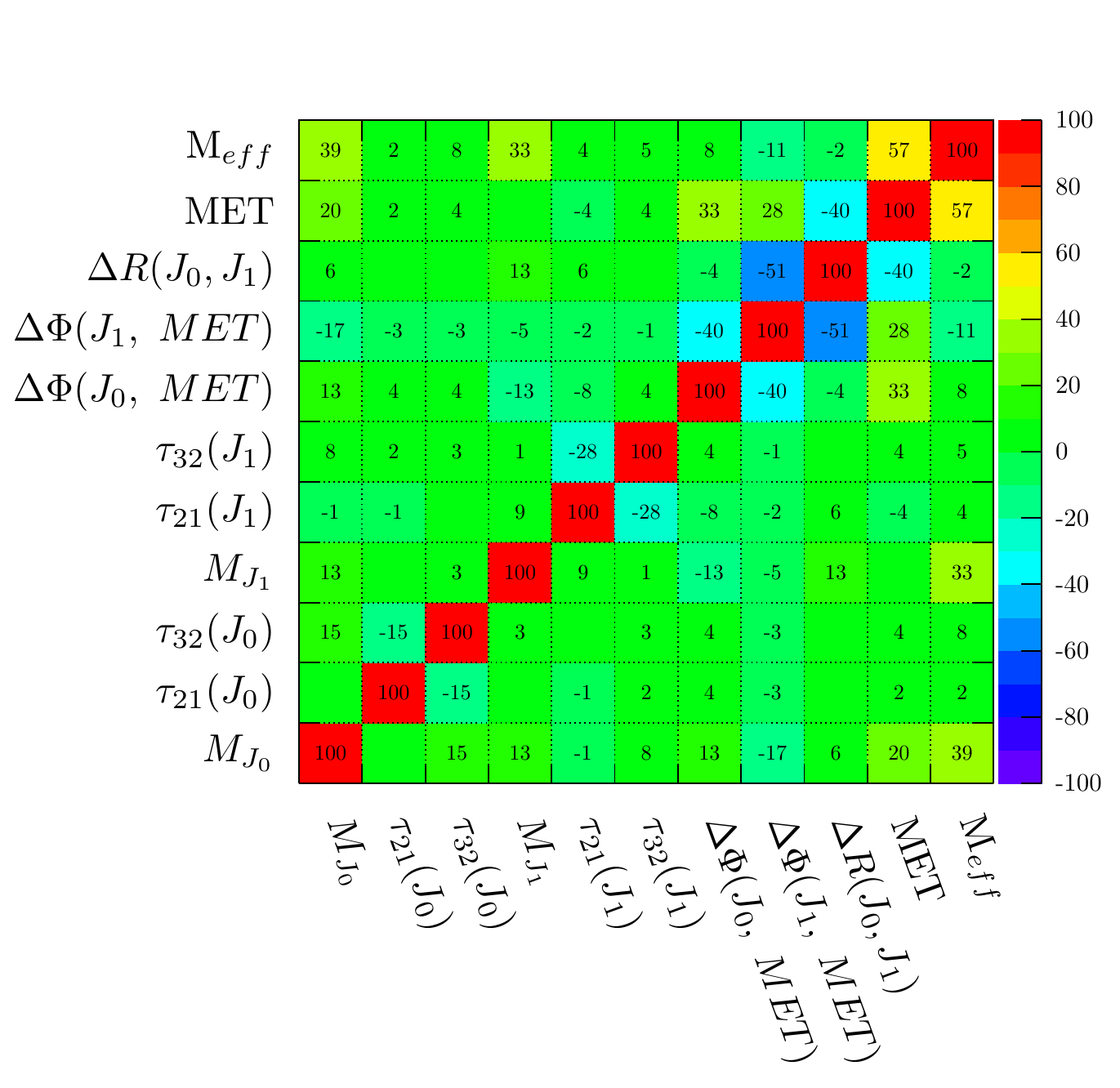}} 
  \subfloat {\label{MVA:correlationB}\includegraphics[width=0.495\textwidth]{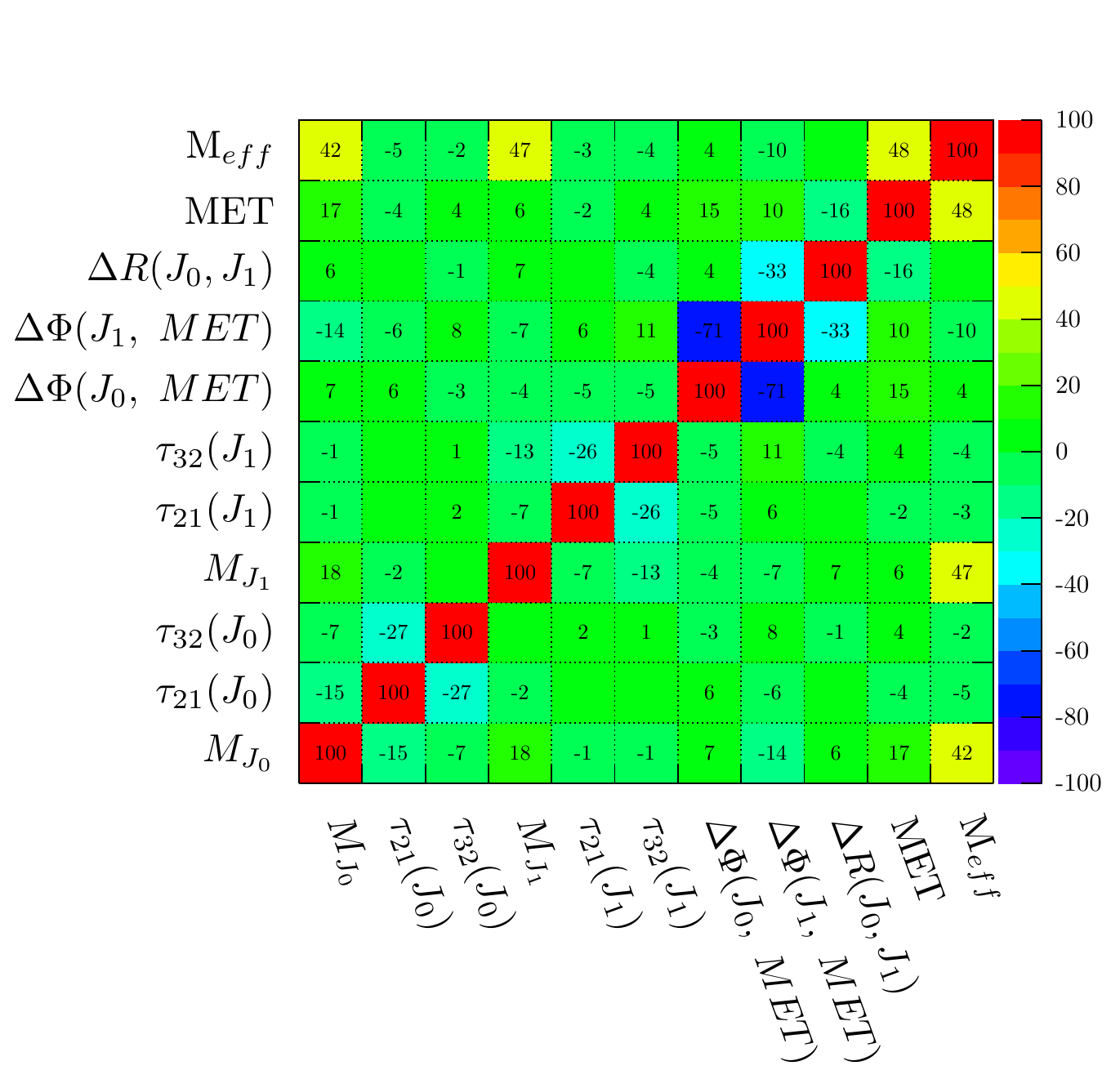}}
  \caption{Linear correlation coefficients (expressed as percentages) between various kinematic variables for the signal (BP1, left panel) and background (right panel). Missing values indicate a negligible correlation or anti-correlation, less than one. Positive coefficients show a correlation, while negative coefficients indicate an anti-correlation.
}\label{correlation}
\end{figure}

\begin{figure}[tb]
\centering
  \subfloat {\label{MVA:sig_bg}\includegraphics[width=0.495\textwidth]{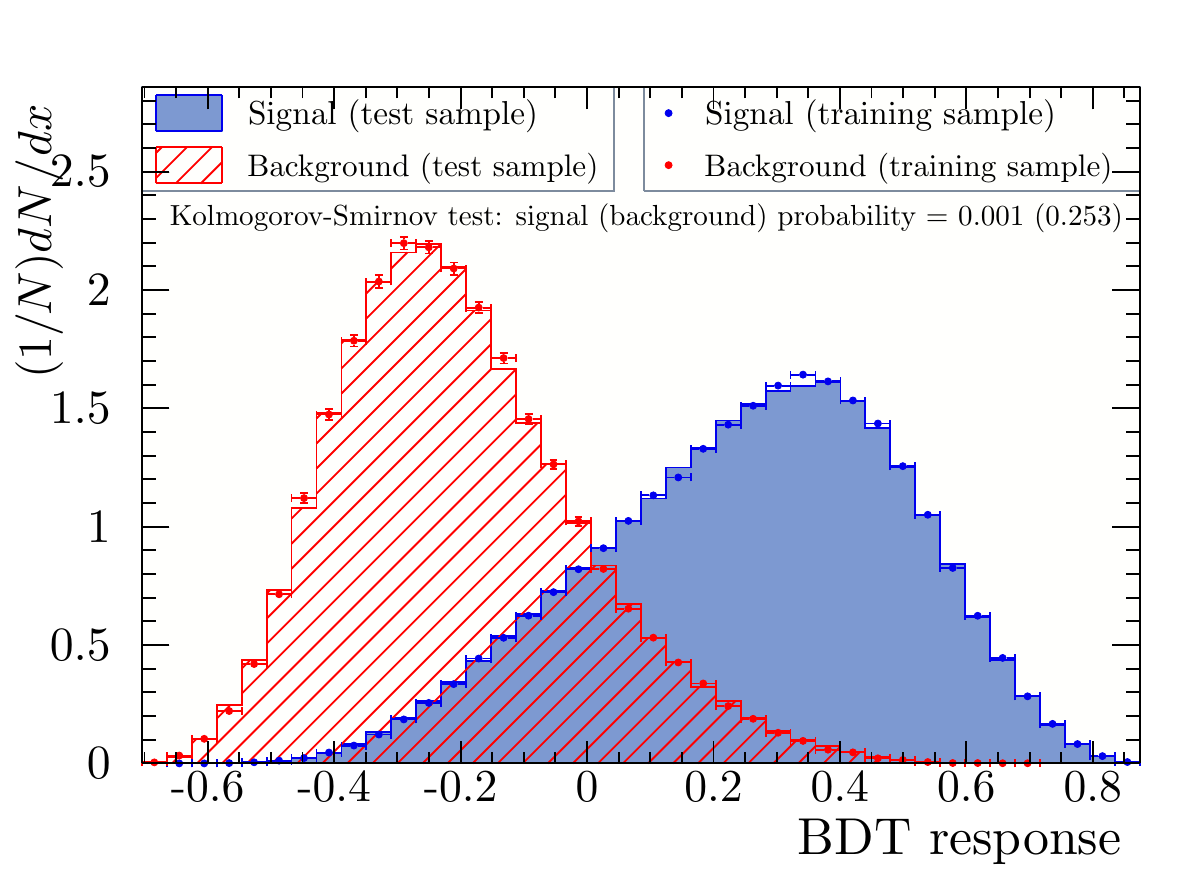}} 
  \subfloat {\label{MVA:significance}\includegraphics[width=0.495\textwidth]{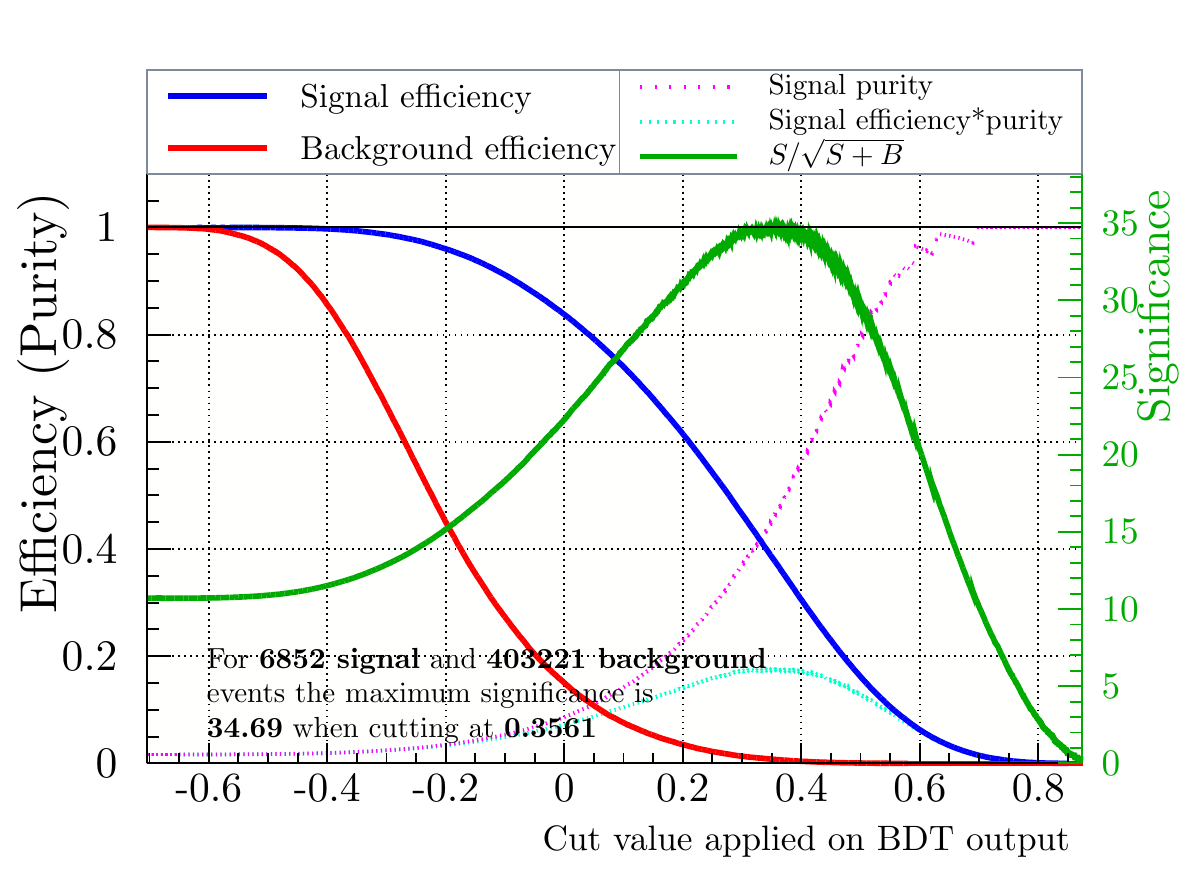}}
  \caption{(Left panel): The normalized distribution of the BDT response is shown for both signal (blue, BP1) and background (red). Dotted histograms represent training samples, while solid histograms denote testing samples. (Right panel): The efficiency of the signal (blue) and background (red), along with the signal's statistical significance (green), are plotted as functions of the BDT output cut applied.}\label{sig_bg_significance}
\end{figure}

The correlation between the variables for the signal (considering benchmark point BP1) and background classes is illustrated in the left and right panels of Figure \ref{correlation}, respectively. Most of these variables exhibit low (anti)correlations. However, $\text{M}_{eff}$ and MET display a moderate correlation in both signal and background classes, and we retain both variables due to their significant relative separation power. In contrast, the azimuthal separations $\Delta \phi(J_0,\text{MET})$ and $\Delta \phi(J_1,\text{MET})$ show strong anti-correlations in the background yet only moderate anti-correlations in the signal. Consequently, we include both variables in our analysis.

Figure \ref{MVA:sig_bg} displays the normalized distributions of the BDT response for both the signal and background classes, encompassing both training and testing samples. The distributions for these two datasets align closely, and there is also a distinct separation between the signal and background classes. The parameters used for training the boosted decision tree are outlined in Appendix \ref{params}. We optimize our training separately for each benchmark point. Although BDT is a robust classifier, it can be susceptible to overtraining. To confirm that the BDT model is not overtrained, we assessed the Kolmogorov-Smirnov probability, a statistical test that measures the difference between the cumulative distribution functions of two datasets (test and train samples), presented in Figure \ref{MVA:sig_bg}.

In the right panel of Figure \ref{sig_bg_significance}, we illustrate how the signal (solid blue) and background (solid red) efficiencies change with respect to the cut applied to the BDT response. Signal (background) efficiency is defined as the ratio of the number of signal (background) events that survive after applying a specific cut on the BDT output to the total number of signal (background) events in the sample. The solid green line represents the variation in statistical significance as a function of the cut on the BDT output. This statistical significance is calculated as $\frac{N_S}{\sqrt{N_S+N_B}}$, where $N_S$ and $N_B$ are the numbers of signal and background events, respectively, that survive after the cuts. The dotted magenta line indicates the signal purity, defined as $\frac{N_S}{N_S+N_B}$. At the far left of the figure, we observe that the signal purity is very low due to the high number of background events relative to the signal. However, as the cut value on the BDT output increases, our model demonstrates its effectiveness, as the signal purity improves significantly.

\section{Results}  \label{sec6}

\begin{table}[tb!]
	\begin{center}
		\scriptsize
		\setlength\tabcolsep{2.9pt} 
		\begin{tabular}{|c|c|c|c|c|c|c|c|c|c|}
			\hline
			& Events & Optimal & Signal & Background & \multirow{2}{*}{$N_S$} & \multirow{2}{*}{$N_B$} & \multirow{2}{*}{$\frac{N_S}{N_B}$} & Significance & Significance  \\ 
			& before BDT & BDT cut& efficiency & efficiency &  &  & & $\Bigl( \frac{N_S}{\sqrt{N_S+N_B}}\Bigr)$ & (cut-and-count)    \\
			\hline\hline
			\multirow{2}{*}{BP1} & \multirow{2}{*}{6852} & \multirow{2}{*}{0.356} & \multirow{2}{*}{0.378} & \multirow{2}{*}{$7.356\times 10^{-3}$} & \multirow{2}{*}{2585} & \multirow{2}{*}{2966} & \multirow{2}{*}{0.87} & \multirow{2}{*}{34.69} &  \multirow{2}{*}{10.7}\\
			&  &  &  &  & &  &  &  &   \\
			\hline
			\multirow{2}{*}{BP2} & \multirow{2}{*}{1233} & \multirow{2}{*}{0.594} & \multirow{2}{*}{0.205} & \multirow{2}{*}{$5.042\times 10^{-4}$} & \multirow{2}{*}{253} & \multirow{2}{*}{203} & \multirow{2}{*}{1.24} & \multirow{2}{*}{11.84} & \multirow{2}{*}{1.94} \\
			&  &  &  &  & &  &  &   &  \\
			\hline
			\multirow{2}{*}{BP3} & \multirow{2}{*}{106} & \multirow{2}{*}{0.683} & \multirow{2}{*}{0.130} & \multirow{2}{*}{$1.292\times 10^{-4}$} & \multirow{2}{*}{14} & \multirow{2}{*}{52} & \multirow{2}{*}{0.27}  & \multirow{2}{*}{1.72} & \multirow{2}{*}{0.17} \\
			&  &  &  &  & &  &  &  &   \\
			\hline
			BG & 403221 & \multicolumn{8}{ c |}{} \\
			\hline
		\end{tabular} 
		\caption{The table presents the effectiveness of the search in terms of statistical significance for signal over background at different benchmark points. $N_S$ and $N_B$ indicate the total surviving signal and background events after an optimal BDT cut. The fourth and fifth columns indicate the efficiencies for the signal and background, respectively. The second-to-last column represents the statistical significance of the MVA analysis for $300 ~\text{fb}^{-1}$ integrated luminosity. The last column, which represents the statistical significance of cut-and-count searches, is provided for comparison.}
		\label{tab:BDT}
	\end{center}
\end{table}

As outlined in the model section, the independent parameters of our simplified model include the mass of the vector-like quark ($m_\Psi$), the coupling ($\tilde y_t$), and the dark matter mass ($m_S$). For the fixed dark matter mass $m_S =12$ KeV, the model's parameter space is represented by a plane with parameters $\tilde y_t$ and $m_\Psi$. We have selected three non-standard benchmark points, as presented in Table \ref{tab:Bemchmark}, to demonstrate the effectiveness of our BDT analysis in different regions of the parameter space. Table \ref{tab:BDT} summarizes the results of the multivariate analysis for these three representative benchmark points. The second column contains the number of signal events for all benchmark scenarios and total background events before doing MVA. The immediate next column is the optimal BDT cut value that should be applied to the BDT response to get the optimal statistical significance. The following columns contain the signal and background efficiencies and the number of the signal ($N_S$) and background ($N_B$) events that survived after applying the optimal BDT cut. The signal-to-background ratio is presented in the eighth column. The second-to-last column shows the statistical significance of the MVA analysis for an integrated luminosity of $300~ \text{fb}^{-1}$. To give a comparison, the last column provides the corresponding value for the traditional cut-and-count search. The MVA analysis achieves a significant improvement over the cut-and-count method.

\begin{figure}[tb!]
\centering
\includegraphics[height=6.3cm,width=11cm]{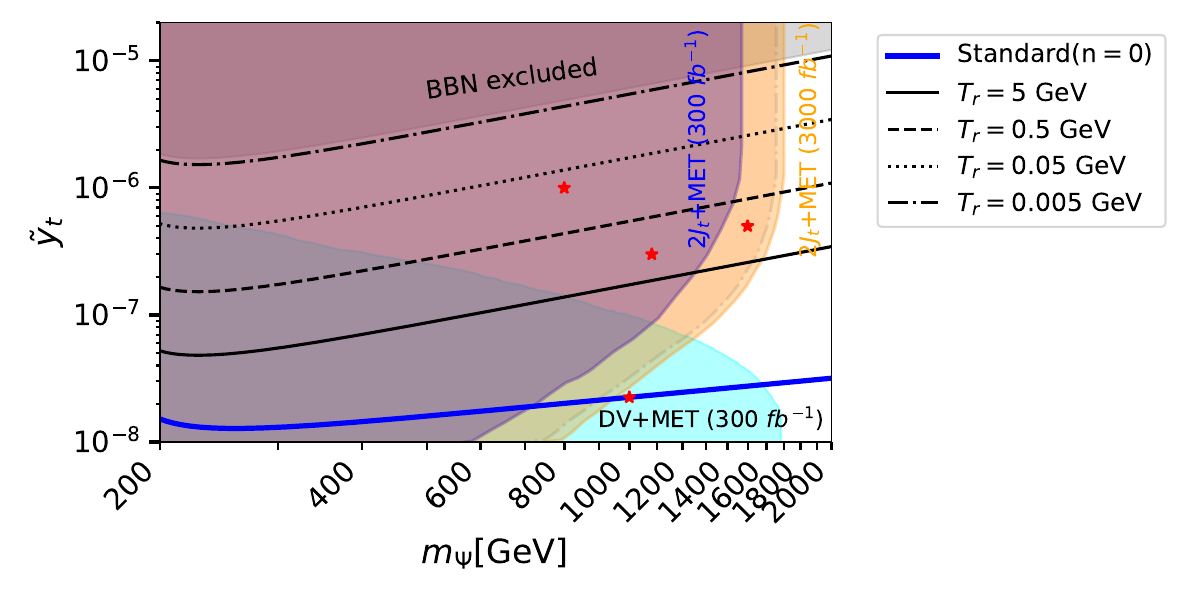}
\caption{
Excluded regions at a 95\% CL are shown in the Yukawa coupling $(\tilde{y}_t)$ and vector-like quark mass $m_\Psi$ plane for the freeze-in dark matter model. Baseline contours satisfy the dark matter relic density constraint $\Omega h^2= 0.12$ for different early Universe cosmological evolution scenarios as discussed in Figure \ref{coupling_mass}.
Model spaces within the blue (orange) region can be excluded for an integrated luminosity  300 $\text{fb}^{-1}$ (3000 $\text{fb}^{-1}$) at LHC, considering a non-standard cosmological scenario. The orange dashed dot contour shows the exclusion contour at 3000 $\text{fb}^{-1}$ integrated luminosity considering 10\% systematics.
The cyan-shaded region shows exclusions from the displaced vertex search (300 $\text{fb}^{-1}$) at the LHC. The grey-shaded region is ruled out since any $T_r$ lower than BBN ($\approx 4$ MeV) conflicts with the cosmological observation. 
}
\label{contour}
\end{figure}

Figure \ref{contour} illustrates the exclusion contours in the $\tilde y_t~\text{vs.}~ m_\Psi$ plane from our prompt search analyses at the LHC. These projected reach can be better appreciated along with contours that satisfy the dark matter relic density constraint for different early Universe cosmological evolution scenarios. 
The $2\sigma$ exclusion contours for the simplified FIMP model are shown in the blue-shaded region for an integrated luminosity of 300 $\text{fb}^{-1}$ and in the orange-shaded region for the projected reach at the 14 TeV High-Luminosity LHC (HL-LHC) with 3 $\text{ab}^{-1}$. The dashed-dotted orange contour represents the HL-LHC exclusion reach when a 10\% systematic uncertainty is considered. At 300 $\text{fb}^{-1}$, the discovery ($5\sigma$) and exclusion ($2\sigma$) reach for the vector-like quark mass extend up to approximately 1305 GeV and 1470 GeV, respectively. The HL-LHC exclusion reach extends up to 1700 GeV. However, when 10\% systematics are included, the exclusion sensitivity at HL-LHC reduces, reaching up to 1655 GeV.

Note that, as long as the BSM coupling $\tilde y_t$ is large enough, the demonstrated exclusion contours in Figure \ref{contour} are primarily insensitive to the coupling due to the model-independent nature of our study. One can follow Footnotes \ref{footnote_1} and \ref{footnote_5} for further discussion. However, this heavy vector-like quark, after its QCD production, can be long-lived if $\tilde y_t$ is substantially small when its decay occurs from a secondary vertex, a few millimetres to meters away from the original primary production vertex, depending upon the smallness of the coupling. For reference, we listed displaced distance before decay for all BP points in Table \ref{tab:Bemchmark}. The crucial point remains that the calorimeter energy deposits in the latter case are not associated with tracks from the hard scattering primary vertex, unlike prompt QCD production. Hence, such hadronic deposits cannot be reconstructed as a (boosted) hadronic jet from the primary vertex and part of our signal. Thus, despite the VLQ having a single decay mode, the exclusion significance reduces their sensitivity with the minuscule coupling values when $\Psi$ fails to decay promptly from the primary vertex. We approximated this reduction in sensitivity by a mathematical function as discussed in Appendix \ref{probability}. We find that if the coupling is considerably large to trigger the prompt decay of the vector-like quark, our collider analysis is capable of probing a broad region of the parameter space. Conversely, if the coupling is small, the displaced vertex + MET search \cite{Calibbi:2018fqf} can explore higher VLQ masses (lower right of Figure \ref{contour}). In contrast, our complementary search can probe larger VLQ masses when the coupling is more significant (top right of Figure \ref{contour}), which is the case for the FIMP scenario with alternative cosmology.

The CMS and ATLAS collaborations have extensively searched for supersymmetric particles. A close comparison can be drawn between our search and stop pair production at the LHC. The stop, the superpartner of the SM top quark, arises from the mixing of left- and right-handed top squarks, forming the lighter and heavier mass eigenstates $\tilde{t}_1$ and $\tilde{t}_2$​. At the LHC, the production of light stop pairs is widely studied through their decay into SM top quarks and neutralinos \cite{ATLAS:2020xzu,ATLAS:2024rcx,CMS:2019ysk,CMS:2020pyk,CMS:2021eha,ATLAS:2020dsf,CMS:2021beq}.
However, a direct comparison between stop and vector-like quark searches may not yield accurate results due to their production mechanisms and spin differences. The leading-order partonic diagrams for stop pair production resemble those in Figure \ref{Feynman_dia} (upper row); an additional gluon-gluon annihilation process via a contact interaction is present for stops. Furthermore, stops are colour scalars (spin-0), whereas VLQs are colour fermions (spin-1/2). The difference in mass dimension of the fields (1 for stop field, 3/2 for VLQ) also leads to distinct Feynman vertex structures. Nonetheless, the result of the stop search can be recast in the VLQ context to derive meaningful constraints. 

A recent analysis by the ATLAS collaboration \cite{ATLAS:2020dsf} considers stop pair production resulting in a final state with two hadronically decaying top quarks and a pair of neutralinos, leading to significant missing transverse momentum and no leptons in the final state. The study uses data collected at the 13 TeV LHC with an integrated luminosity of 139 $\text{fb}^{-1}$. This final state is identical to the one considered in our search. We reinterpret the ATLAS results using \texttt{CheckMATE2}~\cite{Drees:2013wra,Kim:2015wza,Dercks:2016npn} and find that the 95\% confidence level (CL) exclusion reach for the VLQ mass extends to 1230 GeV. In contrast, our dedicated analysis for the 13 TeV LHC with 139 $\text{fb}^{-1}$ of integrated luminosity achieves an exclusion reach of up to 1320 GeV. When a 10\% systematic uncertainty is included, the reach slightly decreases to 1292 GeV. The details about this recasting procedure are provided in Appendix \ref{Reinterpreting}. Our analysis provides improved sensitivity over the reinterpretation of stop searches, highlighting the need for dedicated optimization in VLQ searches.

A significant number of vector-like quark (top partner \cite{Buchkremer:2013bha,Ghosh:2025gdq}) searches have also been conducted at the LHC \cite{ATLAS:2024xne,CMS:2023agg,CMS:2024qdd}. In these searches, the VLQ is typically $\mathcal{Z}_2$ even, allowing it to mix with the SM top quark. As a result, ATLAS and CMS primarily focus on VLQ pair or single production, followed by decays into channels such as $tZ$, $th$, $bW$, or $tg$. In contrast, our model features a $\mathcal{Z}_2$ odd VLQ, which does not mix with the SM top quark. Instead, it interacts with SM quarks and dark matter through a Yukawa interaction.

\section{Summary and Conclusion}\label{sec7}
In this work, we examine a simplified FIMP model where the freeze-in production of dark matter relies on the decay of vector-like quark $\Psi$ into a dark matter particle ($S$) and an up-type quark through a feeble coupling. DM, being a pure singlet, $\Psi$, is forced to have the same gauge quantum number as the SM particle; in particular, the mediator poses the same gauge charge as the right-handed quark since we consider the top-philic model. During the evolution of the early Universe, the parent particle $\Psi$ thermalizes with the primordial soup due to considerable gauge interaction. However, the DM can never become part of the thermal bath owing to the smallness of the Yukawa coupling, and its generation takes place from the decay of $\Psi$ via the freeze-in mechanism.

Since the vector-like quark mediator is already constrained to be rather heavy while the dark matter can be light, it requires an approximate range of Yukawa coupling $\sim(10^{-7}-10^{-13})$ to guarantee the observed dark matter relic through the freeze-in process, considering the standard radiation-dominated early Universe. Hence, the long-live particle searches and displaced vertex signature typically play an essential role in constraining this kind of picture at LHC. On top of that, the typical $\mathcal{O}(10)$ keV dark matter can resolve the small-scale crisis of $\Lambda$CDM model by putting down the structure formation at that scale due to the warmness, in addition to offering the platform to be probed at various astrophysical and cosmological experiments.

However, there is no direct measurement of the Universe's energy budget in the pre-BBN era, so looking at possible alternative background cosmologies is prudent. Here, we adopt a particular scenario with a pre-BBN fast expansion picture, such as a well-motivated quintessence fluid. It alters the dark matter phenomenology and the collider signature by demanding a significant increase in interaction strength compared to the standard radiation-dominated assumption to achieve the DM abundance at the correct ballpark. The characteristic LLP searches and DV signatures fall short of effectively constraining such a scenario. This analysis shows the extended parameter space due to modified cosmology, satisfying the relic density constraint. We built a suitable search strategy at LHC to significantly incarcerate a large part of newly opened regions.

Finally, the generous pair production of vector-like quarks is expected at LHC owing to its strong interaction and promptly decays into the dark matter along with the SM quarks. The production of light dark matter particles from the decay of heavy VLQ ensures the output of top-like boosted fatjets. We exploit the inherent features of these jets and their substructure to construct our search strategy and probe this channel over a vast QCD backdrop. With a precision estimate of the signal at the next-to-leading order QCD, particle-level event simulation, the inclusion of detector effect, and multivariate analysis based on the boosted decision tree and boosted jet variables, a comprehensive computation is done to showcase the possible reach of this model. In this study, our collider analysis explores a broad parameter space of the model in the Yukawa coupling and VLQ mass plane. Our search strategy provides a complementary way to constrain the extended region parameter space for the dark matter, where most regions remain inaccessible to the traditional long-live particle or displaced vertex searches.

\acknowledgments
 This work is supported by the Physical Research Laboratory (PRL), Department of Space, Government of India. The computational works are performed using the Param Vikram-1000 High-Performance Computing Cluster and the TDP project resources of the Physical Research Laboratory (PRL). SS is supported by NPDF grant PDF/2023/002076 from the Science and Engineering Research Board (SERB), Government of India. We sincerely thank the anonymous referee for pointing out the underestimation of the signal efficacy in the calculation of the excluded region plot.

\appendix 
\section{Appendix}
\subsection{Upper bound on $\tilde y_t$ from DM thermalization for fast expanding universe}\label{upper_bound}

\begin{figure}[htb!]
	\centering
	\includegraphics[height=6.3cm,width=11cm]{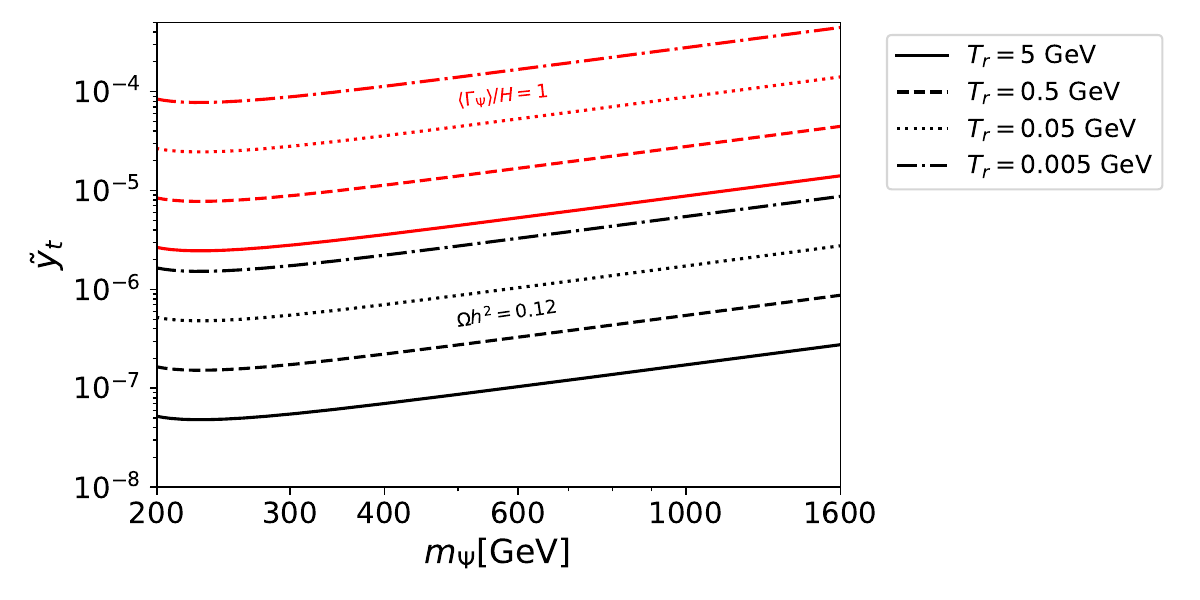}
	\caption{Thermalization contours (defined by the interaction rate equal to the Hubble rate) for different reference temperatures with a fixed value of $n(=2)$ are shown in red in Yukawa coupling vs mediator mass plane. Set of Black lines at the bottom represent the relic density satisfied contours for the same fixed values of $T_r$ and $n$. The corresponding DM is not thermalized (gets thermalized) for a parameter region below (above) each contour.}
	\label{upper_bound_y}
\end{figure}

To understand the theoretical validity of Yukawa coupling $\tilde y_t$ for a given VLQ mass, one is required to calculate the ratio between the corresponding interaction rate with the Hubble expansion rate, i.e. $\langle\Gamma_\Psi\rangle/H$ and check whether this ratio exceeds one. The DM thermalizes for a ratio larger than or equal to one. 
In Figure~\ref{upper_bound_y}, we have shown the contours with the ratio $\langle\Gamma_\Psi\rangle/H=1$ with the red colour lines. This contour for a particular $T_r$ sets a conservative upper bound on $\tilde y_t$ to realize non-thermal DM. Clearly, relic-satisfied points are well within this limit.

\subsection{Distributions after baseline selection}
\label{Appen:baseline}
\begin{figure}[h!]
	\centering
	 {\label{fig:met_pre}               \includegraphics[width=0.24\textwidth]{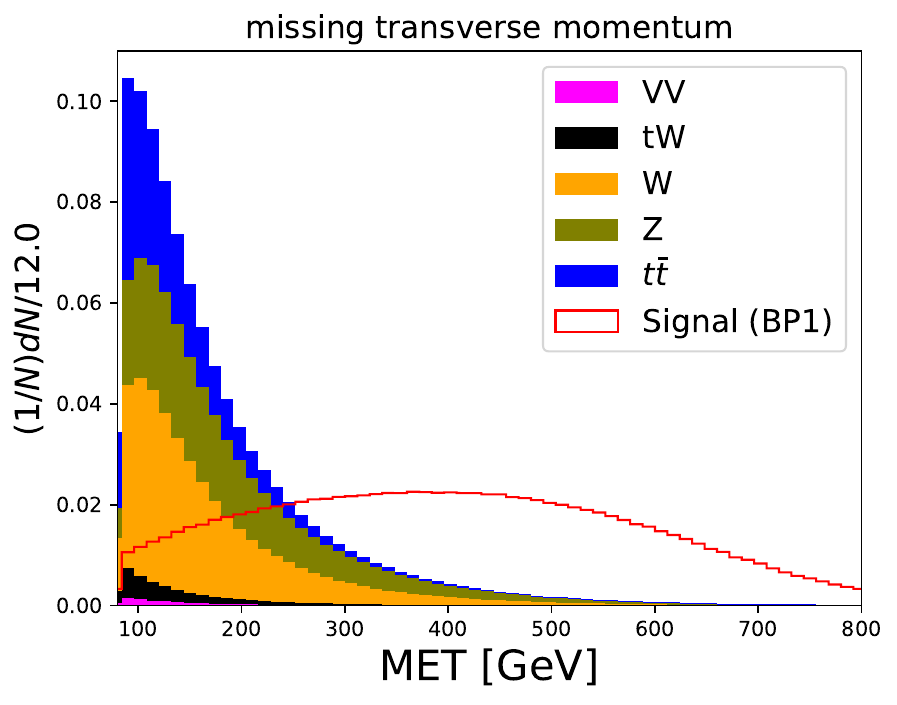}}
         {\label{fig:MJ0_pre}               \includegraphics[width=0.24\textwidth]{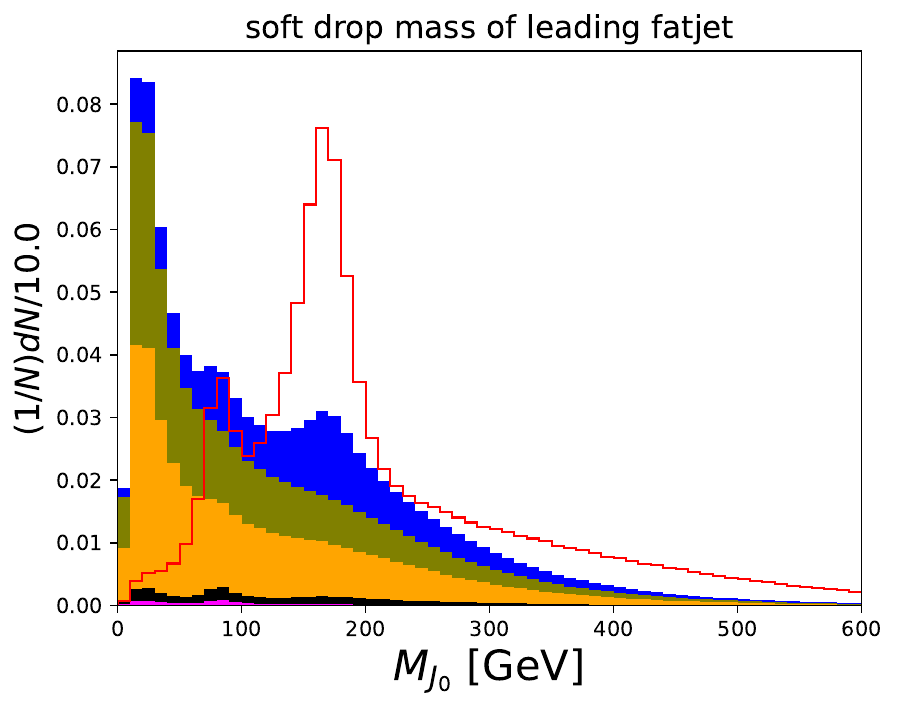}} 
	 {\label{fig:MJ1_pre}               \includegraphics[width=0.24\textwidth]{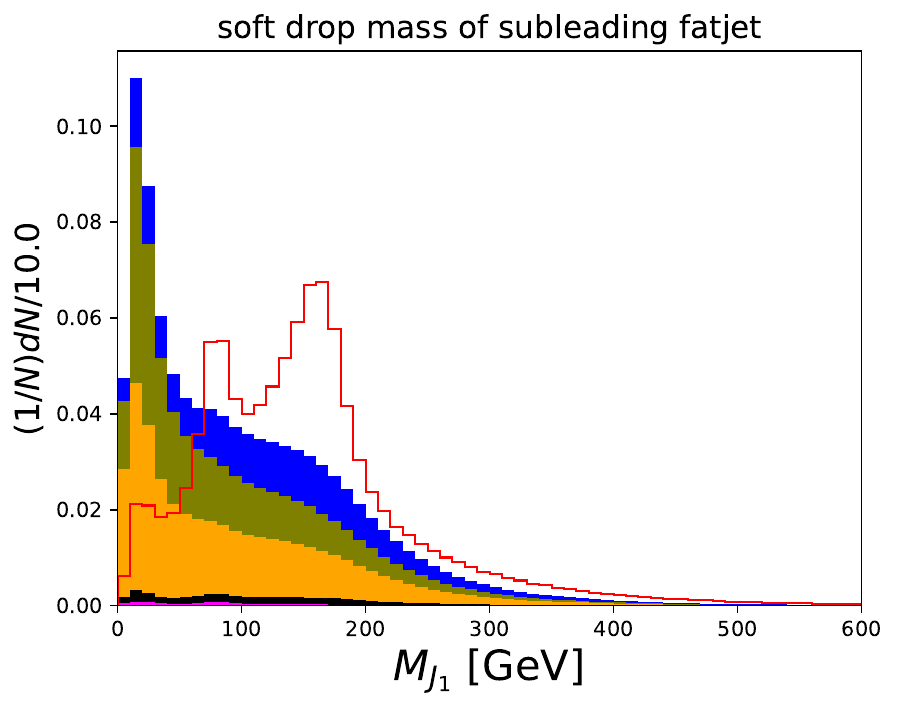}} 
	 {\label{fig:DPhi_J0met_pre} \includegraphics[width=0.24\textwidth]{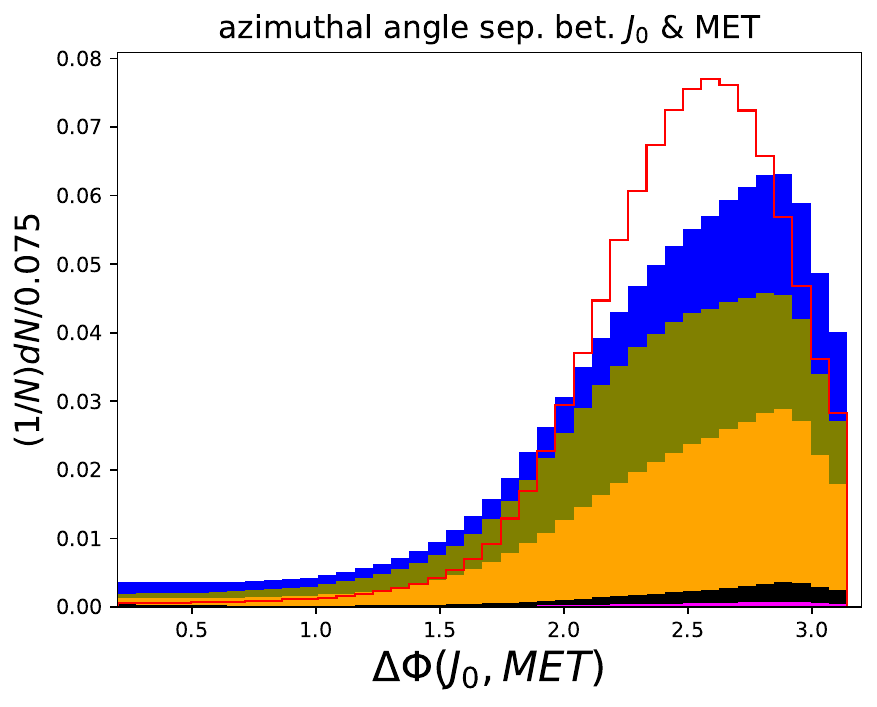}} \\ 
     {\label{fig:HT_pre} \includegraphics[width=0.24\textwidth]{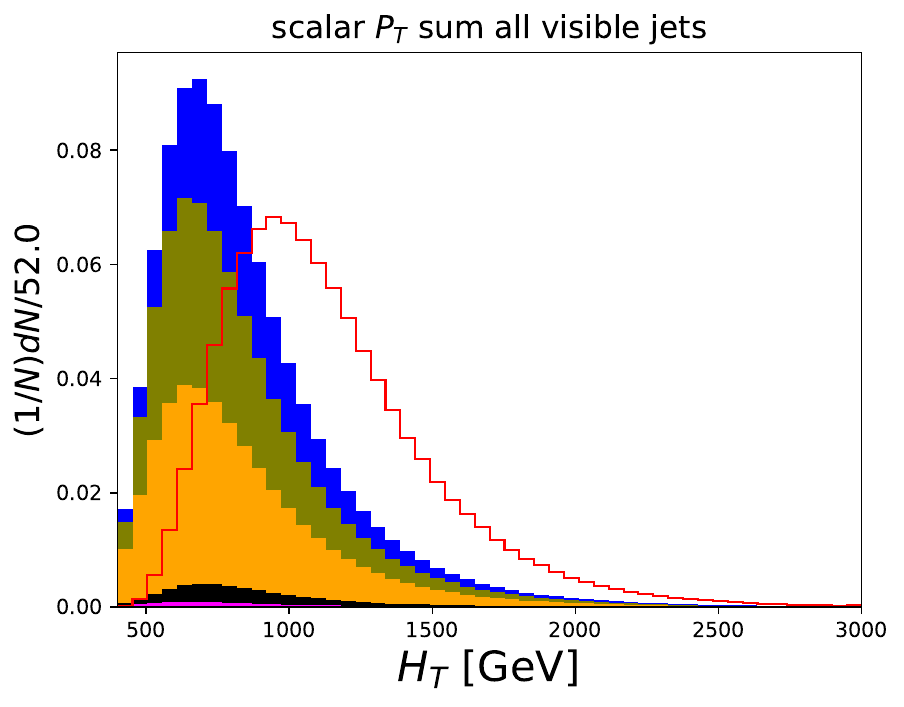}}
	 {\label{fig:shat_pre} \includegraphics[width=0.24\textwidth]{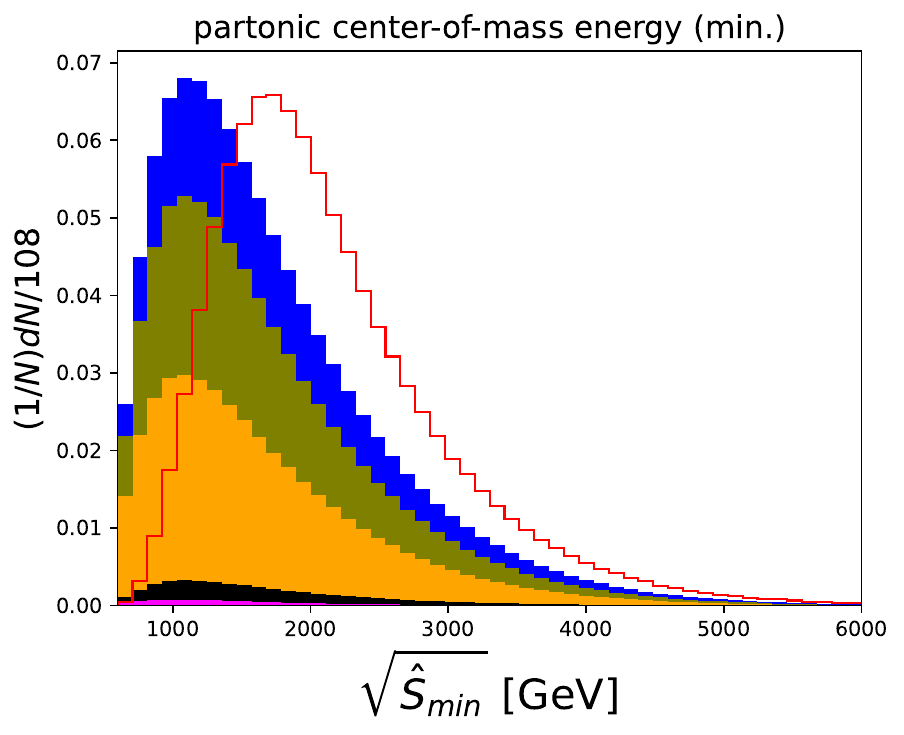}}
	{\label{fig:tau21_J0_pre} \includegraphics[width=0.24\textwidth]{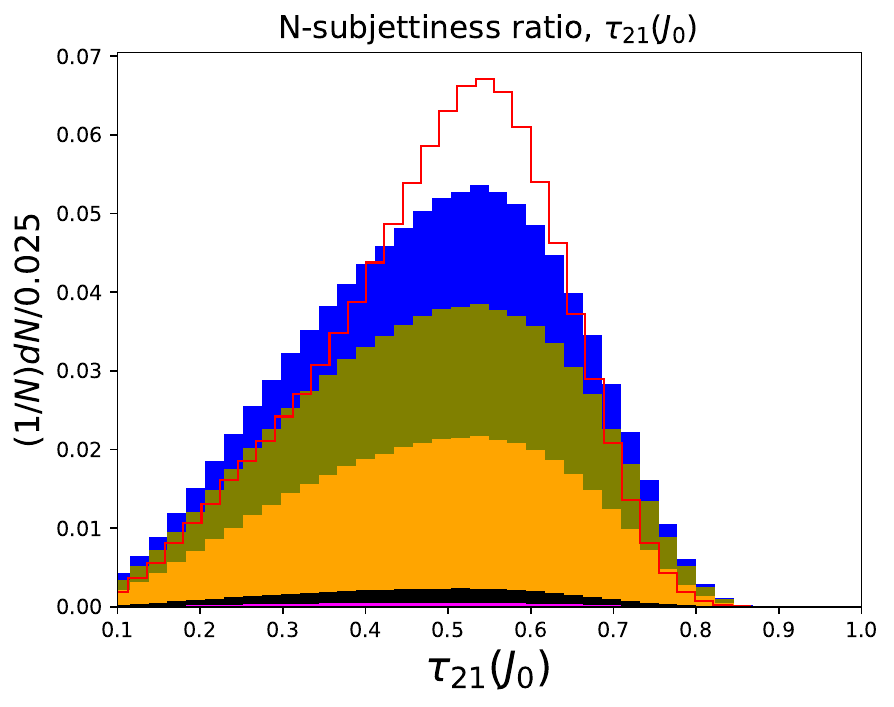}}
	{\label{fig:tau21_J1_pre} \includegraphics[width=0.24\textwidth]{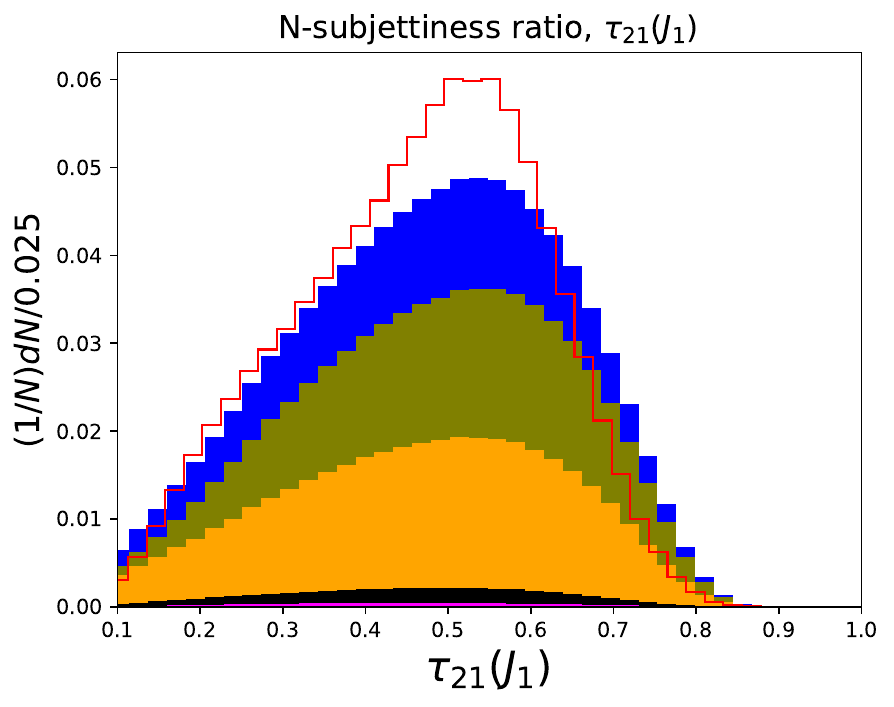}} 
	\caption{Some of the normalized stacked histograms of various kinematic observables for the SM backgrounds and the signal (BP1) are presented after applying only the baseline selection criteria: at least two fatjets with $p_T(J_i)>200$ GeV, MET $> 80$ GeV, and $\Delta \phi(J_i,\text{MET})>0.2~(i=0,1)$.}
	\label{fig:sig_bg_pre}
\end{figure}

After applying only the baseline selection criteria, normalized distributions of various kinematic observables for the signal (BP1) and backgrounds are presented in Figure \ref{fig:sig_bg_pre}. At the baseline selection, the dominant background comes from the $W+$jets channel, with $Z+$jets and $tt+$jets providing the second and third most significant contribution. After applying the full event selection criteria detailed in Subsection \ref{selection}, the primary background shifts to $t\bar{t}+$jets, followed by $Z+$jets and $W+$jets, as shown in Table \ref{tab:cut_flow}.

Figure \ref{fig:sig_bg_pre} reveals that both fatjets from the signal exhibit a second soft-drop mass peak around the masses of the $W$ boson, along with one at the top quark. This low mass peak arises when the fatjets are moderately boosted, capturing only the subjets corresponding to the $W$ boson decay rather than all three subjets of top quark. Additionally, a small shift of the peaks in $\sqrt{\hat{s}_{\text{min}}}$, $\Delta \Phi(J_i,\rm MET)$, and $\tau_{21}(J_i)$ distributions is observed between Figure \ref{fig:sig_bg_pre} and Figure \ref{fig:sig_bg_1} (after applying all event selection cuts).  

\subsection{BDT Hyperparameters}
\label{params}
To enhance the better performance in signal-background discrimination, we perform a multivariate analysis utilizing a gradient boosted decision tree (BDT) method where several characteristics of the reconstructed fatjets and the missing transverse momentum are employed as inputs for the BDT. In Table \ref{tab:opti_param} we list the hyperparameters used in adaptive BDT analysis.

\begin{table}[h!]
\begin{center}
 \begin{tabular}{|c c|}
\hline
BDT hyperparameters & Optimal selection \\ 
\hline\hline
BoostType & AdaBoost  \\
AdaBoostBeta & 0.3  \\
NTrees & 400, 250, 210  \\
MinNodeSize & $5.0\%$  \\
MaxDepth & 4  \\
UseBaggedBoost & True  \\
BaggedSampleFraction & 0.6  \\
SeparationType & GiniIndex  \\
nCuts & 20  \\

\hline
 \end{tabular} 
\caption{Hyperparameters used in adaptive BDT analysis. The number of trees (NTrees) is specified for BP1, BP2, and BP3, respectively.}
\label{tab:opti_param}
\end{center}
\end{table}

\subsection{Exclusion reach as a function of coupling}
\label{probability}

As discussed in the main text, our collider analysis is based on the prompt decay of the vector-like quark ($\Psi$) into the dark matter and the top quark. In our simplified model, $\Psi$ has only one decay channel, and its decay lifetime depends on the coupling $\tilde{y}_t$ and its mass. The decay length of the VLQ in its rest frame is given below. 
\begin{equation}
	c\tau=\dfrac{1.98\times10^{-13}}{\Gamma_\Psi(\tilde{y}_t, m_{\Psi})}~mm,
	\label{decay}
\end{equation}
Here, $\Gamma_\Psi(\tilde{y}_t, m_{\Psi})$ represents the decay width of the VLQ in its rest frame, defined in Equation \ref{DM_phen2}. For our prompt search analysis, estimating the fraction of the $\Psi$ that decays within the tracking system at the LHC is necessary, which can provide us with the boosted fatjet signature. The probability that $\Psi$ can traverse a distance $x$ in the detector from the primary vertex without decaying is given by 
\begin{equation}
	P(x)=\exp(-x/\gamma \beta c \tau),
	\label{prob}
\end{equation}
Here, $\gamma=\dfrac{1}{\sqrt{1-\beta^2}}$ represents the Lorentz factor, and $\beta=v/c$. In natural units ($\hslash$=c=1), some useful relations for the rotation-free Lorentz transformation:
\begin{eqnarray}
\gamma =\frac{E_\Psi}{m_\Psi} ~~~\text{and}~~~\beta =\dfrac{|\vec{p}_\Psi|}{E_\Psi}~.
\end{eqnarray}
The momentum of the VLQ, \( |\vec{p}_\Psi|\), depends on the partonic center-of-mass energy, \( \sqrt{\hat{s}} \), which typically lies above the production threshold, \( \sqrt{\hat{s}} \gtrsim 2m_\Psi \), for VLQ pair production. If $\Psi$ possesses a long lifetime, we must rescale the pair-production cross-section of the vector-like quark for our prompt search as an effective cross-section,
\begin{equation}
	\sigma_{\text{eff}} = \sigma~ (1-P(x))~.
	\label{xsec-prob}
\end{equation}
The surviving probabilities for prompt search ($1 - P(x)$) increase for larger values of $x$.  Hence, for our prompt search, one can get a conservative estimate by setting the detecting scale at one millimetre and boost factor $\gamma \beta = 1.0$. That would reproduce the $\tilde{y}_t$ dependence of the exclusion contour, especially for lower values. In principle, since we know the exact $\sqrt{\hat{s}}$ distributions of our simulated events (for a given mass), the exact distribution of $\gamma \beta$ can be computed trivially to derive this contour. However, this approximation should work equally well, especially since our analysis is not optimized for this region; instead, displaced vertex search is expected to take care of this region.


\subsection{Reinterpreting the stop searches at ATLAS}
\label{Reinterpreting}

We performed a detailed reinterpretation based on the existing 13 TeV LHC data with an integrated luminosity of 139~fb$^{-1}$. A search for top squarks in the all-hadronic final state with a top--antitop pair and missing transverse momentum has been presented by the ATLAS collaboration. The analysis is based on 139~fb$^{-1}$ of proton--proton collision data at a center-of-mass energy of 13~TeV, collected with the ATLAS detector at the LHC~\cite{ATLAS:2020dsf}. We reinterpret this analysis in the context of our top-philic dark matter model using the \texttt{CheckMATE2} framework~\cite{Drees:2013wra,Kim:2015wza,Dercks:2016npn}. 

The ATLAS analysis defines multiple signal regions, labeled SRA to SRD, based on the stop's two- to four-body decay topologies. The SRA regions are specifically optimized for scenarios with two-body stop decays, large mass gaps between the top squark and the neutralino dark matter candidate, and stop masses above 1~TeV. Since our model features a VLQ with a mass at the TeV scale and a light dark matter candidate, we concentrate on the SRA region. In these regions, the dominant background is typically \( Z+\text{jets} \), where the \( Z \)-boson decays invisibly into neutrinos. However, our analysis reveals that the two main background processes—\( Z(\text{inv.})+\text{jets} \) and semi-leptonic \( t\bar{t}+\text{jets} \)—contribute comparably to the total background. In fact, the semi-leptonic \( t\bar{t}+\text{jets} \) channel contributes a larger number of background events than \( Z(\text{inv.})+\text{jets} \). The ATLAS analysis models the stop pair production signal at leading order (LO) in QCD.\\

In the SRA region, the leading reconstructed fat jet ($R=1.2$) is required to have a mass greater than 120 GeV and a $b$-jet is tagged within leading fat jet. Based on the mass of the second fat jet, the SRA region is further divided into three subregions:
\begin{itemize}
	\item SRA-TT: second fat jet mass $> 120$ GeV, and a $b$-jet is tagged within sub-leading fat jet.
	\item SRA-TW: second fat jet mass between $60$\,GeV and $120$\,GeV.
	\item SRA-T0: second fat jet mass $< 60$\,GeV.
\end{itemize}
\textit{We recast our VLQ signal and obtain a \(2\,\sigma\) reach using the signal regions SRA-TT and SRA-TW. For the recasting, the VLQ signal is generated at the 13~TeV LHC using the NNPDF2.3 parton distribution function, consistent with the ATLAS analysis. For a benchmark dark matter mass of 12~keV, the 95\% confidence level exclusion limits obtained from this reinterpretation are summarized below:}

\begin{enumerate}
	\item When using next-to-leading order (NLO) cross sections and NLO signal distributions, the results exclude vector-like quark (VLQ) masses up to 1230\,GeV and 1210\,GeV in the SRA-TT and SRA-TW signal regions, respectively.	
	\item When using leading-order (LO) cross sections while retaining NLO signal distributions, the exclusion reach extends up to 1190\,GeV and 1145\,GeV in the SRA-TT and SRA-TW signal regions, respectively.	
\end{enumerate}
In comparison, our independent analysis with NLO signal events at \(\sqrt{s} = 13\)~TeV LHC with 139~fb\(^{-1}\), incorporating all the optimization strategies and kinematic observables discussed in our manuscript, shows that VLQ masses up to 1320~GeV can be excluded. When a 10\% systematic uncertainty is included, the exclusion reach slightly reduces to 1292~GeV.

Even though existing stop searches in the all-hadronic final state can be recast for the VLQ case, our collider analysis aids the following. In existing searches, the dominant background in the SRA-TT and SRA-TW scenarios is \( Z(\text{inv.})+\text{jets} \). However, our analysis shows that semi-leptonic \( t\bar{t}+\text{jets} \) contributes a comparable--and in fact larger--number of background events than \( Z(\text{inv.})+\text{jets} \). Therefore, the VLQ signal model requires a different optimization strategy compared to stop searches. Furthermore, using NLO cross sections and signal distributions leads to a more stringent bound on the VLQ mass. Finally, our proposed analysis achieves a better reach on the VLQ mass compared to the recasted stop analysis.

\bibliographystyle{JHEP}
\bibliography{Reference}

\providecommand{\href}[2]{#2}\begingroup\raggedright\begin{thebibliography}{100}

\bibitem{Sofue:2000jx}
Y.~Sofue and V.~Rubin, \emph{{Rotation curves of spiral galaxies}},
  \href{http://dx.doi.org/10.1146/annurev.astro.39.1.137}{\emph{Ann. Rev.
  Astron. Astrophys.} {\bfseries 39} (2001) 137--174},
  [\href{https://arxiv.org/abs/astro-ph/0010594}{{\ttfamily
  astro-ph/0010594}}].

\bibitem{Clowe:2006eq}
D.~Clowe, M.~Bradac, A.~H. Gonzalez, M.~Markevitch, S.~W. Randall, C.~Jones
  et~al., \emph{{A direct empirical proof of the existence of dark matter}},
  \href{http://dx.doi.org/10.1086/508162}{\emph{Astrophys. J. Lett.} {\bfseries
  648} (2006) L109--L113},
  [\href{https://arxiv.org/abs/astro-ph/0608407}{{\ttfamily
  astro-ph/0608407}}].

\bibitem{Hinshaw_2013}
G.~Hinshaw, D.~Larson, E.~Komatsu, D.~N. Spergel, C.~L. Bennett, J.~Dunkley
  et~al., \emph{Nine-year wilkinson microwave anisotropy probe ( wmap )
  observations: Cosmological parameter results},
  \href{http://dx.doi.org/10.1088/0067-0049/208/2/19}{\emph{The Astrophysical
  Journal Supplement Series} {\bfseries 208} (Sep, 2013) 19}.

\bibitem{Planck:2018vyg}
{\scshape Planck} collaboration, N.~Aghanim et~al., \emph{{Planck 2018 results.
  VI. Cosmological parameters}},
  \href{http://dx.doi.org/10.1051/0004-6361/201833910}{\emph{Astron.
  Astrophys.} {\bfseries 641} (2020) A6},
  [\href{https://arxiv.org/abs/1807.06209}{{\ttfamily 1807.06209}}].

\bibitem{Drees:2018hzm}
M.~Drees, \emph{{Dark Matter Theory}},
  \href{http://dx.doi.org/10.22323/1.340.0730}{\emph{PoS} {\bfseries ICHEP2018}
  (2019) 730}, [\href{https://arxiv.org/abs/1811.06406}{{\ttfamily
  1811.06406}}].

\bibitem{Kolb:1990vq}
E.~W. Kolb and M.~S. Turner, \emph{{The Early Universe}}, vol.~69.
\newblock 1990.

\bibitem{Arcadi:2017kky}
G.~Arcadi, M.~Dutra, P.~Ghosh, M.~Lindner, Y.~Mambrini, M.~Pierre et~al.,
  \emph{{The waning of the WIMP? A review of models, searches, and
  constraints}},
  \href{http://dx.doi.org/10.1140/epjc/s10052-018-5662-y}{\emph{Eur. Phys. J.
  C} {\bfseries 78} (2018) 203},
  [\href{https://arxiv.org/abs/1703.07364}{{\ttfamily 1703.07364}}].

\bibitem{Ghosh:2021noq}
A.~Ghosh, P.~Konar and S.~Seth, \emph{{Precise probing of the inert
  Higgs-doublet model at the LHC}},
  \href{http://dx.doi.org/10.1103/PhysRevD.105.115038}{\emph{Phys. Rev. D}
  {\bfseries 105} (2022) 115038},
  [\href{https://arxiv.org/abs/2111.15236}{{\ttfamily 2111.15236}}].

\bibitem{Ghosh:2024boo}
A.~Ghosh and P.~Konar, \emph{{Unveiling desert region in inert doublet model
  assisted by Peccei-Quinn symmetry}},
  \href{http://dx.doi.org/10.1007/JHEP09(2024)104}{\emph{JHEP} {\bfseries 09}
  (2024) 104}, [\href{https://arxiv.org/abs/2407.01415}{{\ttfamily
  2407.01415}}].

\bibitem{Ghosh:2022rta}
A.~Ghosh, P.~Konar and R.~Roshan, \emph{{Top-philic dark matter in a hybrid
  KSVZ axion framework}},
  \href{http://dx.doi.org/10.1007/JHEP12(2022)167}{\emph{JHEP} {\bfseries 12}
  (2022) 167}, [\href{https://arxiv.org/abs/2207.00487}{{\ttfamily
  2207.00487}}].

\bibitem{Ghosh:2023xhs}
A.~Ghosh and P.~Konar, \emph{{Precision prediction of a democratic up-family
  philic KSVZ axion model at the LHC}},
  \href{http://dx.doi.org/10.1016/j.dark.2024.101746}{\emph{Phys. Dark Univ.}
  {\bfseries 47} (2025) 101746},
  [\href{https://arxiv.org/abs/2305.08662}{{\ttfamily 2305.08662}}].

\bibitem{Green:2002ht}
A.~M. Green, \emph{{Effect of halo modeling on WIMP exclusion limits}},
  \href{http://dx.doi.org/10.1103/PhysRevD.66.083003}{\emph{Phys. Rev. D}
  {\bfseries 66} (2002) 083003},
  [\href{https://arxiv.org/abs/astro-ph/0207366}{{\ttfamily
  astro-ph/0207366}}].

\bibitem{Bae:2014rfa}
K.~J. Bae, H.~Baer, A.~Lessa and H.~Serce, \emph{{Coupled Boltzmann computation
  of mixed axion neutralino dark matter in the SUSY DFSZ axion model}},
  \href{http://dx.doi.org/10.1088/1475-7516/2014/10/082}{\emph{JCAP} {\bfseries
  10} (2014) 082}, [\href{https://arxiv.org/abs/1406.4138}{{\ttfamily
  1406.4138}}].

\bibitem{Chang:2017gla}
C.-F. Chang, X.-G. He and J.~Tandean, \emph{{Two-Higgs-Doublet-Portal
  Dark-Matter Models in Light of Direct Search and LHC Data}},
  \href{http://dx.doi.org/10.1007/JHEP04(2017)107}{\emph{JHEP} {\bfseries 04}
  (2017) 107}, [\href{https://arxiv.org/abs/1702.02924}{{\ttfamily
  1702.02924}}].

\bibitem{Ghosh:2025agw}
A.~Ghosh, \emph{{Unveiling a natural multicomponent dark sector: an inert
  doublet guided by Peccei{\textendash}Quinn}},
  \href{http://dx.doi.org/10.1140/epjp/s13360-025-06621-5}{\emph{Eur. Phys. J.
  Plus} {\bfseries 140} (2025) 688}.

\bibitem{Srivastava:2025oer}
T.~Srivastava, J.~Das, A.~Ghosh and A.~Chaudhuri, \emph{{Electroweak Phase
  Transition, Gravitational Waves and Collider Probes in Multi-Scalar Dark
  Matter Scenarios}},  \href{https://arxiv.org/abs/2507.05917}{{\ttfamily
  2507.05917}}.

\bibitem{Visinelli:2017qga}
L.~Visinelli, \emph{{(Non-)thermal production of WIMPs during kination}},
  \href{http://dx.doi.org/10.3390/sym10110546}{\emph{Symmetry} {\bfseries 10}
  (2018) 546}, [\href{https://arxiv.org/abs/1710.11006}{{\ttfamily
  1710.11006}}].

\bibitem{Arcadi:2017wqi}
G.~Arcadi, M.~Lindner, F.~S. Queiroz, W.~Rodejohann and S.~Vogl,
  \emph{{Pseudoscalar Mediators: A WIMP model at the Neutrino Floor}},
  \href{http://dx.doi.org/10.1088/1475-7516/2018/03/042}{\emph{JCAP} {\bfseries
  03} (2018) 042}, [\href{https://arxiv.org/abs/1711.02110}{{\ttfamily
  1711.02110}}].

\bibitem{Reinert:2017aga}
A.~Reinert and M.~W. Winkler, \emph{{A Precision Search for WIMPs with Charged
  Cosmic Rays}},
  \href{http://dx.doi.org/10.1088/1475-7516/2018/01/055}{\emph{JCAP} {\bfseries
  01} (2018) 055}, [\href{https://arxiv.org/abs/1712.00002}{{\ttfamily
  1712.00002}}].

\bibitem{Evans:2017kti}
J.~A. Evans, S.~Gori and J.~Shelton, \emph{{Looking for the WIMP Next Door}},
  \href{http://dx.doi.org/10.1007/JHEP02(2018)100}{\emph{JHEP} {\bfseries 02}
  (2018) 100}, [\href{https://arxiv.org/abs/1712.03974}{{\ttfamily
  1712.03974}}].

\bibitem{Garny:2018icg}
M.~Garny, J.~Heisig, M.~Hufnagel and B.~L\"ulf, \emph{{Top-philic dark matter
  within and beyond the WIMP paradigm}},
  \href{http://dx.doi.org/10.1103/PhysRevD.97.075002}{\emph{Phys. Rev. D}
  {\bfseries 97} (2018) 075002},
  [\href{https://arxiv.org/abs/1802.00814}{{\ttfamily 1802.00814}}].

\bibitem{Blanco:2019hah}
C.~Blanco, M.~Escudero, D.~Hooper and S.~J. Witte, \emph{{Z' mediated WIMPs:
  dead, dying, or soon to be detected?}},
  \href{http://dx.doi.org/10.1088/1475-7516/2019/11/024}{\emph{JCAP} {\bfseries
  11} (2019) 024}, [\href{https://arxiv.org/abs/1907.05893}{{\ttfamily
  1907.05893}}].

\bibitem{Bhardwaj:2018lma}
A.~Bhardwaj, A.~Das, P.~Konar and A.~Thalapillil, \emph{{Looking for Minimal
  Inverse Seesaw scenarios at the LHC with Jet Substructure Techniques}},
  \href{http://dx.doi.org/10.1088/1361-6471/ab7769}{\emph{J. Phys. G}
  {\bfseries 47} (2020) 075002},
  [\href{https://arxiv.org/abs/1801.00797}{{\ttfamily 1801.00797}}].

\bibitem{Bhardwaj:2019mts}
A.~Bhardwaj, P.~Konar, T.~Mandal and S.~Sadhukhan, \emph{{Probing the inert
  doublet model using jet substructure with a multivariate analysis}},
  \href{http://dx.doi.org/10.1103/PhysRevD.100.055040}{\emph{Phys. Rev. D}
  {\bfseries 100} (2019) 055040},
  [\href{https://arxiv.org/abs/1905.04195}{{\ttfamily 1905.04195}}].

\bibitem{Konar:2020wvl}
P.~Konar, A.~Mukherjee, A.~K. Saha and S.~Show, \emph{{Linking pseudo-Dirac
  dark matter to radiative neutrino masses in a singlet-doublet scenario}},
  \href{http://dx.doi.org/10.1103/PhysRevD.102.015024}{\emph{Phys. Rev. D}
  {\bfseries 102} (2020) 015024},
  [\href{https://arxiv.org/abs/2001.11325}{{\ttfamily 2001.11325}}].

\bibitem{Konar:2020vuu}
P.~Konar, A.~Mukherjee, A.~K. Saha and S.~Show, \emph{{A dark clue to seesaw
  and leptogenesis in a pseudo-Dirac singlet doublet scenario with
  (non)standard cosmology}},
  \href{http://dx.doi.org/10.1007/JHEP03(2021)044}{\emph{JHEP} {\bfseries 03}
  (2021) 044}, [\href{https://arxiv.org/abs/2007.15608}{{\ttfamily
  2007.15608}}].

\bibitem{Heurtier:2019beu}
L.~Heurtier and H.~Partouche, \emph{{Spontaneous Freeze Out of Dark Matter From
  an Early Thermal Phase Transition}},
  \href{http://dx.doi.org/10.1103/PhysRevD.101.043527}{\emph{Phys. Rev. D}
  {\bfseries 101} (2020) 043527},
  [\href{https://arxiv.org/abs/1912.02828}{{\ttfamily 1912.02828}}].

\bibitem{Habermehl:2020njb}
M.~Habermehl, M.~Berggren and J.~List, \emph{{WIMP Dark Matter at the
  International Linear Collider}},
  \href{http://dx.doi.org/10.1103/PhysRevD.101.075053}{\emph{Phys. Rev. D}
  {\bfseries 101} (2020) 075053},
  [\href{https://arxiv.org/abs/2001.03011}{{\ttfamily 2001.03011}}].

\bibitem{Xing:2021pkb}
C.-Y. Xing and S.-H. Zhu, \emph{{Dark Matter Freeze-Out via Catalyzed
  Annihilation}},
  \href{http://dx.doi.org/10.1103/PhysRevLett.127.061101}{\emph{Phys. Rev.
  Lett.} {\bfseries 127} (2021) 061101},
  [\href{https://arxiv.org/abs/2102.02447}{{\ttfamily 2102.02447}}].

\bibitem{Borah:2022byb}
D.~Borah, S.~Jyoti~Das, A.~K. Saha and R.~Samanta, \emph{{Probing WIMP dark
  matter via gravitational waves\textquoteright{} spectral shapes}},
  \href{http://dx.doi.org/10.1103/PhysRevD.106.L011701}{\emph{Phys. Rev. D}
  {\bfseries 106} (2022) L011701},
  [\href{https://arxiv.org/abs/2202.10474}{{\ttfamily 2202.10474}}].

\bibitem{Belanger:2022qxt}
G.~Belanger, A.~Mjallal and A.~Pukhov, \emph{{WIMP and FIMP dark matter in the
  inert doublet plus singlet model}},
  \href{http://dx.doi.org/10.1103/PhysRevD.106.095019}{\emph{Phys. Rev. D}
  {\bfseries 106} (2022) 095019},
  [\href{https://arxiv.org/abs/2205.04101}{{\ttfamily 2205.04101}}].

\bibitem{Bernal:2022wck}
N.~Bernal and Y.~Xu, \emph{{WIMPs during reheating}},
  \href{http://dx.doi.org/10.1088/1475-7516/2022/12/017}{\emph{JCAP} {\bfseries
  12} (2022) 017}, [\href{https://arxiv.org/abs/2209.07546}{{\ttfamily
  2209.07546}}].

\bibitem{Kundu:2021cmo}
S.~Kundu, A.~Guha, P.~K. Das and P.~S.~B. Dev, \emph{{EFT analysis of
  leptophilic dark matter at future electron-positron colliders in the
  mono-photon and mono-Z channels}},
  \href{http://dx.doi.org/10.1103/PhysRevD.107.015003}{\emph{Phys. Rev. D}
  {\bfseries 107} (2023) 015003},
  [\href{https://arxiv.org/abs/2110.06903}{{\ttfamily 2110.06903}}].

\bibitem{Medina:2021ram}
A.~D. Medina, N.~I. Mileo, A.~Szynkman and S.~A. Tanco, \emph{{Elusive muonic
  WIMP}}, \href{http://dx.doi.org/10.1103/PhysRevD.106.075018}{\emph{Phys. Rev.
  D} {\bfseries 106} (2022) 075018},
  [\href{https://arxiv.org/abs/2112.09103}{{\ttfamily 2112.09103}}].

\bibitem{Tallman:2022nts}
B.~Tallman, A.~Boone, A.~Vijayakumar, F.~Lopez, S.~Apata, J.~Martinez et~al.,
  \emph{{Potential for definitive discovery of a 70 GeV dark matter WIMP with
  only second-order gauge couplings}},
  \href{https://arxiv.org/abs/2210.15019}{{\ttfamily 2210.15019}}.

\bibitem{Kang:2022zqv}
S.~Kang, A.~Kar and S.~Scopel, \emph{{Halo-independent bounds on the
  non-relativistic effective theory of WIMP-nucleon scattering from direct
  detection and neutrino observations}},
  \href{https://arxiv.org/abs/2212.05774}{{\ttfamily 2212.05774}}.

\bibitem{Dutta:2022wdi}
K.~Dutta, A.~Ghosh, A.~Kar and B.~Mukhopadhyaya, \emph{{MeV to multi-TeV
  thermal WIMPs are all observationally allowed}},
  \href{https://arxiv.org/abs/2212.09795}{{\ttfamily 2212.09795}}.

\bibitem{Bernal:2023ura}
N.~Bernal, P.~Konar and S.~Show, \emph{{Unitarity bound on dark matter in
  low-temperature reheating scenarios}},
  \href{http://dx.doi.org/10.1103/PhysRevD.109.035018}{\emph{Phys. Rev. D}
  {\bfseries 109} (2024) 035018},
  [\href{https://arxiv.org/abs/2311.01587}{{\ttfamily 2311.01587}}].

\bibitem{Bernal:2024yhu}
N.~Bernal, K.~Deka and M.~Losada, \emph{{Thermal dark matter with
  low-temperature reheating}},
  \href{http://dx.doi.org/10.1088/1475-7516/2024/09/024}{\emph{JCAP} {\bfseries
  09} (2024) 024}, [\href{https://arxiv.org/abs/2406.17039}{{\ttfamily
  2406.17039}}].

\bibitem{Silva-Malpartida:2024emu}
J.~Silva-Malpartida, N.~Bernal, J.~Jones-P\'erez and R.~A. Lineros, \emph{{From
  WIMPs to FIMPs: Impact of Early Matter Domination}},
  \href{https://arxiv.org/abs/2408.08950}{{\ttfamily 2408.08950}}.

\bibitem{Akerib:2016vxi}
{\scshape LUX} collaboration, D.~S. Akerib et~al., \emph{{Results from a search
  for dark matter in the complete LUX exposure}},
  \href{http://dx.doi.org/10.1103/PhysRevLett.118.021303}{\emph{Phys. Rev.
  Lett.} {\bfseries 118} (2017) 021303},
  [\href{https://arxiv.org/abs/1608.07648}{{\ttfamily 1608.07648}}].

\bibitem{Cui:2017nnn}
{\scshape PandaX-II} collaboration, X.~Cui et~al., \emph{{Dark Matter Results
  From 54-Ton-Day Exposure of PandaX-II Experiment}},
  \href{http://dx.doi.org/10.1103/PhysRevLett.119.181302}{\emph{Phys. Rev.
  Lett.} {\bfseries 119} (2017) 181302},
  [\href{https://arxiv.org/abs/1708.06917}{{\ttfamily 1708.06917}}].

\bibitem{Zhang:2018xdp}
{\scshape PandaX} collaboration, H.~Zhang et~al., \emph{{Dark matter direct
  search sensitivity of the PandaX-4T experiment}},
  \href{http://dx.doi.org/10.1007/s11433-018-9259-0}{\emph{Sci. China Phys.
  Mech. Astron.} {\bfseries 62} (2019) 31011},
  [\href{https://arxiv.org/abs/1806.02229}{{\ttfamily 1806.02229}}].

\bibitem{Aprile:2018dbl}
{\scshape XENON} collaboration, E.~Aprile et~al., \emph{{Dark Matter Search
  Results from a One Ton-Year Exposure of XENON1T}},
  \href{http://dx.doi.org/10.1103/PhysRevLett.121.111302}{\emph{Phys. Rev.
  Lett.} {\bfseries 121} (2018) 111302},
  [\href{https://arxiv.org/abs/1805.12562}{{\ttfamily 1805.12562}}].

\bibitem{MAGIC:2016xys}
{\scshape MAGIC, Fermi-LAT} collaboration, M.~L. Ahnen et~al., \emph{{Limits to
  Dark Matter Annihilation Cross-Section from a Combined Analysis of MAGIC and
  Fermi-LAT Observations of Dwarf Satellite Galaxies}},
  \href{http://dx.doi.org/10.1088/1475-7516/2016/02/039}{\emph{JCAP} {\bfseries
  02} (2016) 039}, [\href{https://arxiv.org/abs/1601.06590}{{\ttfamily
  1601.06590}}].

\bibitem{Chatrchyan:2012xdj}
{\scshape CMS} collaboration, S.~Chatrchyan et~al., \emph{{Observation of a New
  Boson at a Mass of 125 GeV with the CMS Experiment at the LHC}},
  \href{http://dx.doi.org/10.1016/j.physletb.2012.08.021}{\emph{Phys. Lett.}
  {\bfseries B716} (2012) 30--61},
  [\href{https://arxiv.org/abs/1207.7235}{{\ttfamily 1207.7235}}].

\bibitem{Aad:2012tfa}
{\scshape ATLAS} collaboration, G.~Aad et~al., \emph{{Observation of a new
  particle in the search for the Standard Model Higgs boson with the ATLAS
  detector at the LHC}},
  \href{http://dx.doi.org/10.1016/j.physletb.2012.08.020}{\emph{Phys. Lett.}
  {\bfseries B716} (2012) 1--29},
  [\href{https://arxiv.org/abs/1207.7214}{{\ttfamily 1207.7214}}].

\bibitem{ATLAS:2020kdi}
{\scshape ATLAS} collaboration, \emph{{Combination of searches for invisible
  Higgs boson decays with the ATLAS experiment}}, .

\bibitem{Hall:2009bx}
L.~J. Hall, K.~Jedamzik, J.~March-Russell and S.~M. West, \emph{{Freeze-In
  Production of FIMP Dark Matter}},
  \href{http://dx.doi.org/10.1007/JHEP03(2010)080}{\emph{JHEP} {\bfseries 03}
  (2010) 080}, [\href{https://arxiv.org/abs/0911.1120}{{\ttfamily 0911.1120}}].

\bibitem{Co:2015pka}
R.~T. Co, F.~D'Eramo, L.~J. Hall and D.~Pappadopulo, \emph{{Freeze-In Dark
  Matter with Displaced Signatures at Colliders}},
  \href{http://dx.doi.org/10.1088/1475-7516/2015/12/024}{\emph{JCAP} {\bfseries
  12} (2015) 024}, [\href{https://arxiv.org/abs/1506.07532}{{\ttfamily
  1506.07532}}].

\bibitem{Hessler:2016kwm}
A.~G. Hessler, A.~Ibarra, E.~Molinaro and S.~Vogl, \emph{{Probing the
  scotogenic FIMP at the LHC}},
  \href{http://dx.doi.org/10.1007/JHEP01(2017)100}{\emph{JHEP} {\bfseries 01}
  (2017) 100}, [\href{https://arxiv.org/abs/1611.09540}{{\ttfamily
  1611.09540}}].

\bibitem{Ghosh:2017vhe}
A.~Ghosh, T.~Mondal and B.~Mukhopadhyaya, \emph{{Heavy stable charged tracks as
  signatures of non-thermal dark matter at the LHC : a study in some
  non-supersymmetric scenarios}},
  \href{http://dx.doi.org/10.1007/JHEP12(2017)136}{\emph{JHEP} {\bfseries 12}
  (2017) 136}, [\href{https://arxiv.org/abs/1706.06815}{{\ttfamily
  1706.06815}}].

\bibitem{No:2019gvl}
J.~M. No, P.~Tunney and B.~Zaldivar, \emph{{Probing Dark Matter freeze-in with
  long-lived particle signatures: MATHUSLA, HL-LHC and FCC-hh}},
  \href{http://dx.doi.org/10.1007/JHEP03(2020)022}{\emph{JHEP} {\bfseries 03}
  (2020) 022}, [\href{https://arxiv.org/abs/1908.11387}{{\ttfamily
  1908.11387}}].

\bibitem{Konar:2021oye}
P.~Konar, R.~Roshan and S.~Show, \emph{{Freeze-in dark matter through forbidden
  channel in U(1)B-L}},
  \href{http://dx.doi.org/10.1088/1475-7516/2022/03/021}{\emph{JCAP} {\bfseries
  03} (2022) 021}, [\href{https://arxiv.org/abs/2110.14411}{{\ttfamily
  2110.14411}}].

\bibitem{Ghosh:2021wrk}
P.~Ghosh, P.~Konar, A.~K. Saha and S.~Show, \emph{{Self-interacting freeze-in
  dark matter in a singlet doublet scenario}},
  \href{http://dx.doi.org/10.1088/1475-7516/2022/10/017}{\emph{JCAP} {\bfseries
  10} (2022) 017}, [\href{https://arxiv.org/abs/2112.09057}{{\ttfamily
  2112.09057}}].

\bibitem{DeRomeri:2020wng}
V.~De~Romeri, D.~Karamitros, O.~Lebedev and T.~Toma, \emph{{Neutrino dark
  matter and the Higgs portal: improved freeze-in analysis}},
  \href{http://dx.doi.org/10.1007/JHEP10(2020)137}{\emph{JHEP} {\bfseries 10}
  (2020) 137}, [\href{https://arxiv.org/abs/2003.12606}{{\ttfamily
  2003.12606}}].

\bibitem{Kim:2017mtc}
J.~Kim and J.~McDonald, \emph{{Clockwork Higgs portal model for freeze-in dark
  matter}}, \href{http://dx.doi.org/10.1103/PhysRevD.98.023533}{\emph{Phys.
  Rev. D} {\bfseries 98} (2018) 023533},
  [\href{https://arxiv.org/abs/1709.04105}{{\ttfamily 1709.04105}}].

\bibitem{Kim:2018xsp}
J.~Kim and J.~Mcdonald, \emph{{Freeze-In Dark Matter from a sub-Higgs Mass
  Clockwork Sector via the Higgs Portal}},
  \href{http://dx.doi.org/10.1103/PhysRevD.98.123503}{\emph{Phys. Rev. D}
  {\bfseries 98} (2018) 123503},
  [\href{https://arxiv.org/abs/1804.02661}{{\ttfamily 1804.02661}}].

\bibitem{Im:2019iwd}
S.~H. Im and K.~S. Jeong, \emph{{Freeze-in Axion-like Dark Matter}},
  \href{http://dx.doi.org/10.1016/j.physletb.2019.135044}{\emph{Phys. Lett. B}
  {\bfseries 799} (2019) 135044},
  [\href{https://arxiv.org/abs/1907.07383}{{\ttfamily 1907.07383}}].

\bibitem{Hambye:2018}
T.~Hambye, M.~H.~G. Tytgat, J.~Vandecasteele and L.~Vanderheyden, \emph{Dark
  matter direct detection is testing freeze-in},
  \href{http://dx.doi.org/10.1103/PhysRevD.98.075017}{\emph{Phys. Rev. D}
  {\bfseries 98} (Oct, 2018) 075017}.

\bibitem{Banerjee_2019}
A.~Banerjee, G.~Bhattacharyya, D.~Chowdhury and Y.~Mambrini, \emph{Dark matter
  seeping through dynamic gauge kinetic mixing},
  \href{http://dx.doi.org/10.1088/1475-7516/2019/12/009}{\emph{Journal of
  Cosmology and Astroparticle Physics} {\bfseries 2019} (dec, 2019) 009}.

\bibitem{Dutra:2018gmv}
M.~Dutra, M.~Lindner, S.~Profumo, F.~S. Queiroz, W.~Rodejohann and C.~Siqueira,
  \emph{{MeV Dark Matter Complementarity and the Dark Photon Portal}},
  \href{http://dx.doi.org/10.1088/1475-7516/2018/03/037}{\emph{JCAP} {\bfseries
  03} (2018) 037}, [\href{https://arxiv.org/abs/1801.05447}{{\ttfamily
  1801.05447}}].

\bibitem{Chu:2011be}
X.~Chu, T.~Hambye and M.~H.~G. Tytgat, \emph{{The Four Basic Ways of Creating
  Dark Matter Through a Portal}},
  \href{http://dx.doi.org/10.1088/1475-7516/2012/05/034}{\emph{JCAP} {\bfseries
  05} (2012) 034}, [\href{https://arxiv.org/abs/1112.0493}{{\ttfamily
  1112.0493}}].

\bibitem{Shakya:2015xnx}
B.~Shakya, \emph{{Sterile Neutrino Dark Matter from Freeze-In}},
  \href{http://dx.doi.org/10.1142/S0217732316300056}{\emph{Mod. Phys. Lett. A}
  {\bfseries 31} (2016) 1630005},
  [\href{https://arxiv.org/abs/1512.02751}{{\ttfamily 1512.02751}}].

\bibitem{Coy:2022unt}
R.~Coy and M.~A. Schmidt, \emph{{Freeze-in and freeze-out of sterile neutrino
  dark matter}},
  \href{http://dx.doi.org/10.1088/1475-7516/2022/08/070}{\emph{JCAP} {\bfseries
  08} (2022) 070}, [\href{https://arxiv.org/abs/2204.08795}{{\ttfamily
  2204.08795}}].

\bibitem{Chakrabarty:2022bcn}
N.~Chakrabarty, P.~Konar, R.~Roshan~and and S.~Show, \emph{{Thermally corrected
  masses and freeze-in dark matter: a case study}},
  \href{https://arxiv.org/abs/2206.02233}{{\ttfamily 2206.02233}}.

\bibitem{Cheung:2011mg}
C.~Cheung, G.~Elor and L.~J. Hall, \emph{{The Cosmological Axino Problem}},
  \href{http://dx.doi.org/10.1103/PhysRevD.85.015008}{\emph{Phys. Rev. D}
  {\bfseries 85} (2012) 015008},
  [\href{https://arxiv.org/abs/1104.0692}{{\ttfamily 1104.0692}}].

\bibitem{Medina:2014bga}
A.~D. Medina, \emph{{Higgsino-like Dark Matter From Sneutrino Late Decays}},
  \href{http://dx.doi.org/10.1016/j.physletb.2017.04.054}{\emph{Phys. Lett. B}
  {\bfseries 770} (2017) 161--165},
  [\href{https://arxiv.org/abs/1409.2560}{{\ttfamily 1409.2560}}].

\bibitem{Bernal:2017kxu}
N.~Bernal, M.~Heikinheimo, T.~Tenkanen, K.~Tuominen and V.~Vaskonen, \emph{{The
  Dawn of FIMP Dark Matter: A Review of Models and Constraints}},
  \href{http://dx.doi.org/10.1142/S0217751X1730023X}{\emph{Int. J. Mod. Phys.
  A} {\bfseries 32} (2017) 1730023},
  [\href{https://arxiv.org/abs/1706.07442}{{\ttfamily 1706.07442}}].

\bibitem{Chowdhury:2023jft}
D.~Chowdhury and A.~Hait, \emph{{Thermalization in the presence of a
  time-dependent dissipation and its impact on dark matter production}},
  \href{http://dx.doi.org/10.1007/JHEP09(2023)085}{\emph{JHEP} {\bfseries 09}
  (2023) 085}, [\href{https://arxiv.org/abs/2302.06654}{{\ttfamily
  2302.06654}}].

\bibitem{Banerjee:2024caa}
A.~Banerjee, D.~Chowdhury, A.~Hait and M.~S. Islam, \emph{{Dark matter cooling
  during early matter-domination boosts sub-earth halos}},
  \href{https://arxiv.org/abs/2408.08360}{{\ttfamily 2408.08360}}.

\bibitem{Barman:2020plp}
B.~Barman, D.~Borah and R.~Roshan, \emph{{Effective Theory of Freeze-in Dark
  Matter}}, \href{http://dx.doi.org/10.1088/1475-7516/2020/11/021}{\emph{JCAP}
  {\bfseries 11} (2020) 021},
  [\href{https://arxiv.org/abs/2007.08768}{{\ttfamily 2007.08768}}].

\bibitem{Barman:2022njh}
B.~Barman and A.~Ghoshal, \emph{{Probing pre-BBN era with scale invariant
  FIMP}}, \href{http://dx.doi.org/10.1088/1475-7516/2022/10/082}{\emph{JCAP}
  {\bfseries 10} (2022) 082},
  [\href{https://arxiv.org/abs/2203.13269}{{\ttfamily 2203.13269}}].

\bibitem{Barman:2024lxy}
B.~Barman, A.~Das and S.~Mandal, \emph{{Dark matter-electron scattering and
  freeze-in scenarios in the light of Z' mediation}},
  \href{http://dx.doi.org/10.1103/PhysRevD.110.055029}{\emph{Phys. Rev. D}
  {\bfseries 110} (2024) 055029},
  [\href{https://arxiv.org/abs/2407.00969}{{\ttfamily 2407.00969}}].

\bibitem{Freese:2024ogj}
K.~Freese, G.~Montefalcone and B.~Shams Es~Haghi, \emph{{Dark Matter Production
  during Warm Inflation via Freeze-In}},
  \href{http://dx.doi.org/10.1103/PhysRevLett.133.211001}{\emph{Phys. Rev.
  Lett.} {\bfseries 133} (2024) 211001},
  [\href{https://arxiv.org/abs/2401.17371}{{\ttfamily 2401.17371}}].

\bibitem{Sakurai:2024apm}
K.~Sakurai and W.~Yin, \emph{{Stimulated Emission of Dark Matter via Thermal
  Scattering: Novel Limits for Freeze-In and eV Cold Dark Matter}},
  \href{https://arxiv.org/abs/2410.18968}{{\ttfamily 2410.18968}}.

\bibitem{Barman:2024nhr}
B.~Barman, S.~Bhattacharya, S.~Jahedi, D.~Pradhan and A.~Sarkar, \emph{{Lepton
  Collider as a window to Reheating}},
  \href{https://arxiv.org/abs/2406.11963}{{\ttfamily 2406.11963}}.

\bibitem{Barman:2024tjt}
B.~Barman, S.~Bhattacharya, S.~Jahedi, D.~Pradhan and A.~Sarkar, \emph{{Lepton
  Collider as a window to Reheating: II}},
  \href{https://arxiv.org/abs/2410.18198}{{\ttfamily 2410.18198}}.

\bibitem{Das:2023owa}
P.~K. Das, P.~Konar, S.~Kundu and S.~Show, \emph{{Jet substructure probe to
  unfold singlet-doublet dark matter in the presence of non-standard
  cosmology}}, \href{http://dx.doi.org/10.1007/JHEP06(2023)198}{\emph{JHEP}
  {\bfseries 06} (2023) 198},
  [\href{https://arxiv.org/abs/2301.02514}{{\ttfamily 2301.02514}}].

\bibitem{DEramo:2017gpl}
F.~D'Eramo, N.~Fernandez and S.~Profumo, \emph{{When the Universe Expands Too
  Fast: Relentless Dark Matter}},
  \href{http://dx.doi.org/10.1088/1475-7516/2017/05/012}{\emph{JCAP} {\bfseries
  05} (2017) 012}, [\href{https://arxiv.org/abs/1703.04793}{{\ttfamily
  1703.04793}}].

\bibitem{DEramo:2017ecx}
F.~D'Eramo, N.~Fernandez and S.~Profumo, \emph{{Dark Matter Freeze-in
  Production in Fast-Expanding Universes}},
  \href{http://dx.doi.org/10.1088/1475-7516/2018/02/046}{\emph{JCAP} {\bfseries
  02} (2018) 046}, [\href{https://arxiv.org/abs/1712.07453}{{\ttfamily
  1712.07453}}].

\bibitem{Viel:2013fqw}
M.~Viel, G.~D. Becker, J.~S. Bolton and M.~G. Haehnelt, \emph{{Warm dark matter
  as a solution to the small scale crisis: New constraints from high redshift
  Lyman-\ensuremath{\alpha} forest data}},
  \href{http://dx.doi.org/10.1103/PhysRevD.88.043502}{\emph{Phys. Rev. D}
  {\bfseries 88} (2013) 043502},
  [\href{https://arxiv.org/abs/1306.2314}{{\ttfamily 1306.2314}}].

\bibitem{Yeche:2017upn}
C.~Y\`eche, N.~Palanque-Delabrouille, J.~Baur and H.~du~Mas~des Bourboux,
  \emph{{Constraints on neutrino masses from Lyman-alpha forest power spectrum
  with BOSS and XQ-100}},
  \href{http://dx.doi.org/10.1088/1475-7516/2017/06/047}{\emph{JCAP} {\bfseries
  06} (2017) 047}, [\href{https://arxiv.org/abs/1702.03314}{{\ttfamily
  1702.03314}}].

\bibitem{Irsic:2017ixq}
V.~Ir\v{s}i\v{c} et~al., \emph{{New Constraints on the free-streaming of warm
  dark matter from intermediate and small scale Lyman-$\alpha$ forest data}},
  \href{http://dx.doi.org/10.1103/PhysRevD.96.023522}{\emph{Phys. Rev. D}
  {\bfseries 96} (2017) 023522},
  [\href{https://arxiv.org/abs/1702.01764}{{\ttfamily 1702.01764}}].

\bibitem{Belanger:2018sti}
G.~B\'elanger et~al., \emph{{LHC-friendly minimal freeze-in models}},
  \href{http://dx.doi.org/10.1007/JHEP02(2019)186}{\emph{JHEP} {\bfseries 02}
  (2019) 186}, [\href{https://arxiv.org/abs/1811.05478}{{\ttfamily
  1811.05478}}].

\bibitem{Ghosh:2023ocz}
A.~Ghosh, P.~Konar, D.~Saha and S.~Seth, \emph{{Precise probing and
  discrimination of third-generation scalar leptoquarks}},
  \href{http://dx.doi.org/10.1103/PhysRevD.108.035030}{\emph{Phys. Rev. D}
  {\bfseries 108} (2023) 035030},
  [\href{https://arxiv.org/abs/2304.02890}{{\ttfamily 2304.02890}}].

\bibitem{Ghosh:2025gue}
A.~Ghosh, P.~Konar, T.~Samui and R.~K. Singh, \emph{{Jet substructure probe on
  scalar leptoquark models via top polarization}},
  \href{http://dx.doi.org/10.1007/JHEP07(2025)145}{\emph{JHEP} {\bfseries 07}
  (2025) 145}, [\href{https://arxiv.org/abs/2505.16328}{{\ttfamily
  2505.16328}}].

\bibitem{Alloul:2013bka}
A.~Alloul, N.~D. Christensen, C.~Degrande, C.~Duhr and B.~Fuks,
  \emph{{FeynRules 2.0 - A complete toolbox for tree-level phenomenology}},
  \href{http://dx.doi.org/10.1016/j.cpc.2014.04.012}{\emph{Comput. Phys.
  Commun.} {\bfseries 185} (2014) 2250--2300},
  [\href{https://arxiv.org/abs/1310.1921}{{\ttfamily 1310.1921}}].

\bibitem{Degrande:2014vpa}
C.~Degrande, \emph{{Automatic evaluation of UV and R2 terms for beyond the
  Standard Model Lagrangians: a proof-of-principle}},
  \href{http://dx.doi.org/10.1016/j.cpc.2015.08.015}{\emph{Comput. Phys.
  Commun.} {\bfseries 197} (2015) 239--262},
  [\href{https://arxiv.org/abs/1406.3030}{{\ttfamily 1406.3030}}].

\bibitem{Alwall:2014hca}
J.~Alwall, R.~Frederix, S.~Frixione, V.~Hirschi, F.~Maltoni, O.~Mattelaer
  et~al., \emph{{The automated computation of tree-level and next-to-leading
  order differential cross sections, and their matching to parton shower
  simulations}}, \href{http://dx.doi.org/10.1007/JHEP07(2014)079}{\emph{JHEP}
  {\bfseries 07} (2014) 079},
  [\href{https://arxiv.org/abs/1405.0301}{{\ttfamily 1405.0301}}].

\bibitem{Mangano:2006rw}
M.~L. Mangano, M.~Moretti, F.~Piccinini and M.~Treccani, \emph{{Matching matrix
  elements and shower evolution for top-quark production in hadronic
  collisions}},
  \href{http://dx.doi.org/10.1088/1126-6708/2007/01/013}{\emph{JHEP} {\bfseries
  01} (2007) 013}, [\href{https://arxiv.org/abs/hep-ph/0611129}{{\ttfamily
  hep-ph/0611129}}].

\bibitem{Hoeche:2005vzu}
S.~Hoeche, F.~Krauss, N.~Lavesson, L.~Lonnblad, M.~Mangano, A.~Schalicke
  et~al., \emph{{Matching parton showers and matrix elements}},  in \emph{{HERA
  and the LHC: A Workshop on the Implications of HERA for LHC Physics: CERN -
  DESY Workshop 2004/2005 (Midterm Meeting, CERN, 11-13 October 2004; Final
  Meeting, DESY, 17-21 January 2005)}}, pp.~288--289, 2005.
\newblock \href{https://arxiv.org/abs/hep-ph/0602031}{{\ttfamily
  hep-ph/0602031}}.
\newblock \href{http://dx.doi.org/10.5170/CERN-2005-014.288}{DOI}.

\bibitem{Sjostrand:2001yu}
T.~Sjostrand, L.~Lonnblad and S.~Mrenna, \emph{{PYTHIA 6.2: Physics and
  manual}},  \href{https://arxiv.org/abs/hep-ph/0108264}{{\ttfamily
  hep-ph/0108264}}.

\bibitem{Sjostrand:2014zea}
T.~Sj\"ostrand, S.~Ask, J.~R. Christiansen, R.~Corke, N.~Desai, P.~Ilten
  et~al., \emph{{An introduction to PYTHIA 8.2}},
  \href{http://dx.doi.org/10.1016/j.cpc.2015.01.024}{\emph{Comput. Phys.
  Commun.} {\bfseries 191} (2015) 159--177},
  [\href{https://arxiv.org/abs/1410.3012}{{\ttfamily 1410.3012}}].

\bibitem{NNPDF:2014otw}
{\scshape NNPDF} collaboration, R.~D. Ball et~al., \emph{{Parton distributions
  for the LHC Run II}},
  \href{http://dx.doi.org/10.1007/JHEP04(2015)040}{\emph{JHEP} {\bfseries 04}
  (2015) 040}, [\href{https://arxiv.org/abs/1410.8849}{{\ttfamily 1410.8849}}].

\bibitem{deFavereau:2013fsa}
{\scshape DELPHES 3} collaboration, J.~de~Favereau, C.~Delaere, P.~Demin,
  A.~Giammanco, V.~Lema\^\i{}tre, A.~Mertens et~al., \emph{{DELPHES 3, A
  modular framework for fast simulation of a generic collider experiment}},
  \href{http://dx.doi.org/10.1007/JHEP02(2014)057}{\emph{JHEP} {\bfseries 02}
  (2014) 057}, [\href{https://arxiv.org/abs/1307.6346}{{\ttfamily 1307.6346}}].

\bibitem{Cacciari:2008gp}
M.~Cacciari, G.~P. Salam and G.~Soyez, \emph{{The anti-$k_t$ jet clustering
  algorithm}},
  \href{http://dx.doi.org/10.1088/1126-6708/2008/04/063}{\emph{JHEP} {\bfseries
  04} (2008) 063}, [\href{https://arxiv.org/abs/0802.1189}{{\ttfamily
  0802.1189}}].

\bibitem{Cacciari:2011ma}
M.~Cacciari, G.~P. Salam and G.~Soyez, \emph{{FastJet User Manual}},
  \href{http://dx.doi.org/10.1140/epjc/s10052-012-1896-2}{\emph{Eur. Phys. J.
  C} {\bfseries 72} (2012) 1896},
  [\href{https://arxiv.org/abs/1111.6097}{{\ttfamily 1111.6097}}].

\bibitem{Dokshitzer:1997in}
Y.~L. Dokshitzer, G.~D. Leder, S.~Moretti and B.~R. Webber, \emph{{Better jet
  clustering algorithms}},
  \href{http://dx.doi.org/10.1088/1126-6708/1997/08/001}{\emph{JHEP} {\bfseries
  08} (1997) 001}, [\href{https://arxiv.org/abs/hep-ph/9707323}{{\ttfamily
  hep-ph/9707323}}].

\bibitem{Larkoski:2014wba}
A.~J. Larkoski, S.~Marzani, G.~Soyez and J.~Thaler, \emph{{Soft Drop}},
  \href{http://dx.doi.org/10.1007/JHEP05(2014)146}{\emph{JHEP} {\bfseries 05}
  (2014) 146}, [\href{https://arxiv.org/abs/1402.2657}{{\ttfamily 1402.2657}}].

\bibitem{Roe:2004na}
B.~P. Roe, H.-J. Yang, J.~Zhu, Y.~Liu, I.~Stancu and G.~McGregor,
  \emph{{Boosted decision trees, an alternative to artificial neural
  networks}}, \href{http://dx.doi.org/10.1016/j.nima.2004.12.018}{\emph{Nucl.
  Instrum. Meth. A} {\bfseries 543} (2005) 577--584},
  [\href{https://arxiv.org/abs/physics/0408124}{{\ttfamily physics/0408124}}].

\bibitem{FREUND1995256}
Y.~Freund, \emph{Boosting a weak learning algorithm by majority},
  \href{http://dx.doi.org/https://doi.org/10.1006/inco.1995.1136}{\emph{Information
  and Computation} {\bfseries 121} (1995) 256--285}.

\bibitem{Freund:1997xna}
Y.~Freund and R.~E. Schapire, \emph{{A Decision-Theoretic Generalization of
  On-Line Learning and an Application to Boosting}},
  \href{http://dx.doi.org/10.1006/jcss.1997.1504}{\emph{J. Comput. Syst. Sci.}
  {\bfseries 55} (1997) 119--139}.

\bibitem{Hocker:2007ht}
A.~Hocker et~al., \emph{{TMVA - Toolkit for Multivariate Data Analysis}},
  \href{https://arxiv.org/abs/physics/0703039}{{\ttfamily physics/0703039}}.

\bibitem{Muselli:2015kba}
C.~Muselli, M.~Bonvini, S.~Forte, S.~Marzani and G.~Ridolfi, \emph{{Top Quark
  Pair Production beyond NNLO}},
  \href{http://dx.doi.org/10.1007/JHEP08(2015)076}{\emph{JHEP} {\bfseries 08}
  (2015) 076}, [\href{https://arxiv.org/abs/1505.02006}{{\ttfamily
  1505.02006}}].

\bibitem{Kidonakis:2015nna}
N.~Kidonakis, \emph{{Theoretical results for electroweak-boson and single-top
  production}}, \href{http://dx.doi.org/10.22323/1.247.0170}{\emph{PoS}
  {\bfseries DIS2015} (2015) 170},
  [\href{https://arxiv.org/abs/1506.04072}{{\ttfamily 1506.04072}}].

\bibitem{Campbell:2011bn}
J.~M. Campbell, R.~K. Ellis and C.~Williams, \emph{{Vector boson pair
  production at the LHC}},
  \href{http://dx.doi.org/10.1007/JHEP07(2011)018}{\emph{JHEP} {\bfseries 07}
  (2011) 018}, [\href{https://arxiv.org/abs/1105.0020}{{\ttfamily 1105.0020}}].

\bibitem{Catani:2009sm}
S.~Catani, L.~Cieri, G.~Ferrera, D.~de~Florian and M.~Grazzini, \emph{{Vector
  boson production at hadron colliders: a fully exclusive QCD calculation at
  NNLO}}, \href{http://dx.doi.org/10.1103/PhysRevLett.103.082001}{\emph{Phys.
  Rev. Lett.} {\bfseries 103} (2009) 082001},
  [\href{https://arxiv.org/abs/0903.2120}{{\ttfamily 0903.2120}}].

\bibitem{Balossini:2009sa}
G.~Balossini, G.~Montagna, C.~M. Carloni~Calame, M.~Moretti, O.~Nicrosini,
  F.~Piccinini et~al., \emph{{Combination of electroweak and QCD corrections to
  single W production at the Fermilab Tevatron and the CERN LHC}},
  \href{http://dx.doi.org/10.1007/JHEP01(2010)013}{\emph{JHEP} {\bfseries 01}
  (2010) 013}, [\href{https://arxiv.org/abs/0907.0276}{{\ttfamily 0907.0276}}].

\bibitem{Thaler:2010tr}
J.~Thaler and K.~Van~Tilburg, \emph{{Identifying Boosted Objects with
  N-subjettiness}},
  \href{http://dx.doi.org/10.1007/JHEP03(2011)015}{\emph{JHEP} {\bfseries 03}
  (2011) 015}, [\href{https://arxiv.org/abs/1011.2268}{{\ttfamily 1011.2268}}].

\bibitem{Lloyd:1982zni}
S.~Lloyd, \emph{{Least squares quantization in PCM}},
  \href{http://dx.doi.org/10.1109/TIT.1982.1056489}{\emph{IEEE Trans. Info.
  Theor.} {\bfseries 28} (1982) 129--137}.

\bibitem{Larkoski:2014uqa}
A.~J. Larkoski, D.~Neill and J.~Thaler, \emph{{Jet Shapes with the Broadening
  Axis}}, \href{http://dx.doi.org/10.1007/JHEP04(2014)017}{\emph{JHEP}
  {\bfseries 04} (2014) 017},
  [\href{https://arxiv.org/abs/1401.2158}{{\ttfamily 1401.2158}}].

\bibitem{Bertolini:2013iqa}
D.~Bertolini, T.~Chan and J.~Thaler, \emph{{Jet Observables Without Jet
  Algorithms}}, \href{http://dx.doi.org/10.1007/JHEP04(2014)013}{\emph{JHEP}
  {\bfseries 04} (2014) 013},
  [\href{https://arxiv.org/abs/1310.7584}{{\ttfamily 1310.7584}}].

\bibitem{Stewart:2015waa}
I.~W. Stewart, F.~J. Tackmann, J.~Thaler, C.~K. Vermilion and T.~F. Wilkason,
  \emph{{XCone: N-jettiness as an Exclusive Cone Jet Algorithm}},
  \href{http://dx.doi.org/10.1007/JHEP11(2015)072}{\emph{JHEP} {\bfseries 11}
  (2015) 072}, [\href{https://arxiv.org/abs/1508.01516}{{\ttfamily
  1508.01516}}].

\bibitem{Thaler:2011gf}
J.~Thaler and K.~Van~Tilburg, \emph{{Maximizing Boosted Top Identification by
  Minimizing N-subjettiness}},
  \href{http://dx.doi.org/10.1007/JHEP02(2012)093}{\emph{JHEP} {\bfseries 02}
  (2012) 093}, [\href{https://arxiv.org/abs/1108.2701}{{\ttfamily 1108.2701}}].

\bibitem{CMS:2012feb}
{\scshape CMS} collaboration, S.~Chatrchyan et~al., \emph{{Identification of
  b-Quark Jets with the CMS Experiment}},
  \href{http://dx.doi.org/10.1088/1748-0221/8/04/P04013}{\emph{JINST}
  {\bfseries 8} (2013) P04013},
  [\href{https://arxiv.org/abs/1211.4462}{{\ttfamily 1211.4462}}].

\bibitem{Calibbi:2018fqf}
L.~Calibbi, L.~Lopez-Honorez, S.~Lowette and A.~Mariotti,
  \emph{{Singlet-Doublet Dark Matter Freeze-in: LHC displaced signatures versus
  cosmology}}, \href{http://dx.doi.org/10.1007/JHEP09(2018)037}{\emph{JHEP}
  {\bfseries 09} (2018) 037},
  [\href{https://arxiv.org/abs/1805.04423}{{\ttfamily 1805.04423}}].

\bibitem{ATLAS:2020xzu}
{\scshape ATLAS} collaboration, G.~Aad et~al., \emph{{Search for new phenomena
  with top quark pairs in final states with one lepton, jets, and missing
  transverse momentum in $pp$ collisions at $ \sqrt{s} $ = 13 TeV with the
  ATLAS detector}},
  \href{http://dx.doi.org/10.1007/JHEP04(2021)174}{\emph{JHEP} {\bfseries 04}
  (2021) 174}, [\href{https://arxiv.org/abs/2012.03799}{{\ttfamily
  2012.03799}}].

\bibitem{ATLAS:2024rcx}
{\scshape ATLAS} collaboration, G.~Aad et~al., \emph{{Search for new phenomena
  with top-quark pairs and large missing transverse momentum using 140
  fb$^{−1}$ of pp collision data at $ \sqrt{s} $ = 13 TeV with the ATLAS
  detector}}, \href{http://dx.doi.org/10.1007/JHEP03(2024)139}{\emph{JHEP}
  {\bfseries 03} (2024) 139},
  [\href{https://arxiv.org/abs/2401.13430}{{\ttfamily 2401.13430}}].

\bibitem{CMS:2019ysk}
{\scshape CMS} collaboration, A.~M. Sirunyan et~al., \emph{{Search for direct
  top squark pair production in events with one lepton, jets, and missing
  transverse momentum at 13 TeV with the CMS experiment}},
  \href{http://dx.doi.org/10.1007/JHEP05(2020)032}{\emph{JHEP} {\bfseries 05}
  (2020) 032}, [\href{https://arxiv.org/abs/1912.08887}{{\ttfamily
  1912.08887}}].

\bibitem{CMS:2020pyk}
{\scshape CMS} collaboration, A.~M. Sirunyan et~al., \emph{{Search for top
  squark pair production using dilepton final states in ${\text {p}}{\text
  {p}}$ collision data collected at $\sqrt{s}=13\,\text {TeV} $}},
  \href{http://dx.doi.org/10.1140/epjc/s10052-020-08701-5}{\emph{Eur. Phys. J.
  C} {\bfseries 81} (2021) 3},
  [\href{https://arxiv.org/abs/2008.05936}{{\ttfamily 2008.05936}}].

\bibitem{CMS:2021eha}
{\scshape CMS} collaboration, A.~Tumasyan et~al., \emph{{Combined searches for
  the production of supersymmetric top quark partners in
  proton\textendash{}proton collisions at $\sqrt{s} = 13\,\text {Te}\text {V}
  $}}, \href{http://dx.doi.org/10.1140/epjc/s10052-021-09721-5}{\emph{Eur.
  Phys. J. C} {\bfseries 81} (2021) 970},
  [\href{https://arxiv.org/abs/2107.10892}{{\ttfamily 2107.10892}}].

\bibitem{ATLAS:2020dsf}
{\scshape ATLAS} collaboration, G.~Aad et~al., \emph{{Search for a scalar
  partner of the top quark in the all-hadronic $t{\bar{t}}$ plus missing
  transverse momentum final state at $\sqrt{s}=13$ TeV with the ATLAS
  detector}},
  \href{http://dx.doi.org/10.1140/epjc/s10052-020-8102-8}{\emph{Eur. Phys. J.
  C} {\bfseries 80} (2020) 737},
  [\href{https://arxiv.org/abs/2004.14060}{{\ttfamily 2004.14060}}].

\bibitem{CMS:2021beq}
{\scshape CMS} collaboration, A.~M. Sirunyan et~al., \emph{{Search for top
  squark production in fully-hadronic final states in proton-proton collisions
  at $\sqrt{s} =$ 13 TeV}},
  \href{http://dx.doi.org/10.1103/PhysRevD.104.052001}{\emph{Phys. Rev. D}
  {\bfseries 104} (2021) 052001},
  [\href{https://arxiv.org/abs/2103.01290}{{\ttfamily 2103.01290}}].

\bibitem{Buchkremer:2013bha}
M.~Buchkremer, G.~Cacciapaglia, A.~Deandrea and L.~Panizzi, \emph{{Model
  Independent Framework for Searches of Top Partners}},
  \href{http://dx.doi.org/10.1016/j.nuclphysb.2013.08.010}{\emph{Nucl. Phys. B}
  {\bfseries 876} (2013) 376--417},
  [\href{https://arxiv.org/abs/1305.4172}{{\ttfamily 1305.4172}}].

\bibitem{Ghosh:2025gdq}
A.~Ghosh, S.~Ghosh, S.~Mitra, T.~Samui and R.~K. Singh, \emph{{Improving
  Sensitivity of Vector-like Top Partner Searches with Jet Substructure}},
  \href{https://arxiv.org/abs/2507.03199}{{\ttfamily 2507.03199}}.

\bibitem{ATLAS:2024xne}
{\scshape ATLAS} collaboration, G.~Aad et~al., \emph{{Search for new particles
  in final states with a boosted top quark and missing transverse momentum in
  proton-proton collisions at $ \sqrt{s} $ = 13 TeV with the ATLAS detector}},
  \href{http://dx.doi.org/10.1007/JHEP05(2024)263}{\emph{JHEP} {\bfseries 05}
  (2024) 263}, [\href{https://arxiv.org/abs/2402.16561}{{\ttfamily
  2402.16561}}].

\bibitem{CMS:2023agg}
{\scshape CMS} collaboration, A.~Tumasyan et~al., \emph{{Search for a
  vector-like quark T$'$$\to$ tH via the diphoton decay mode of the Higgs boson
  in proton-proton collisions at $\sqrt{s}$ = 13 TeV}},
  \href{http://dx.doi.org/10.1007/JHEP09(2023)057}{\emph{JHEP} {\bfseries 09}
  (2023) 057}, [\href{https://arxiv.org/abs/2302.12802}{{\ttfamily
  2302.12802}}].

\bibitem{CMS:2024qdd}
{\scshape CMS} collaboration, A.~Hayrapetyan et~al., \emph{{Search for
  production of a single vectorlike quark decaying to tH or tZ in the
  all-hadronic final state in pp collisions at s=13\,\,TeV}},
  \href{http://dx.doi.org/10.1103/PhysRevD.110.072012}{\emph{Phys. Rev. D}
  {\bfseries 110} (2024) 072012},
  [\href{https://arxiv.org/abs/2405.05071}{{\ttfamily 2405.05071}}].

\bibitem{Drees:2013wra}
M.~Drees, H.~Dreiner, D.~Schmeier, J.~Tattersall and J.~S. Kim,
  \emph{{CheckMATE: Confronting your Favourite New Physics Model with LHC
  Data}}, \href{http://dx.doi.org/10.1016/j.cpc.2014.10.018}{\emph{Comput.
  Phys. Commun.} {\bfseries 187} (2015) 227--265},
  [\href{https://arxiv.org/abs/1312.2591}{{\ttfamily 1312.2591}}].

\bibitem{Kim:2015wza}
J.~S. Kim, D.~Schmeier, J.~Tattersall and K.~Rolbiecki, \emph{{A framework to
  create customised LHC analyses within CheckMATE}},
  \href{http://dx.doi.org/10.1016/j.cpc.2015.06.002}{\emph{Comput. Phys.
  Commun.} {\bfseries 196} (2015) 535--562},
  [\href{https://arxiv.org/abs/1503.01123}{{\ttfamily 1503.01123}}].

\bibitem{Dercks:2016npn}
D.~Dercks, N.~Desai, J.~S. Kim, K.~Rolbiecki, J.~Tattersall and T.~Weber,
  \emph{{CheckMATE 2: From the model to the limit}},
  \href{http://dx.doi.org/10.1016/j.cpc.2017.08.021}{\emph{Comput. Phys.
  Commun.} {\bfseries 221} (2017) 383--418},
  [\href{https://arxiv.org/abs/1611.09856}{{\ttfamily 1611.09856}}].

\end{thebibliography}\endgroup
\end{document}